\documentclass[aps,twocolumn,amsmath,amssymb,preprintnumbers]{revtex4}
\usepackage{amsmath} \usepackage{amsfonts} \usepackage{amssymb}
\usepackage{titlesec}
\usepackage{bbm}
\usepackage{epsfig}
\usepackage{graphics}
\usepackage{graphicx}
\newcommand{\be}{\begin{equation}}
\newcommand{\ee}{\end{equation}}
\newcommand{\ba}{\begin{eqnarray}}
\newcommand{\ea}{\end{eqnarray}}
\newcommand{\nn}{\nonumber}
\newcommand{\kr}{\rangle}
\newcommand{\kl}{\langle}
\newcommand{\ri}{_{\rm in}}
\newcommand{\ro}{_{\rm out}}
\newcommand{\mn}{_{\mu\nu}}
\newcommand{\z}{z_{\rm max}}

\titleformat{\subsection}[block]{\normalfont\bfseries}{\thesubsection.}{1ex}{}
\titlespacing{\subsection}{0pt}{10pt}{1pt}[0pt]
\titleformat*{\section}{\large\bfseries}
\renewcommand{\thesubsection}{\arabic{subsection}}

\begin{document}

\title[ ]{Inflation, quintessence, and the origin of mass}

\author{C. Wetterich}
\affiliation{Institut  f\"ur Theoretische Physik\\
Universit\"at Heidelberg\\
Philosophenweg 16, D-69120 Heidelberg}

\begin{abstract}
In a unified picture both inflation and present dynamical dark energy arise from the same scalar field. The history of the Universe describes a crossover from a scale invariant ``past fixed point'' where all particles are massless, to a ``future fixed point'' for which spontaneous breaking of the exact scale symmetry generates the particle masses. The cosmological solution can be extrapolated to the infinite past in physical time - the universe has no beginning. This is seen most easily in a frame where particle masses and the Planck mass are field-dependent and increase with time. In this ``freeze frame'' the Universe shrinks and heats up during radiation and matter domination. In the equivalent, but singular Einstein frame cosmic history finds the familiar big bang description. The vicinity of the past fixed point corresponds to inflation. It ends at a first stage of the crossover. A simple model with no more free parameters than $\Lambda$CDM predicts for the primordial fluctuations a relation between the 
tensor amplitude $r$ and the spectral index $n,r=8.19(1-n)-0.137$. The crossover is completed by a second stage where the beyond-standard-model sector undergoes the transition to the future fixed point. The resulting increase of neutrino masses stops a cosmological scaling solution, relating the present dark energy density to the present neutrino mass. At present our simple model seems compatible with all observational tests. We discuss how the fixed points can be rooted within quantum gravity in a crossover between ultraviolet and infrared fixed points. Then quantum properties of gravity could be tested both by very early and late cosmology.

\end{abstract}

\maketitle

\section{Introduction}
\label{Introduction}

A scalar field plays a dominant role both for inflation in primordial cosmology and dynamical dark energy in the present epoch. The potential of this field constitutes primordial or late dark energy, driving an accelerated expansion in the big bang picture. Quintessential inflation \cite{PV,BM} identifies the inflaton field for inflation with the scalar field of quintessence or cosmon which is responsible for present dynamical dark energy. In particular, cosmon inflation \cite{CI} formulates this unification in the context of variable gravity \cite{VG}, where the strength of gravity depends on the value of the cosmon field. 

Both inflation and quintessence can be closely related to approximate dilatation or scale symmetry. For inflation this symmetry is at the origin of the observed approximate scale invariance of the spectrum of primordial fluctuations. For present dynamical dark energy the cosmon plays the role of the pseudo Goldstone boson of spontaneously broken dilatation symmetry \cite{CW3}. In case of exact dilatation symmetry it would be an exactly massless dilaton, while a tiny mass and potential are generated by a ``scale symmetry violation'' or ``dilatation anomaly''. Scale symmetry is intimately related to fixed points of ``running'' dimensionless couplings or mass ratios. At a fixed point any information about intrinsic mass or length scales is lost. Quantum scale symmetry is then realized even if the underlying quantum field theory is not scale invariant. 

The presence of approximate scale symmetry both for the primordial and late cosmology suggests that the infinite past and infinite future of the universe correspond to fixed points. We propose here that the two fixed points have different properties. For the fixed point in the infinite past scale symmetry is not spontaneously broken. All masses vanish. In contrast, the fixed point that will be approached in the infinite future is characterized by spontaneous symmetry breaking of dilatation symmetry, resulting in a spectrum of massive particles and a massless dilaton. 

The way how scale symmetry is realized and explicitly or spontaneously broken is directly related to the basic origin of mass. All particle masses are generated either by explicit or spontaneous breaking of scale symmetry. The explicit breaking by an intrinsic mass scale plays a crucial role for the crossover between the two fixed points. It is responsible for scale violation in the primordial fluctuation spectrum and for the end of inflation. Spontaneous breaking characterizes the ``future fixed point'' and our present universe. The presently observed particle masses are dominated by spontaneous scale symmetry breaking, while dark energy reflects the tiny explicit breaking. The basic mechanisms that generate the particle masses thus provide the physical ``raison d'\^{e}tre'' for inflation and late dark energy, such that these key cosmological ingredients appear less ``ad hoc''. 

This work is motivated by a central assumption about the properties of quantum gravity that we call ``crossover gravity''. The running of dimensionless couplings or mass ratios as a function of some intrinsic mass scale $\mu$ is assumed to exhibit two fixed points for $\mu\to\infty$ and $\mu\to 0$, with a crossover between the fixed points for finite nonzero $\mu$. Dimensionless functions can only depend on dimensionless quantities. If time and space gradients or momenta are proportional to $\mu$ (or can be neglected), the renormalized dimensionless functions can still depend on the ratio $\mu/\chi$, with $\chi$ the value of the scalar cosmon field which equals the variable Planck mass in our normalization. The ultraviolet (UV) field point is realized for $\mu\to\infty$ at fixed $\chi$ or $\chi\to 0$ at fixed $\mu$. Indeed, with all particle masses proportional to $\chi$ this fixed point realizes unbroken scale symmetry. All excitations are massless. The infrared (IR) fixed point occurs for $\mu\to 0$ or $\
chi\to\infty$. A nonvanishing value of $\chi$ spontaneously breaks scale symmetry. We will see that the cosmological solutions of our model realize an evolution where $\chi$ vanishes in the infinite past and diverges in the infinite future. The cosmological evolution therefore interpolates between the UV-fixed point in the past and the IR-fixed point in the future. 

Inflation describes the vicinity of the past fixed point. It can extend to the infinite past in physical time. The inflationary epoch has to end, however. ''Late cosmology'' comprises epochs of radiation-, matter- and dark energy domination. It is characterized by the approach towards the future fixed point. We will describe the transition from inflation to late cosmology as a first stage of the crossover between the two fixed points. In the crossover region couplings have to run from one fixed point to the other. Scale symmetry is therefore necessarily violated in the crossover region. This is the basic reason for the qualitative change in the dynamics of the cosmon that occurs at the end of inflation. 

If there is more than one relevant or marginal deviation from the ``past fixed point'' the crossover may occur in different stages. In case of a slow running (e.g. logarithmic) the scales associated to these stages can be separated by many orders of magnitude. We assume here that in the beyond standard model sector of particle physics the crossover is completed only in a second stage. This sector influences the masses of the neutrinos by ``non-renormalizable operators'' according to the see-saw or cascade mechanism. While the mass ratios of all particles except for neutrinos reach fixed values already at the end of inflation, the ratio of neutrino mass to electron mass makes the transition to the future constant value only in the present epoch. The relative increase of the neutrino masses realizes ``growing neutrino quintessence'' \cite{ABW,CWNEU} and explains the ``why now problem'' by relating the present dark energy density to the present neutrino mass. 

The history of dark energy reflects the two stages of the crossover. A primordial scaling solution corresponds to dominant dark energy during inflation. The first stage of crossover ends this scaling solution, triggering a transition to a different scaling solution during the radiation and matter dominated epochs. As a consequence of this scaling solution dark energy decreases proportional to the dominant radiation or matter component \cite{CW3}, constituting a small fraction of ``early dark energy''. Neutrinos are relativistic during this epoch and their masses play no role. The second stage of the crossover takes place in the present cosmological epoch. A substantial increase of the neutrino masses ends the second scaling solution once neutrinos become non-relativistic. This cosmic ``trigger event'' has happened around redshift $z\approx 5$, inducing a transition epoch with dominant dark energy and accelerated expansion. Once the second stage of the crossover is completed, the ratio between neutrino and 
electron 
mass approaches a constant value according to the future fixed point. Cosmology in the far future may correspond to a new scaling solution for which dark energy needs not to remain dominant.

A crossover in two steps can be associated with a flow trajectory in the vicinity of an intermediate (approximate) fixed point. We may refer to this fixed point as the ``standard model fixed point''. For this fixed point the renormalizable dimensionless couplings of all particles are the ones observed in present experiments.~ Neutrino masses,~ however, are typically substantially smaller than their present value. The standard model fixed point may be unstable in the sector of heavy particles with masses much larger than the Fermi scale, or in a sector of standard model singlets coupled only very weakly to the particles of the standard model. Such an instability will finally drive the flow trajectory away from the standard model (SM) fixed point and towards the infrared fixed point. 

On the other hand, the zeros of the $\beta$-functions for the renormalizable couplings of the standard model are stable for decreasing $\mu$, such that the presently measured values hold to high accuracy for the entire matter and 
radiation dominated epochs. The second step of the crossover affects first only the neutrino masses. Nevertheless, when the second step of the crossover will be completed in the far future, it is possible that the changes in the beyond standard model sector also affect the renormalizable couplings of the standard model. Their values at the infrared fixed point could be different from the present ones. 

\begin{figure}[h!tb]
\centering
\includegraphics[scale=1.2]{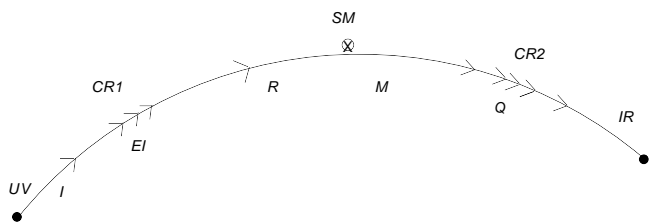}
\caption{Schematic view of the crossover from the UV-fixed point (UV) to the infrared fixed point (IR). Arrows indicate the direction of decreasing $\mu$ or increasing $\chi$. This direction corresponds to the flow of cosmic time from the infinite past (UV) to the infinite future (IR). The crossover trajectory passes near an (approximate) fixed point (SM) that characterizes the present standard model. The two regimes of fast changes, CR1 and CR2, correspond to the two steps of the crossover. We also indicate the corresponding cosmological epochs: inflation (I), end of inflation (EI), radiation domination (R), matter domination (M), dark energy domination (Q). }
\label{IQOM1}
\end{figure}

We have depicted the flow trajectory in some abstract ``coupling space'' or ``theory space'' in Fig. \ref{IQOM1}. It shows the first stage of the crossover from the UV-fixed point to the vicinity of the standard model fixed point, and the subsequent second step of the crossover to the infrared fixed point. We can associate the different cosmological epochs to the corresponding parts of the flow trajectory.

This paper is organized as follows: In sect. \ref{Fixed points and crossover} we introduce the flow equations underlying our approach. They describe the change of couplings as an intrinsic overall mass scale $\mu$ is varied. We discuss the properties of the ultraviolet and infrared fixed points. For this purpose we choose a frame of variable gravity where the crossover is described by the flow equation for the ``kinetial'', e.g. the coefficient of the scalar kinetic term. In particular, we investigate settings where the kinetial diverges at the ultraviolet fixed point with a large anomalous dimension.

In sect. \ref{Primordial cosmology and inflation} we turn to the cosmological solution for values of the cosmon field $\chi$ close to the ultraviolet fixed point and the first step of crossover away from it. It describes an epoch of inflation and its end. We compute the properties of the primordial density fluctuations. Both the spectral index $n$ and the tensor to scalar ratio $r$ are determined by the anomalous dimension $\sigma$ and therefore related, $1-n=r(2+\sigma)/16$. Computing $r$ and $n$ in terms of $\sigma$ and the number of $e$-foldings $N$ between horizon crossing of the observable fluctuations and the end of inflation, we establish the relation $r=8.19(1-n)-0.137$. The crossover provides for a natural explanation of the small amplitude of primordial fluctuations. This amplitude is suppressed by the ratio of the intrinsic mass scale $\mu$ over the crossover scale $m$, which is exponentially small due to the slow running near the fixed point. 

Sect. \ref{Late cosmology and dark energy} discusses ``late cosmology'' after the end of inflation. It starts with a scaling solution for the 
radiation and matter dominated epochs that is characterized by a small almost constant fraction of early dark energy \cite{CW3,CW2,EDE,DR}. This scaling explains why the present dark energy density is of the same order as the present matter energy density. In particular, we discuss models where the infrared fixed point corresponds to a ``conformal kinetic term''. The deviation from the fixed point is characterized by a function $B(\chi/\mu)$ that decreases with an inverse logarithm for large $\chi/\mu,B^{-1}=\kappa\ln(\chi/\mu)$. The fraction in early dark energy is proportional to $B$ and therefore naturally small for the large values of $\chi$ relevant for late cosmology. The slow flow of $B$ induces small scaling violations for the cosmological solution that we discuss in terms of an approximate analytic solution. We find a low value for the dark energy fraction at last scattering, close to the observational bounds. As the second step of the crossover sets in the neutrino masses start to increase 
substantially. Once neutrinos become non-relativistic they stop the scaling solution, ``freezing'' the dark energy density at the value it has reached at this moment. This leads to a phenomenology very close to a cosmological constant, with a value determined by the present average neutrino mass. Such a scenario solves the ``why now?'' problem. 

In sect. \ref{Past fixed point} we turn more closely to the particle physics aspects of the ultraviolet fixed point. For an anomalous dimension in the range $\sigma>1$ the couplings of the renormalized cosmon field are asymptotically free. The assumed fixed point that would provide for non-perturbative renormalizability (asymptotic safety) of quantum gravity has then the simple structure of a massless renormalized scalar field (with standard kinetic term) coupled to fourth order gravity. The non-perturbative character is related to anomalous dimensions for deviations from this fixed point that are of the order one. We argue that large anomalous dimensions can lead to a natural explanation of the small ratio Fermi scale/Planck scale and therefore provide for a possible solution of the gauge hierarchy problem. The gauge hierarchy and the small amplitude of primordial fluctuations could become related. 

Sect. \ref{Field relativity} describes the ultraviolet fixed point in different frames (different choices of field variables for metric and scalar field). We show that our ansatz with a simple quadratic cosmon potential and crossover described by the kinetial can be obtained by field transformations from a very large class of variable gravity models. It is therefore rather generic. The description of the crossover in terms of the kinetial is a convenience rather than a fundamental feature. We also discuss the asymptotic solutions for the infinite past in an equivalent ``primordial flat frame''. In this frame cosmology approaches flat space in the infinite past and the eternity of the universe is particularly apparent. Our conclusions are presented in sect. \ref{Conclusions}.

Several parts of a more detailed discussion are displayed in a series of appendices. In appendix A we discuss the status of the flow equations in view of a future quantum gravity computation, e.g.~within functional renormalization. We relate the $\mu$-dependence of couplings in the quantum effective action to the scaling solutions for the effective average action. We discuss the appearance of relevant parameters at the UV-fixed point as free integration constants in the scaling solution. In appendix B we illustrate the crossover between two fixed points in the flow of dimensionless couplings or mass ratios. We discuss the time variation of couplings in the standard model and neutrino masses. 

Appendix C contains the field equations derived from the quantum effective action of crossover gravity. We include higher order curvature invariants for the discussion of asymptotic solutions extending to the infinite past. Numerical solutions show the approach of a large class of solutions towards a family of scaling solutions. Some of the solutions correspond in the Einstein frame to a transition from pre-big-bang to big-bang cosmology, while they are completely smooth in the freeze frame. Appendix D reformulates the model with a curvature squared invariant in terms of an explicit additional scalar field. This helps to understand the properties of the solutions discussed in appendix C. Appendix E enlarges the class of crossover models and maps them to the freeze frame. 

\section{Fixed points and crossover}
\label{Fixed points and crossover}

In this section we display our model of ``crossover gravity''. We discuss the ultraviolet and infrared fixed point for a system of gravity coupled to a scalar field. Away from the fixed point the dimensionless couplings are scale-dependent and realize a crossover between the two fixed points. Our main tool is the quantum effective action $\Gamma$ from which the exact field equations follow by variation. The fixed points and the crossover are reflected in the properties of $\Gamma$. Besides the masses and couplings of other particles our model involves only four parameters which describe cosmology from inflation to present dark energy domination. 

\subsection{Running couplings and fixed points}

In quantum field theories the renormalized dimensionless couplings ``run'' as functions of an intrinsic mass scale $\mu$. Here we consider all intrinsic mass parameters as being proportional to $\mu$, with ratios of intrinsic mass scales associated to dimensionless couplings. For a fixed point this flow stops and dimensionless couplings become independent of $\mu$. An ultraviolet (UV) fixed point is reached if suitable dimensionless couplings reach constant values for $\mu\to\infty$. Such a fixed point renders gravity non-perturbatively renormalizable (asymptotic safety \cite{Wei,Rev,Per,HPRW}). Dilatation symmetry is an exact quantum symmetry at the UV fixed point. An infrared (IR) fixed point corresponds to the stop of the flow of dimensionless couplings for $\mu\to 0$. All intrinsic mass parameters vanish in this limit.  With dimensionless couplings independent of $\mu$ scale symmetry is again realized. In general, the existence of an IR fixed point is not compulsory - alternatives are diverging 
dimensionless couplings for $\mu\to 0$ or even a breakdown of the model at a critical value $\mu_c>0$. We assume here that such divergencies do not happen and an IR-fixed point therefore exists. A first functional renormalization investigation of such a possible IR fixed point can be found in ref. \cite{HPRW}. 

The flow of couplings as a function of $\mu$ is similar but not identical to the running as a function of $\tilde \mu\sim$ momentum divided by particle mass (say the electron mass). It is this running as a function of momentum/mass that is described by the usual $\beta$-functions of the standard model of particle physics. There a non-trivial running typically occurs in the range where $\tilde\mu$ is larger than the relevant particle masses, while it stops once $\tilde\mu$ is below those masses. In our setting, the flow as a function of $\mu$ describes the effect of a simultaneous change of all intrinsic mass scales $\sim \mu$.  Besides the change in momentum scale this also includes the change due to a mass parameter in the effective cosmon potential. The $\mu$-flow equations need a separate computation which has not yet been performed. They partly are similar in spirit to the running of couplings as a function of a mass parameter investigated by Symanzik \cite{SY}. Different $\mu$ correspond conceptually to 
a family of different 
theories. These theories cannot be distinguished by observation, however. Since only dimensionless ratios can be observed the value of $\mu$ just sets the unit for quantities with dimension of mass or inverse length or time. (We use $\hbar=c=k_B=1$). We will employ here
\be\label{1AA}
\mu^{-1}=10^{10}{\rm yr},
\ee
such that the present value of the variable Planck mass takes its usual value, cf. sect. \ref{Late cosmology and dark energy}.  There is also some analogy to the functional renormalization flow of the effective average action \cite{CWFE,RW}, with IR-cutoff $k$ associated to $\mu$. We discuss the conceptual setting of the $\mu$-flow equation in more detail in the appendix A and give examples in appendix B.

\subsection{Variable gravity}

We will work within variable gravity \cite{VG} and investigate the cosmological solutions of the field equations derived from the quantum effective action for the coupled cosmon-gravity system
\be\label{1}
\Gamma=\int_x\sqrt{g}
\left\{-\frac12\chi^2 R+\mu^2\chi^2+\frac12\big(B(\chi/\mu)-6\big)\partial^\mu\chi\partial_\mu\chi\right\}.
\ee
The variable Planck mass is given by the value of the cosmon field $\chi$. The quadratic cosmon potential $V=\mu^2\chi^2$ involves the intrinsic mass scale $\mu$. A large family of effective actions can be brought by field transformations to a form where the coefficient of the curvature scalar $R$ is $-\frac12 \chi^2$ and the scalar potential is quadratic, $V(\chi)=\mu^2\chi^2$. We will discuss this issue in sect. \ref{Field relativity}. We then remain with the dimensionless function $B(\chi/\mu)$. Its dependence on $\mu$ is described by the $\mu$-flow equation. Stability requires $B\geq 0$. Conformal symmetry is realized for $B=0, \mu=0$. 

The quantum effective action should be supplemented by higher order curvature invariants,
\ba\label{1A}
\Delta\Gamma&=&\int_x\sqrt{g}
\left\{ -\frac12 C(\chi/\mu)R^2\right.\nn\\
&&\left.+D(\chi/\mu)\left(R^{\mu\nu}R_{\mu\nu}-\frac13 R^2\right)\right.\\
&&\left.+E(\chi/\mu)(R_{\mu\nu\rho\sigma}R^{\mu\nu\rho\sigma}-4R_{\mu\nu}R^{\mu\nu}+R^2)
\right\}.\nn
\ea
These terms will play a role for graviton-graviton scattering at and near the UV-fixed point and for the approach of the cosmological solution to the infinite past, $\chi/\mu\to 0$. For the cosmological epochs discussed in this paper they are subleading and will be omitted in the explicit calculations of the main text. For constant $E$ the last term in eq. \eqref{1A} is the topological Gauss-Bonnet invariant which does not contribute to the field equations. The invariant multiplying $D$ can be written as a linear combination of the squared Weyl tensor and the Gauss-Bonnet invariant. For constant $D$ the term $\sim D$ does not contribute to the field equations for a spatially flat Robertson-Walker metric. The influence of the higher curvature invariant with constant $C$ is discussed in the appendices C and D. 

We do not include a possible scale invariant contribution to the cosmon potential $\Delta V=\lambda \chi^4$. Indeed, the functional renormalization investigation \cite{HPRW} of the behavior of a possible fixed point suggests that the cosmon potential cannot increase $\sim\chi^4$ for $\chi\to\infty$. The infrared fixed point proposed in ref. \cite{HPRW} has indeed $\lambda=0$. A term $\sim\lambda\chi^4$ is scale invariant but not conformal invariant. If the scale invariance of the IR-fixed point for $\chi\to\infty$ implies conformal symmetry, as advocated in ref. \cite{SICI1,SICI2}, such a term is not allowed. This situation is suggested by the investigation of higher dimensional theories with scale invariance \cite{WHD1,WHD2}. It is precisely the behavior $V(\chi\to\infty)\sim\chi^4$ that would not allow a field redefinition to the freeze frame $V=\mu^2\chi^2$. (Other powers are allowed.) The absence of a term $\sim\lambda\chi^4$ implies the asymptotic vanishing of the observable cosmological constant \cite{
CW3}.

Dimensionless functions as $B$ (or $C$, $D$ and $E$) can only depend on the dimensionless ratio $\chi/\mu$. This links their $\mu$-dependence according to the flow equation to their dependence on $\chi$. The UV-fixed point for $\mu/\chi\to\infty$ can also be seen as the limit $\chi\to 0$, while the IR-fixed point corresponds to the limit $\chi\to\infty$. We will find cosmological solutions where $\chi$ varies from $\chi\to 0$ in the infinite past to $\chi\to\infty$ in the infinite future. This is how cosmology can describe the crossover between two fixed points. The UV-fixed point for $\chi\to 0$ will often be called the ``past fixed point'', and the IR-fixed point for $\chi\to \infty$ is associated with the ``future fixed point''. 

In the context of variable gravity the flow equations for dimensionless couplings concern the quantities $B,C$, $D$ and $E$. More on the conceptual status of the flow equations for these quantities can be found in appendix A. In the present paper we assume a specific form of $B(\chi/\mu)$ and investigate the cosmological consequences of such a setting. 

\subsection{Infrared and ultraviolet fixed points}

For the IR-fixed point $\mu$ vanishes and $B$ reaches a constant, $\lim\limits_{\mu\to 0} B(\chi/\mu)=B_\infty$. The term $\sim\mu^2\chi^2$ in eq. \eqref{1} is absent for $\mu\to 0$, and the quantum effective action contains no longer any parameter with dimension mass. It is invariant under scale symmetry, with a scaling of $\chi$ according to its canonical dimension. At the IR-fixed point the effective action is scale invariant and takes the simple form of a free scalar field coupled to gravity,
\be\label{2AA}
\Gamma_{\rm IR}=\int_x\sqrt{g}
\left\{-\frac12\chi^2 R+\frac12(B_\infty-6)\partial^\mu \chi\partial_\mu\chi\right\}.
\ee
(For $\chi\to\infty$ we can neglect the higher order curvature terms.) For $B_\infty=0$ the effective action is also invariant under conformal transformations. The scalar is no longer a propagating degree of freedom. 

For the realization of an UV-fixed point the anomalous dimension of the cosmon will be crucial. Indeed, for a canonical scaling of $\chi$ the ``mass term'' $\sim \mu^2\chi^2$ would spoil scale invariance for $\mu\to\infty$. An anomalous dimension for $\mu\to\infty$ is realized if $B(\chi/\mu)$ diverges for $\mu\to\infty$ with a power law,
\be\label{2B}
B=b\left(\frac\mu\chi\right)^\sigma
=\left(\frac m\chi\right)^\sigma.
\ee
(We take $\chi=0$, negative $\chi$ being covered by the symmetry $\chi\to-\chi$.) For $\sigma\neq 2$ the gravitational higher order invariants \eqref{1A} (with constant $C,D,E$) and the scalar kinetic term are then invariant under the scaling
\be\label{2C}
g_{\mu\nu}\to \alpha^2 g_{\mu\nu}~,~\chi\to \alpha^{-\frac{2}{2-\sigma}}\chi.
\ee
At the UV-fixed point the effective action can be written in terms of a renormalized scalar field,
\be\label{2D}
\chi_R=b^{\frac12}
\left(1-\frac\sigma2\right)^{-1}\mu^{\frac\sigma 2}\chi^{1-\frac\sigma2},
\ee
as
\ba\label{2E}
\Gamma_{UV}&=&\int_x\sqrt{g}
\left\{\frac12 \partial^\mu\chi_R\partial_\mu\chi_R-\frac12 CR^2\right.\nn\\
&&\left.+D\left(R^{\mu\nu}R_{\mu\nu}-\frac13R^3\right)\right\}.
\ea
For constant $C$ and $D$ scale invariance is manifest - it is the renormalized field that shows the standard scaling $\chi_R\to \alpha^{-1}\chi_R$. It is possible that $C$ and $D$ vanish at the fixed point, such that only a kinetic term for the renormalized scalar field is left. 

The remaining terms for the cosmon potential and $\sim R$ read
\ba\label{2F}
\Delta\Gamma_{UV}&=&\int_x\sqrt{g}c_{\rm UV}
\left(\mu^2-\frac R2\right)\mu^{-\frac{2\sigma}{2-\sigma}}|\chi_{R}|{^{\frac{4}{2-\sigma}}},\nn\\
c_{\rm UV}&=&b^{-\frac{2}{2-\sigma}}
\left|1-\frac\sigma2\right|^{\frac{4}{2-\sigma}}.
\ea
The limit $\mu\to\infty$ and fixed $\chi$ corresponds to $\chi_R/\mu\to 0$ for $\sigma<2$, and $\chi_R/\mu\to -\infty$ for $\sigma>2$. In sect. \ref{Past fixed point} we will discuss that for $\sigma>1$ the term $\Delta\Gamma_{UV}$ accounts for deviations from the fixed point and can be neglected at the $UV$-fixed point. At the fixed point one finds a free massless scalar field coupled to higher order gravity. For the boundary case $\sigma=1$ the renormalized scalar field has a non-vanishing self-interaction, with scale invariant potential
\be\label{2G}
V(\chi_R)=\frac{1}{16b^2}\chi^4_R.
\ee
Also the limiting case $\sigma=2$ can be associated with a fixed point - see sect. \ref{Past fixed point}. An UV-fixed point is therefore realized for
\be\label{2H}
\sigma\geq 1.
\ee

Besides the cosmon-gravity part of the effective action \eqref{1}, \eqref{1A} we also have to specify the part for matter and radiation. We will take the standard model of particle physics and assume that all renormalized dimensionless couplings (e.g. gauge couplings, Yukawa couplings, Higgs-boson self interaction), normalized at momenta $\sim\mu$, are functions of $\chi/\mu$ that reach fixed constant values for $\mu\to\infty$ and $\mu\to 0$. The UV-values are typically different from the IR-values. For example, one may imagine that all renormalized couplings vanish for $\mu\to\infty$, leaving only free particles at the UV-fixed point. 

We write the coefficient of the quadratic term in the Higgs potential as $-\epsilon_H(\chi/\mu)\chi^2$. For an IR-fixed point the dimensionless coupling $\epsilon_H$ goes to a (very small) constant. It actually becomes independent of $\mu$ for small enough ratios $\mu/\chi$, cf. app. A. No memory of the scale $\mu$ is then left for $\chi\to\infty$ - the Fermi scale is proportional to 
$\chi$ such that the charged lepton and quark masses as well as the gauge boson masses are proportional to $\chi$  \cite{CW3}. Consider next the strong gauge coupling normalized at momenta given by the Fermi scale, 
$\bar g_s=g_s(Q^2=\epsilon_H\chi^2)$. It can only depend on the ratio $Q^2/\chi^2=\epsilon_H$. For $\epsilon_H$ independent of $\mu$ also $\bar g_s$ is independent of $\mu$. Therefore $\Lambda_{{\rm QCD}}$ scales $\sim\chi$, such that hadron masses are $\sim\chi$ as well. 

On the opposite end we may assume that the renormalized coupling corresponding to $\epsilon_H$ does not diverge for $\chi\to 0$. With masses $\sim \chi$ all particles are then massless at the UV-fixed point. Massless particles at the UV-fixed point are also realized for diverging $\epsilon_H$, provided the renormalized masses scale with a positive power of $\chi$. 

The same general picture applies for particles beyond the standard model, in particular the sector of heavy singlets which influence the neutrino masses by the seesaw \cite{ss1,ss2,ss3} or cascade \cite{MCW1,MCW2,MCW3,MCW4} mechanism. The only difference to the standard model sector will be the relevant value of $\chi/\mu$ for which the crossover between the two fixed points takes place. 

The cosmology at the fixed points is not per se very interesting. For the past fixed point matter and radiation may be negligible. The field equations for the cosmon-gravity system derived from the effective action \eqref{1}, \eqref{1A} admit for constant $C$ and $D$ the simple solution 
\be\label{2I}
R_{\mu\nu}=\mu^2 g_{\mu\nu}~,~\chi=0.
\ee
Another solution is simply flat Minkowski space,
\be\label{10A}
g_{\mu\nu}=\eta_{\mu\nu}~,~\chi=0.
\ee

The cosmology for the future fixed point is again of a simple type. A scale invariant model that obtains by omitting in eq. \eqref{1} the potential $\mu^2\chi^2$ has been proposed by Fujii \cite{FU1,FU2}. After Weyl scaling it describes standard cosmology plus a massless dilaton with derivative couplings. The dilaton settles to a fixed value after a short period of initial  damping of its motion, and plays no role for the subsequent ``late'' cosmology of the present epoch \cite{CW3}. While interesting in its own right, such a scale invariant model cannot describe dynamical dark energy or quintessence. In our setting a cosmology similar to this model is reached for the future fixed point.

The interesting cosmological features of inflation and dynamical dark energy are a consequence of small scaling violations in the vicinity of the fixed points. Close to the past fixed point the scale symmetry violating terms (``dilatation anomaly'') $\Delta\Gamma_{UV}$, cf. eq. \eqref{2F}, render the cosmological solution \eqref{2I} unstable, such that any small value of $\chi$ slowly increases with increasing time $t$. This slow increase will be associated with the almost scale invariant epoch of inflation. As $\chi$ grows large enough the crossover to the future fixed point starts and inflation ends. 

The subsequent radiation and matter dominated epochs belong already to the neighborhood of the standard model fixed point. For this fixed point the dominant scaling violation arises from the cosmon potential $\sim\mu^2\chi^2$. This will describe dynamical dark energy, according to an approximate scaling solution with dark energy proportional to the dominant radiation or matter component. Indeed, the ratio of the potential $V=\mu^2\chi^2$ divided by the fourth power of the effective Planck mass $\chi^4$ decreases $\sim \mu^2/\chi^2$ and reaches tiny values as $\chi$ moves to very large values in late cosmology. Such a behavior amounts to a solution of the ``cosmological constant problem''. At the second step of the crossover an additional violation of dilatation symmetry in the neutrino sector stops the scaling evolution of the cosmon. 

Realistic scaling solutions with a small fraction of early dark energy have been extensively discussed \cite{VG,CW3,CW2} for the case where $B$ reaches for $\mu\to 0$ a small value
\be\label{2J}
B(\chi\to\infty)=B_\infty=\frac{4}{\alpha^2}.
\ee
In the context of crossover gravity this would be the value of $B$ at the IR-fixed point. Observational bounds on early dark energy restrict the allowed values to $\alpha \gtrsim 10$ \cite{A2b,Re,Sievers:2013wk,A2d,PL,PEDE}. (In view of possible degeneracies in the parameter space of our model we take here a conservative bound.) While this setting is perfectly viable, we investigate in the present paper the possible alternative that $B$ vanishes at the IR-fixed point, $B_\infty=0$.  For the ``conformal value'' $B=0$ the cosmon is no propagating degree of freedom. Furthermore, for $B<0$ the model becomes unstable. The flow of couplings typically avoids to cross from a stable to an unstable situation. It seems therefore reasonable to assume that $B=0$ is a fixed point of the flow of $B$, and we will assume that it is reached for $\chi\to\infty$. As mentioned before, the IR-fixed point realizes conformal symmetry in this case. For finite $\chi$ one has $B>0$. 

\subsection{Crossover}

The crossover that leads to the end of inflation is related to the flow of the dimensionless function $B(\chi/\mu)$. As a first example we take $\sigma=1$ and discuss a one parameter flow equation
\be\label{2}
\mu\frac{\partial B}{\partial \mu}=\frac{\kappa B^2}{1+\kappa B}.
\ee
The approach to the fixed point at $B=0$ is quadratic (vanishing anomalous dimension)
\be\label{3}
\mu\partial_\mu B=\kappa B^2\qquad \text{for}\qquad B\to 0,
\ee
while the approach to the fixed point $B^{-1}=0$ involves an anomalous dimension
\be\label{4}
\mu\partial_\mu B^{-1}=-B^{-1}.
\ee
The finer details of the crossover will be less important.

The UV-fixed point $B^{-1}=0$ is approached for $\mu\to\infty$ or $\chi\to 0$. This fixed point is relevant for the infinite past of our universe. The  IR-fixed point $B=0$ is approached for $\mu\to 0$ or $\chi\to\infty$. It governs the infinite future. The solution of eq. \eqref{2} involves an integration constant $c_t$ which determines the particular trajectory of the flow according to the implicit solution
\be\label{5}
B^{-1}-\kappa \ln B=\kappa
\left[\ln \left(\frac{\chi}{\mu}\right)-c_t\right]=\kappa\ln \left(\frac\chi m\right).
\ee
It is related to a mass scale $m$ by dimensional transmutation, $m=\mu\exp(c_t)$. The crossover between the two fixed points occurs in the region $\chi\approx m$ and we will see that this coincides with the end of inflation. Late cosmology corresponds to $\chi\gg m$, while primordial cosmology is characterized by $\chi\ll m$. With $B(\chi)$  determined by eq. \eqref{5} our model \eqref{1} contains two free dimensionless parameters in the scalar-gravity sector, namely $\kappa$ and $c_t$. We will find below that realistic cosmology can be obtained in the region
\be\label{5AA}
\kappa=0.5 \qquad,~ c_t=14.
\ee
No tiny or huge dimensionless parameters appear in our setting. 

The flow equation \eqref{1A} is is only a particular example for a crossover between two fixed points for which $B^{-1}$ or $B$ vanish, respectively. For an arbitrary anomalous dimension $\sigma$ it is generalized to an extended family of models,
\be\label{7A}
\mu\frac{\partial B}{\partial \mu}=\frac{\kappa\sigma B^2}{\sigma+\kappa B},
\ee
with solution 
\be\label{7B}
B^{-1}-\frac{\kappa}{\sigma}\ln B=\kappa
\left[\ln \left(\frac\chi \mu\right)-c_t\right] =\kappa \ln 
\left(\frac\chi m\right).
\ee
Eqs. \eqref{2}, \eqref{5} correspond to $\sigma=1$, while eq. \eqref{7A} indeed accounts for an arbitrary anomalous dimension $\sigma$ at the UV-fixed point, 
\be\label{18A}
\sigma=\lim_{\mu\to\infty}\mu\partial_\mu\ln B.
\ee 
The values of realistic parameters do not depend strongly on $\sigma$. For $\sigma=3$ one has $c_t\approx 12$. 

The crossover behavior needs also to be specified for the particle physics sector of our model. We present details in the appendix B and outline here only a few characteristics. We will assume that for the large values of $\chi/m$ relevant for nucleosynthesis and later epochs the dimensionless couplings are already very close to their fixed point values, such that their dependence on $\chi$ can be neglected for the purpose of cosmology. Similarly, we assume for these periods that the masses of all particles except for neutrinos have reached the scaling behavior $m_p\sim \chi$ appropriate for the fixed point. With this simple assumption the severe observational bounds on the time variation of fundamental couplings and apparent violations of the equivalence principle are obeyed \cite{CW3}.

Neutrino masses also involve a sector of superheavy particles by virtue of the seesaw or cascade mechanism. These particles are part of the beyond standard model sector of particle physics.  For this sector we postulate that the crossover is happening in the region of $\chi/\mu$ relevant for present cosmology, such that the present variation of the average neutrino mass with $\chi$,
\be\label{6}
\frac{\partial\ln m_\nu}{\partial\ln \chi}_{|_{{\rm today}}}=2\tilde\gamma+1,
\ee
involves a parameter $\tilde\gamma>0$. This parameter only matters for a rather recent cosmological epoch when neutrinos have become non-relativistic. It plays no role as long as neutrinos are relativistic. Together with the present values for the masses and couplings of particles, including some dark matter candidate, the four parameters $\kappa,\sigma,c_t$ and $\tilde \gamma$ will be sufficient to describe a realistic cosmological sequence of inflation, radiation- and matter-domination, as well as the present transition to a new dark energy dominated epoch. All four parameters are of the order one and no particular fine-tuning is needed. At present, it seems that our model is compatible with all cosmological observations. We will see that the anomalous dimension $\sigma$ is closely related to the spectral index of the primordial fluctuations. 

The field equations derived from the effective action \eqref{1}, together with the crossover of the kinetial \eqref{7B} and the crossover parameter in the neutrino sector $\tilde \gamma$ (eq. \eqref{6}) are the practical basis for computing our cosmological results. Besides the deeper motivation from quantum gravity our model stands alone as a simple phenomenological description of cosmology. 

\section{Primordial cosmology and inflation}
\label{Primordial cosmology and inflation}

In this section we discuss the early cosmology of our model. It is governed by the proximity of the ultraviolet fixed point and describes an epoch of inflation. After a brief discussion of this epoch within the freeze frame of variable gravity, we perform a field transformation to the Einstein frame with a standard exponential form of the inflaton potential. This is most suitable for a simple detailed discussion of the properties of primordial fluctuations. For the observable aspects of the inflationary epoch the higher order curvature invariants play no role. We can therefore limit the discussion to the effective action \eqref{1}, for which the crossover is described by the kinetial \eqref{7B}.

\subsection{Primordial cosmology} 

We begin with a brief discussion of primordial cosmology  in the freeze frame. The field equations derived from the effective action \eqref{1}, \eqref{1A} are displayed and discussed for constant $C$ and $D$ in the appendix C. For $C>0$ an equivalent description of our model in terms of two scalar fields is discussed in the appendix D. We find that the higher order curvature term $\sim CR^2$ only affects the remote past of the universe before observable density fluctuations left the inflationary horizon. We refer the discussion of the interesting properties of the infinite past, where the higher curvature terms play a role, to the appendices C and D. 

For our discussion of the ``observable epoch'' of inflation we can omit the higher order curvature invariants \eqref{1A}. As an example, we start with $\sigma=1$ and approximate $\chi\ll m,~B=m/\chi$. The primordial epoch will correspond to an inflationary universe and we can neglect matter and radiation. The cosmon field equation obtains by variation of the effective action \eqref{1} and reads \cite{VG}
\be\label{7}
\ddot{\chi}+\left(3H+\frac12\frac{\dot\chi}{\chi}\right)\dot\chi=
\frac{2\mu^2\chi^2}{m}.
\ee
Here we have inserted the expression for $R$ according to the gravitational field equation. The Hubble parameter is given by
\be\label{8}
H=\sqrt{\frac{\mu^2}{3}+\frac{m\dot\chi^2}{6\chi^3}}-\frac{\dot\chi}{\chi}.
\ee
The evolution of the mean value of the inflaton and geometry until the end of inflation are described by the two equations \eqref{7}, \eqref{8}. 

We may use dimensionless variables $y=mt,w=\chi/m$, $h=H/m,\lambda=\mu^2/m^2$, such that
\ba\label{9}
&&\partial^2_yw+(3h+\frac12\partial_y\ln w)\partial_y w=2\lambda w^2,\nn\\
&&h=\sqrt{\frac\lambda3+\frac{(\partial_y w)^2}{6w^3}}-\partial_y\ln w.
\ea
One finds an approximate solution that approaches a constant $h$ for $w\to 0$ 
\ba\label{10}
h&=&\sqrt{\frac\lambda3}-\frac{5}{6(y_c-y)},\nn\\
w&=&w_0(y)
+\frac{\bar{w_1}}{(y_c-y)^2}.
\ea
The function $w_0(y)$ vanishes in the infinite past for $y\to-\infty$,
\be\label{11}
w_0=\frac12\sqrt{\frac3\lambda}(y_c-y)^{-1},
\ee
and $y_c,\bar{w_1}$ integration constants. Restoring dimensions yields in leading order
\be\label{12}
H=\frac{\mu}{\sqrt{3}}~,~\chi=\frac{\sqrt{3}m}{2\mu(t_c-t)}.
\ee

We conclude that time can be continued in this approximation to the infinite past, $t\to-\infty$. In this limit geometry approaches de Sitter space and the cosmon field vanishes. The limiting solution $H=\mu/\sqrt{3}$, $\chi=0$ is unstable, however. A small deviation $\chi$ increases with $t$ according to eq. \eqref{12} or \eqref{10}, and the Hubble parameter decreases. Primordial cosmology describes an inflationary epoch. This will end if the increase of $\chi$ or decrease of $H$ becomes too fast. A quantitative estimate for the end of inflation will be given later in the Einstein frame. 

A similar qualitative behavior extends to other values of the anomalous dimension $\sigma$. For our second example $\sigma=2$ we use $B=m^2/\chi^2$, such that the field equations take the form
\ba\label{25Aa}
\ddot{\chi}+3H\dot\chi&=&\frac{2\mu^2\chi^3}{m^2},\nn\\
H&=&\sqrt{\frac{\mu^2}{3}
+\frac{m^2\dot\chi^2}{6\chi^4}}-\frac{\dot\chi}{\chi}.
\ea
The leading order solution becomes now
\be\label{25Bb}
H=\frac{\mu}{\sqrt{3}}~,~
\chi=\frac{3^{\frac14}m}{2\sqrt{\mu}}(t_c-t)^{-\frac12}.
\ee
For $\sigma>1$ one finds a decrease of $\chi$ towards the infinite past, cf. appendix C, eq. \eqref{B.7C},
\be\label{31A}
\chi \sim (t_c-t)^{\frac1\sigma}.
\ee

We will see below that the solutions \eqref{12}, \eqref{25Bb} correspond to a standard inflationary scenario in the Einstein frame or ``big bang frame''. Horizon crossing of the observable primordial fluctuations occurs when $\chi/\mu$ is already large, $\chi\approx 1.5\cdot 10^4\mu$. For these values the relative contribution of higher order curvature invariants from eq. \eqref{1A} is suppressed by a factor $\sim C\mu^2/\chi^2$, which is tiny for any moderate $C$. We can therefore indeed neglect such terms for the discussion of observable signals from inflation. Nevertheless, as $\chi$ becomes much smaller than $\mu$ for $t\to-\infty$, the role of the higher order curvature invariants becomes important. We give more details of the behavior of cosmology in the infinite past in appendix C. This includes the role of the higher order curvature invariants. For constant $C\neq 0$ and $\sigma<2$ the asymptotic solutions for the infinite past are extended to a whole family for which a constant Hubble parameter $H=H_
0$ is a free 
parameter.

We will show next that our crossover model predicts a rather large ratio between primordial tensor and scalar fluctuations. Since the value of the cosmon field $\chi$ (which plays the role of the inflaton) equals the dynamical Planck mass, the Lyth bound \cite{Ly,LR} plays no role in our setting of variable gravity \cite{HMSS3}.

\subsection{Cosmon inflation} 

The association of primordial cosmology with an inflationary epoch is most easily understood in the Einstein frame. Also a quantitative discussion of the generation of primordial density fluctuations and the end of inflation is best done in this frame. With
\be\label{I1}
g'_{\mu\nu}=\frac{\chi^2}{M^2}g_{\mu\nu}~,~\varphi=\frac{2M}{\alpha}\ln \left(\frac{\chi}{\mu}\right)
\ee
the quantum effective action \eqref{1} reads
\ba\label{I2}
\Gamma&=&\int_x\sqrt{g'}
\left\{-\frac12M^2R'+V'(\varphi)+\frac12k^2(\varphi)\partial^\mu\varphi\partial_\mu\varphi\right\},\nn\\
V'(\varphi)&=&
M^4\exp\left(-\frac{\alpha\varphi}{M}\right).
\ea
We identify $M=2.44\cdot 10^{18}$GeV with the Planck mass and observe that the cosmon potential $V'$ decays exponentially to zero \cite{CW3,CW2}. The absence of an additional constant for $\varphi\to\infty$ in the Einstein frame is a direct consequence of the vanishing ratio between potential and fourth power of the dynamical Planck mass, $V/\chi^4=\mu^2/\chi^2$, for $\chi\to\infty$ in the freeze frame. 

It has been advocated \cite{CI,VG,Lin} that it is advantageous to use a field basis where the potential takes a fixed form while the detailed model information appears in the form of the coefficient of the kinetic term, the kinetial. The reason is that the association of the value of the scalar field with the value of the potential energy is universal for a standardized potential. This makes it easy to compare different models. For our choice of a standard exponential potential the slow roll parameter $\epsilon$ and $\eta$ reflect indeed very simple properties of $k(\varphi)$. We could choose $\alpha=1$ as far as inflation is concerned, but we prefer here a different value in order to match the notation of quintessence potentials for late cosmology. 

The kinetial $k$ is related to $B$ by 
\be\label{I3}
k^2=\frac{\alpha^2B}{4}.
\ee
Since the parameter $\alpha$ appears only in the definition of $\varphi$, eq. \eqref{I1}, one is free to choose it at will and we could indeed have set $\alpha=1$. Instead, we find it convenient to adopt a definition of $\alpha$ such that the field $\varphi$ has a standard normalization for the present cosmological epoch, $k^2(\varphi_0)=1$, or 
\be\label{I4}
\alpha^2=\frac{4}{B(\chi=M)}\approx 4\kappa\ln(M/m).
\ee
Typical values of $\alpha$ will be around ten or somewhat larger, see below. The normalization of $\varphi$ and the precise value of $\alpha$ do not matter for the physics of inflation, however. For $\chi\to 0$ eq. \eqref{7B} yields
\be\label{33A}
B=\left(\frac{m}{\mu}\right)^\sigma\exp 
\left\{-\frac{\sigma\alpha\varphi}{2M}\right\}.
\ee

A slow roll period for inflation is realized for large enough $k^2$. We consider here a general function $B(\chi)$ and specialize to eq. \eqref{7B} later. The usual slow roll parameters $\epsilon$ and $\eta$ obtain as \cite{CI,VG}
\be\label{I5}
\epsilon=\frac{\alpha^2}{2k^2}=\frac2B~,~\eta=\frac1B(4-\sigma)~,~\sigma=-\frac{\partial\ln B}{\partial\ln \chi}.
\ee
Inflation ends when $\epsilon$ or $|\eta|$ are of order one. We define the end of inflation by the field value $\chi_f$ determined by $B(\chi_f)=6$, with $\epsilon_f=1/3,~\eta_f=2/3-\sigma/6$. This is the value where the kinetic term in eq. \eqref{1} changes sign. Inflation is realized for a rather generic shape of the function $B(\chi)$. It is sufficient that $B$ is large enough for small $\chi$ in order to induce an epoch of slow roll, and that $B-6$ reaches negative values as $\chi$ increases in order to end inflation. 

The definition of $\sigma$ employed in the present section, given by eq. \eqref{I5}, differs slightly from the preceding section. In the present section, $\sigma$ is considered as a function of $\chi$. It agrees with the parameter $\sigma$ in the preceding section for $\chi\to 0$. For the inflationary period this difference is minor (except possibly for the end of inflation), justifying the use of the same symbol.

\subsection{Spectral index and tensor ratio of primordial fluctuations}

The spectrum of primordial scalar density fluctuations is characterized by the spectral index $n=1-6\epsilon+2\eta$, while the relative amplitude of tensor fluctuations over scalar fluctuations reads $r=16\epsilon$. Here $\epsilon$ and $\eta$ have to be evaluated for the value of $\chi$ at horizon crossing, $N~e$-foldings before the end of inflation. One finds the relations
\be\label{I6}
r=\frac{32}{B(N)}~,~1-n=\frac r8
\left(1+\frac12\sigma(N)\right).
\ee
We observe an interesting general relation between $n$ and $r$. Horizon crossing occurs in the region $\chi\ll m~,~B\gg 6$. The particular models with $\sigma=1$, $\sigma=2$ or $\sigma=3$ predict
\be\label{I7}
1-n=\frac{3r}{16}~,~\frac r4~,~\frac{5r}{16},
\ee
respectively. For a spectral index $n=0.97$ this implies a rather high amplitude, $r=0.16,~0.12,~0.096$, in the range claimed originally by BICEP \cite{BI}. In the view of recent CMB-results \cite{BIPL,PCP} the range $\sigma>2$ seems to be preferred as compared to the range $\sigma<2$. 

We next compute the relation between the value of $\chi(N)$ at horizon crossing and the number $N$ of $e$-foldings before the end of inflation,
\be\label{I8}
N=\frac{1}{\alpha M}\int^{\varphi_f}_\varphi d\varphi'k^2(\varphi')=\frac12
\int^{\chi_f}_{\chi(N)}
\frac{d\chi}{\chi}B(\chi).
\ee
In the range of interest $B(\chi)$ is typically a strongly decreasing function. The integral is dominated by the region around $\chi(N)$ where we may approximate $B=(m/\chi)^{\sigma(N)}$. This relates $B(N)=B\big (\chi(N)\big)$ to $N$,
\be\label{I9}
N=\frac{B(N)-B(\chi_f)}{2\sigma(N)}.
\ee
With $B(\chi_f)=6$ one finds
\be\label{21A}
r=\frac{16}{N\sigma(N)+3}
\ee
and 
\be\label{23A}
n=1-\frac{2+\sigma(N)}{N\sigma(N)+3}.
\ee
These two central formulae express both $n$ and $r$ in terms of $\sigma(N)$ and $N$. 

\begin{figure}[tb]
\centering
\includegraphics[scale=0.7]{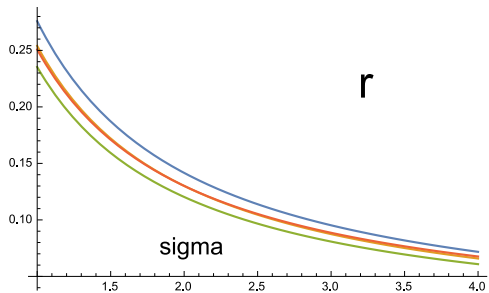}
\caption{Tensor ratio $r$ for primordial fluctuations as function of the anomalous dimension $\sigma$. The curves from top to down are for $N=55,60,65$. We also show the result for eq. \eqref{54B} which almost coincides with $N=60$. }
\label{IQOM2}
\end{figure}

\begin{figure}[h!tb]
\centering
\includegraphics[scale=0.7]{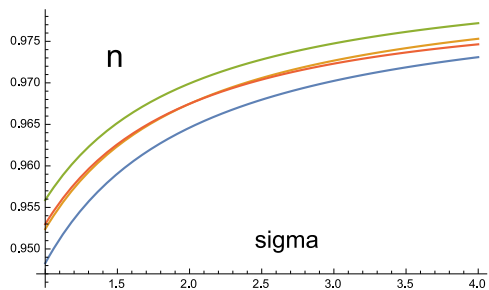}
\caption{Spectral index $n$ for primordial fluctuations as function of the anomalous dimension $\sigma$. The curves from top to down are for $N=65,60,55$. We also show the result of eq. \eqref{54B} which almost coincides with $N=60$.}
\label{IQOM3}
\end{figure}

For the particular model with $\sigma=1$ one obtains $B(N)=2N+6$ and predicts
\be\label{I10}
r=\frac{16}{N+3}~,~1-n=\frac{3}{N+3}.
\ee
while $\sigma=2$ yields
\be\label{23B}
r=\frac{16}{2N+3}~,~1-n=\frac{4}{2N+3},
\ee
and $\sigma=3$ results in 
\be\label{45A}
r=\frac{16}{3N+3}~,~1-n=\frac{5}{3N+3}.
\ee
We show the spectral index and the tensor ratio as a function of $\sigma$ for various $N$ in Figs. \ref{IQOM2}, \ref{IQOM3}.

We will see below that $N$ depends only very mildly on $\sigma$. Its precise value shows some influence of the details of the entropy production after the end of inflation. A typical value is $N=60$. For a given $N$ both $n$ and $r$ are uniquely determined by $\sigma$. More precisely, $\sigma=\sigma(N)$ is defined in terms of the function $B(\chi)$ by 
\be\label{23C}
\sigma=-\frac{\partial\ln B}{\partial \ln \chi}\begin{array}{l}~\\|\end{array}_{B=2\sigma N+6}.
\ee
Thus only the logarithmic derivative of $\ln B$ at a particular value of $B$ matters for the computation of $n$ and $r$! For $N=60$ we display the values of $n$ and $r$ for three values of $\sigma$ in table 1.
\begin{center}
\begin{table}[h!tb]
\begin{tabular}{|l|l|l|l|}
\hline
$\sigma$&1&2&3\\ \hline
$n$&0.952&0.967&0.973\\
$r$&0.254&0.13& 0.087\\ \hline
\end{tabular}
\caption{ Properties of primordial fluctuations for different values of $\sigma, N=60$.}
\end{table}
\end{center}

\bigskip

We can use eqs. \eqref{21A}, \eqref{23A} in order to relate $r$ to $1-n$. Within the relation
\be\label{48A}
\frac r8=1-n-\frac{1}{N+1}-\frac{1}{(N+1)^2}
\left(1-\frac{3}{\sigma(N)}\right)
\ee
we can use the lowest order relation for $\sigma(N)$, evaluated for $N=60$,
\be\label{40B}
\frac{3}{\sigma(N)}=90(1-n)-1.5.
\ee
This yields
\be\label{48C}
r=8.193(1-n)-0.1365+0.0021(N-60).
\ee
For $n=0.97$ this predicts $r=0.109$, while for $n=0.965 ~(0.975)$ one has $r=0.15 ~(0.068)$. An upper bound on $r\lesssim 0.1$ prefers a value of $n$ close to one, $n\gtrsim 0.97$, corresponding to $\sigma\gtrsim 2.5$. 

We conclude that our model can be falsified by precision observations of the CMB. Since inflation lasts for an extremely long time before horizon crossing of the observable fluctuations, perhaps even since the infinite past, there seems to be no issue that memory of the initial conditions could spoil its predictivity \cite{CWF1,CWF2}. If $r$ and $n$ can be established in accordance with the relation \eqref{48C} this will constitute a measurement of the anomalous dimension $\sigma$. Hopefully, this anomalous dimension is computable in quantum gravity, leading to a direct observational test. 

The relation \eqref{48C} is approximately valid for a large class of inflationary models beyond our particular setting. The slow roll parameters $\epsilon$ and $\eta$ only involve the value and the derivative of the kinetial at the value of $\varphi$ corresponding to horizon crossing of the observable fluctuations. The relation \eqref{48C} follows qualitatively whenever the $\varphi'$-integral in eq. \eqref{I8} is dominated by the large value of $k^2$ at $\varphi$, while the decrease of $k^2$ is well approximated by its first derivative. Within eq. \eqref{I9} the model-uncertainty can be cast into values of $B(\chi_f$) deviating from the value $B(\chi_f)=6$ for our setting. 

\subsection{Amplitude of primordial fluctuations}

The amplitude of the primordial scalar fluctuations can be related to the value of the potential at horizon crossing and the tensor to scalar ratio
\be\label{22AA}
{\cal A}=\frac{V\ro}{rM^4}=3.56\cdot 10^{-8},
\ee
where the last equation employs the observed amplitude of the spectrum of CMB-anisotropies. This measurement determines the ratio
\be\label{25A}
\frac m\mu=\frac{\chi(N)}{\mu}\frac{m}{\chi (N)}=
\frac{M^2}{\sqrt{V\ro}} \frac{m}{\chi(N)} =\frac{1}{\sqrt{{\cal A}r}}\frac{m}{\chi(N)}.
\ee
We next employ the approximate form $B=(m/\chi)^\sigma$ or 
\be\label{25B}
\frac{m}{\chi(N)}=B(N)^{\frac{1}{\sigma(N)}}=\left(\frac{r}{32}\right)^{-\frac{1}{\sigma(N)}}, 
\ee
such that 
\ba\label{25C}
\frac m\mu&=&\frac{1}{4\sqrt{2{\cal A}}}
B(N)^{\frac12+\frac{1}{\sigma(N)}}\nn\\
&=&2^{\frac{1}{\sigma(N)}-2}
\big(N\sigma(N)+3\big)^{\frac12+\frac{1}{\sigma(N)}}{\cal A}^{-\frac12}.
\ea

For the particular model with $\sigma=1$ one finds
\be\label{22AB}
\frac m\mu=\frac{(N+3)^{\frac32}}{2\sqrt{{\cal A}}}=1.32\cdot 10^6\left(\frac{N}{60}\right)^{\frac32}.
\ee
For the constant $c_t$ in eq. \eqref{5} one infers
\be\label{22AC}
c_t=\ln\left(\frac m\mu\right)=14.1.
\ee
(For the numerical value we have taken $N=60$, see below.) Due to the exponential dependence on $c_t$ no very large or small parameter is needed in order to obtain a small fluctuation amplitude
\be\label{22AD}
{\cal A}=\frac{(N+3)^3}{4}e^{-2c_t}.
\ee
The flow equation \eqref{2} generates the scale $m$ by dimensional transmutation. The small amplitude ${\cal A}$ indicates that this scale is larger than the ``intrinsic scale'' $\mu$. The situation is similar for other values of $\sigma$. For $\sigma=2$ the ratio $m/\mu$ decreases by a factor $1/\sqrt{30}$ as compared to $\sigma=1$. For $\sigma=3$ one has
\be\label{58A}
\frac{m}{\mu}=\frac{2^{1/3}}{4}\big[3(N+1)\big]^{\frac56}{\cal A}^{-1/2}
\ee
and 
\be\label{58B}
c_t=11.8.
\ee

We may turn this argument around and state that crossover models provide for a natural explanation of a small fluctuation amplitude ${\cal A}$. We can relate the dimensionless cosmon potential $V/\mu^4=\chi^2/\mu^2$ to a dimensionless flow parameter by
\be\label{28A}
\tilde\mu
=\ln \left(\frac{\mu}{\chi}\right)
=-\frac12 \ln\left(\frac{V}{\mu^4}\right).
\ee
We have associated the scale $m$ with the crossover value $\mu_{{\rm cr}}$ where the flow moves away from the behavior dictated by the ``past fixed point'' for $\tilde\mu\to\infty$,
\be\label{28B}
\frac m\mu=e^{-\tilde\mu_{{\rm cr}}}.
\ee
Different trajectories (solutions of the flow equations) can be characterized by how close to the fixed point they are for $\tilde\mu=1$. The larger $m/\mu$, the closer a trajectory is to the fixed point. In view of the exponential behavior of eq. \eqref{28B}, already moderate negative values of $\tilde\mu_{{\rm cr}}$ are sufficient to induce large values of $m/\mu$, and therefore a small amplitude ${\cal A}\sim(\mu/m)^2\sim e^{2\tilde\mu_{{\rm cr}}}$. 

We may also evaluate the dimensionless ratio
\be\label{28C}
\frac{V}{\chi^4}=\frac{\mu^2}{\chi^2}=e^{2\tilde\mu}.
\ee
For $\chi=m$ this quantity measures the potential in units of the Planck mass at the crossover. For many of our models the corresponding scale of the potential in the Einstein frame $V^{\frac14}$ is of the order where spontaneous symmetry breaking is expected in a grand unified theory. This suggests that the crossover could be associated with grand unified symmetry breaking. 

Finally, we may compare the value of $\chi(N)$ at horizon crossing with $m$ using eq. \eqref{25B}, 
\be\label{49AA}
x(N)=\frac{\chi^2(N)}{m^2}=
\left(\frac{r}{32}\right)^{\frac2\sigma}.
\ee
For all models one finds a small value $x(N)\ll 1$, justifying the approximation \eqref{2B}. On the other hand, we observe that $\chi(N)$ is much larger than $\mu$, cf. eq. \eqref{25A}
\be\label{49AB}
\frac{\chi^2(N)}{\mu^2}=\frac{1}{{\cal A}r}.
\ee
A simple picture arises. Horizon crossing happens when $\chi$ is already much larger than $\mu$, but still smaller than $m$. Inflation ends when $\chi$ reaches $m$. 

\subsection{Horizon crossing} 
We finally need to evaluate the value of $N$ for our type of crossover models. We present here a detailed treatment that allows one to estimate where various uncertainties come from. Horizon crossing of a mode with comoving wave vector $k$ occurs for
\be\label{I11}
k=a\ro H\ro=a\ri H\ri,
\ee
where $a\ro$ or $a\ri$ corresponds to the scale factor when the mode leaves the horizon after inflation or enters again in the more recent past. 

We use 
\be\label{I12}
1=\frac{a\ro H\ro}{a\ri H\ri}=
\frac{a\ro}{a_f}
\frac{a_f}{a_r}
\frac{a_r}{a\ri}
\frac{H\ro}{H_r}
\frac{H_r}{H\ri},
\ee
with $a_f$ and $a_r$ the scale factors at the end of inflation and at a time when the universe begins to be dominated by radiation, respectively, and $H_r=H(a_r)$. For $a\ro/a_f=e^{-N}$ one finds the relation
\be\label{I13}
N=\ln
\left(\frac{H_r}{H\ri}\right)+\ln
\left(\frac{H\ro}{H_r}\right)-\ln
\left(\frac{a\ri}{a_r}\right)-\ln
\left(\frac{a_r}{a_f}\right).
\ee
Neglecting entropy production for photons for $a>a_r$ we use $a\ri/a_r=T_r/T\ri$ with $T$ the photon temperature. We relate $T$ to the total energy density in radiation
\be\label{I14}
\rho_r=3M^2H^2_r=f_r T^4_r~,~
\rho^{(\gamma)}\ri=3M^2H^2\ri\Omega^{(\gamma)}\ri=f\ri T^4\ri,
\ee
with $\Omega^{(\gamma)}$ the photon fraction of energy density and $f_r(f\ri)$ the number of degrees of freedom in radiation (photons). These relations allow us to express $a_{{\rm in}}/a_r$ in terms of $H_{{\rm in}}$ and $H_r$.  We further approximate $H\ro/H_f\approx \sqrt{V\ro/V_f}=\chi_f/\chi(N)$, resulting in 
\ba\label{I15}
N&=&\frac12\ln\left(\frac{H\ro}{H\ri}\right)-\frac14\ln
\left(\frac{f\ri}{f_r\Omega\ri^{(\gamma)}}\right)\nn\\
&&+\frac14\ln
\left(\frac{V\ro}{V_f}\right)+\Delta N,\nn\\
\Delta N&=&\frac12 \ln \left(\frac{H_f}{H_r}\right)-\ln
\left(\frac{a_r}{a_f}\right).
\ea

We first evaluate $N_0$ for modes that come into the horizon today, and subsequently extrapolate to larger $k$. The dominant contribution is the first term $\sim \ln(H\ro/H\ri)$. We can relate $H\ro$ to the tensor amplitude of the primordial fluctuations
\ba\label{I16}
3M^2H^2\ro=V\ro={\cal A}r(N_0)M^4,
\ea
and use $3M^2H^2\ri=\rho_c=(2 \cdot 10^{-3}{\rm eV})^4$, which yields
\be\label{I17}
\frac{H\ro}{H\ri}=1.5\sqrt{{\cal A}r}\cdot 10^{60}.
\ee
With $\Omega^{(\gamma)}\ri=5\cdot 10^{-5}$ and $f_r/f\ri=100$ one has $\ln(f_r\Omega^{(\gamma)}\ri/f\ri)/4=-1.3$. (Note that the neglected entropy production for photons can be incorporated into a modification of the poorly known ratio $f_r/f_{{\rm in}}$.) With $\ln {\cal A}/4=-4.3$ from eq. \eqref{22AA} we obtain
\be\label{44}
N_0=63.7+\frac14\ln r+\frac14\ln \left(\frac{V_{\rm out}}{V_f}\right)+\Delta N.
\ee
The two last terms involve the details of the epochs between $a_{\rm out}$ and $a_f$, or between $a_f$ and $a_r$, respectively.

For an estimate of $V_f$ we employ $B(\chi_f)=6$ and eq. \eqref{7B},
\be\label{A3A}
\ln\left(\frac{\chi_f}{m}\right)=\frac{1}{6\kappa}-\frac{\ln 6}{\sigma}~,~
\frac{V_f}{M^4}=\frac{\mu^2}{\chi^2_f}.
\ee
With eq. \eqref{25B} one finds
\be\label{46}
\frac{V\ro}{V_f}=\frac{\chi^2_f}{\chi^2(N)}=\exp \left(\frac{1}{3\kappa}\right)\left(\frac{3r}{16}\right)^{-\frac2\sigma},
\ee
or
\be\label{47}
N_0=63.7+\frac{1}{12\kappa}+\frac{0.84}{\sigma}+\frac14
\left(1-\frac{2}{\sigma}\right)\ln r+\Delta N,
\ee
with $r$ depending on $N$ and $\sigma$ according to eq. \eqref{21A}. Inserting $N_0\approx 65$ in the subleading term $\sim \ln r$, and $\kappa=\frac12$ (see below) yields for $\sigma=1(\sigma=2)$
\be\label{A3F}
N_0= 65.1(64.3) +\Delta N.
\ee

The remaining piece $\Delta N$ reflects the details of entropy production between the end of inflation $(a_f)$ and the beginning of the radiation dominated universe $(a_r)$. We may parametrize this epoch by two parameters, the number of $e$-foldings $N_{fr}$ for the duration of this period 
\be\label{A3G}
N_{fr}=\ln\frac{a_r}{a_f},
\ee
and the averaged equation of state $\bar w$ which governs the evolution of the total energy density,
\be\label{A3H}
\partial_t\rho=-3H(1+\bar w)\rho.
\ee
With 
\be\label{A3I}
\rho=\bar\rho a^{-3(1+\bar w)}=3M^2H^2~,~Ha^2\sim a^{\frac{1-3\bar w}{2}},
\ee
one finds
\be\label{A3J}
\Delta N=\frac{3\bar w-1}{4}N_{fr}.
\ee
For a fast entropy production $N_{fr}$ is of the order one. The parameter $\bar w$ is a suitable average of a function $w(a)$ that starts close to $w(a_f)\approx -1$ for $a=a_f$, may then be given for a period of domination of scalar kinetic energy, $w(a)\approx 1$, and finally end with $w(a_r) \approx 1/3$. For not too large $N_{fr}$ and $\bar w$ close to $1/3$ one may simply neglect $\Delta N$, and we will use this approximation in the following.

We may finally extrapolate to modes with present wavelength smaller than the horizon. As compared to $N_0$ the dominant correction factor is
\ba\label{54A}
N&=&N_0+\delta N,\nn\\
\delta N&=&\ln
\left(\frac{H_0a_0}{H\ri a\ri}\right)=\ln \frac{k_0}{k}=-\ln \frac{L_0}{L},
\ea
with $k$ and $L$ the wave number or wave length of the mode, and index zero denoting the ones corresponding to the present horizon ($L_0\approx 3000 M{pc}$). In the range where primordial gravitational waves may be detected $(k/k_0\approx 80)$ one has $\delta N\approx -4.4$, such that a reasonable overall estimate is
\be\label{54B}
N\approx 60-0.8(\sigma-2).
\ee
We can neglect the $\sigma$-dependence of $N$ and use $N=60$. Up to small calculable corrections for $\sigma\neq 2$ this entails the predictions 
\ba\label{54C}
r&=&\frac{0.26}{\sigma}\nn\\
n&=&1-\frac{0.065}{\sigma}\cdot 
\left(1+\frac{\sigma-2}{4}\right).
\ea
In particular, one obtains for $\sigma=2$
\be\label{54D}
r=0.13~,~n=0.967,
\ee
in accordance with the prediction in ref. \cite{CWU} (see also refs. \cite{HMSS1,HMSS2,SaSt}). For $\sigma=1$ one finds $r=0.25,n=0.953$.

Due to the dependence of $N$ on the wave number $k$ the spectral index and tensor amplitude depend on $k$ according to 
\ba\label{54E}
\frac{\partial r}{\partial \ln k}&=&-\frac{\partial r}{\partial N}=\frac{r^2}{16}
\left(1+\frac{\partial\sigma}{\partial \ln N}\right),\\
\frac{\partial n}{\partial \ln k}&=&-\frac{\partial n}{\partial N}=-\frac{r(1-n)}{16}
\left(1+\frac{\partial \sigma}{\partial \ln N}\left(1-\sigma-\frac3N\right)\right).\nn
\ea
Since $|\partial \sigma/\partial \ln N|$ is typically of the order one or smaller the running of the spectral index is very slow. Indeed, for our model $\sigma$ changes over $60$ e-foldings only from $\sigma(N)$ to 
\be\label{54F}
\sigma_f=-\frac{\partial \ln B}{\partial \ln \chi}
\begin{array}{r}
~\\|
\end{array}_{B=6}
\approx \kappa B\approx 3,
\ee
and this change occurs towards the end of inflation. We conclude that our UV-fixed point scenario provides for a rather simple and predictive model of inflation.

\section{Late cosmology and dark energy}
\label{Late cosmology and dark energy}

The crossover in the kinetial $K(\chi)=B(\chi)-6$ from positive to negative values triggers the end of the inflationary slow roll solution. Subsequently, radiation and entropy are produced by various mechanisms \cite{CI,HMSS1}. We recall that during the crossover the dimensionless couplings and mass ratios of the standard model of particle physics are supposed to change from their values for the past fixed point to the ones for the standard model fixed point. In particular, the effective quartic cosmon-Higgs-coupling $\epsilon_H$ could be much larger than the tiny value for the standard model fixed point. A fluctuating Higgs doublet or a similar field related to spontaneous symmetry breaking in a grand unified setting could play a major role for the heating \cite{CI}. More precisely, a $\chi$-dependence of $\epsilon_H$ results in the Einstein frame in an effective coupling between the cosmon and the Higgs doublet. This generalizes to other $\chi$-dependent dimensionless couplings. In a grand unified theory 
the heating period may be associated with the onset of spontaneous symmetry breaking of the GUT-gauge group. Rather generically, the $\chi$-dependence of couplings is large precisely in the crossover region. Thus at the end of inflation the cosmon coupling to other particles is large in the Einstein frame, in contrast  to the tiny couplings close to the standard model fixed point. The large couplings provide for rather efficient heating mechanisms. 

After the heating and entropy production have occurred the universe enters its ``late epoch'', beginning with radiation domination. The late universe is characterized by the approach to the 
future fixed point. During the radiation and matter dominated periods this approach is slow, as accounted for by the (approximate) standard model fixed point, recall Fig. \ref{IQOM1}.

In the freeze frame the particle masses increase with increasing $\chi$, while the universe shrinks, in contrast to the usual big bang picture \cite{CWU}. (For early cosmological models with varying particle masses see. refs. \cite{Na1,Na2,Na3}.) Indeed, only the dimensionless ratio of the distance between galaxies divided by the atom radius is observable \cite{CWVN,FR3,Cat1,DeS}. The overall picture of late cosmology in the freeze frame has been described in detail in ref. \cite{VG} for the case where the kinetial takes a constant value $K_\infty=B_\infty-6$ at the IR-fixed point. Radiation and matter domination are characterized by a negative constant Hubble parameter $H=b\mu$, while particle masses increase exponentially, $\chi\sim\exp(c\mu t$). The characteristic time scale for both epochs is given by $\mu^{-1}=10^{10}$yr, such that the evolution is always very slow. Temperature increases $T\sim \sqrt{\chi}$ due to the shrinking of the universe. Particle masses $m_p\sim\chi$ increase even faster, however,
 such 
that the relevant ratio $T/m_p$ decreases 
as in the usual big bang picture.

In the present crossover model $B$ depends only mildly on $\chi$ for $\chi\gg m$,
\be\label{L1}
B=\frac{1}{\kappa\ln \left(\frac\chi m\right)}.
\ee
The cosmology with constant $B_\infty=B(\chi\to\infty)=4/\alpha^2$ is therefore a good approximation. A priori, both the behavior \eqref{L1} and a small fixed value of $B_\infty$ are perfectly viable candidates for realistic late cosmology. In the present paper we supplement the earlier discussion with constant $B_\infty$ by a quantitative investigation of a slowly varying $B(\chi)$ according to eq. \eqref{L1}. We employ the Einstein frame in order to facilitate the embedding of this model in standard scenarios of quintessence. 

\subsection{Late cosmology in the Einstein frame} 

In the Einstein frame \eqref{I2} the kinetial is given for late cosmology by 
\be\label{L2}
k^2=\frac{M\alpha}{2\kappa(\varphi-\bar\varphi)}
=\left(1+\frac{2\kappa(\varphi-\varphi_0)}{M\alpha}\right)^{-1},
\ee
with $\varphi_0$ the present value of the cosmon
\be\label{L3}
\varphi_0=\frac{2M}{\alpha}\ln\frac{M}{\mu}~,
~\bar\varphi=\frac{2M}{\alpha}\ln
\left(\frac m\mu\right).
\ee
With the exponential potential \eqref{I2}, this is a typical model of dynamical dark energy. (The divergence for $\varphi\to\bar \varphi$ is outside the vicinity of the approximation.) Except for neutrinos the standard model particles and dark matter do not couple to $\varphi$. The cosmon field equation
\be\label{F1}
k^2(\ddot{\varphi}+3H\dot{\varphi})+\frac12
\frac{\partial k^2}{\partial\varphi}\dot{\varphi}^2=\alpha M^3\exp 
\left(-\frac{\alpha\varphi}{M}\right)+\frac{\beta}{M}
(\rho_\nu-3p_\nu)
\ee
involves, however, the cosmon-neutrino coupling
\be\label{F2}
\beta(\varphi)=-M\frac{\partial \ln m_\nu(\varphi)}{\partial\varphi},
\ee
with $m_\nu$ the $\varphi$-dependent average neutrino mass. The coupling $\beta$ is large only in the range of $\varphi$ which corresponds to the second step of the crossover (CR2 in Fig. 1.) It plays no role as long as neutrinos are relativistic. 

The Hubble parameter obeys 
\be\label{F3}
H^2=\frac{\rho}{3M^2}~,~\rho=\rho_h+\rho_r+\rho_m+\rho_\nu,
\ee
with 
\be\label{F4}
\rho_h=V+\frac{k^2}{2}\dot{\varphi}^2~,~p_h=-V+\frac{k^2}{2}\dot{\varphi}^2,
\ee
and $\rho_{r,m,\nu}$ the energy densities of radiation, matter and neutrinos, respectively. While $\rho_r$ and $\rho_m$ obey the usual conservation equations, $\dot{\rho}_r=-4H\rho_r,~\dot\rho_m=-3H\rho_m$, the neutrinos exchange energy momentum with the cosmon due to the variable mass
\ba\label{F5}
\dot\rho_\nu+3H(\rho_\nu+p_\nu)=-\frac\beta M
(\rho_\nu-3p_\nu)\dot\varphi,\nn\\
\dot\rho_h+3H(\rho_h+p_h)=\frac{\beta}{M}(\rho_\nu-3 p_\nu)\dot{\varphi}.
\ea
(The second equation follows from eqs. \eqref{F1}, \eqref{F4}.)

We may follow the evolution in terms of $y=\ln a+y_0$ instead of time \cite{HW,WR,CWNEU},
\ba\label{F6}
\partial_y\ln V&=&-\frac\alpha M \partial_y\varphi =-\alpha\sqrt{\frac{6(\Omega_h-\Omega_V)}{k^2}},\nn\\
\partial_y\ln \rho_h&=&-6\left(1-\frac{\Omega_V}{\Omega_h}\right)+\tilde \gamma(1-3w_\nu)
\frac{\Omega_\nu}{\Omega_h}\partial_y\ln V,\nn\\
\partial_y\ln \rho_\nu&=&-3(1+w_\nu)-\tilde\gamma(1-3w_\nu)\partial_y\ln V,\nn\\
\partial_y\ln \rho_r&=&-4~,~\partial_y\ln \rho_m=-3,
\ea
with $\Omega_V=V/\rho,~\Omega_{r,m,\nu,h}=\rho_{r,m,\nu,h}/\rho,~w_\nu=p_\nu/\rho_\nu$. The parameter
\be\label{84A}
\tilde\gamma=-\beta/\alpha
\ee
may depend on $\varphi$. Evaluated at the present value of $\varphi$ it is the same as in eq. \eqref{6}. It will determine the precise timing of the crossover to dark energy domination. The system of differential equations \eqref{F6} can be solved numerically \cite{CWNEU}. 

As long as neutrinos are relativistic one has $w_\nu=1/3$ and the terms $\sim\tilde\gamma$ can be neglected. For the radiation dominated epoch we can neglect $\rho_m$ and incorporate $\rho_\nu$ into $\rho_r,~\rho_r=\bar \rho_r M^4\exp (-4y)$. For the matter dominated period neutrinos can be neglected as long as they are relativistic, similar to radiation. We only need to keep $\rho_m=\bar \rho_m M^4\exp (-3y)$. We may combine the discussion of these periods by taking $\rho_d=\bar\rho M^4\exp (-ny)$ for the energy density of all other components except the cosmon, with $n=4(3)$ for radiation (matter) domination. 

The last epoch in the cosmological evolution starts when neutrinos become non-relativistic. The terms proportional to the cosmon-neutrino couplings $\beta$ in eq. \eqref{F5} can no longer be neglected. For large enough $\tilde\gamma$ they stop effectively the further change of $\varphi$, such that $V'(\varphi)$ acts like a cosmological constant. This scenario of ``growing neutrino quintessence'' \cite{ABW,CWNEU,GNQ2,GNQ8,GNQ10,GNQ9} relates the present dark energy density to the average neutrino mass
\be\label{F28}
\rho^{\frac14}_h(t_0)=1.27
\left(\frac{\tilde\gamma m_\nu(t_0)}{{\rm eV}}\right)^{\frac14}
10^{-3}{\rm eV}.
\ee 
(Here $\tilde\gamma$ is evaluated today.) A realistic present dark energy fraction $\Omega_h(t_0)\approx 0.7$ is found for 
\be\label{106A}
\tilde\gamma m_\nu(t_0)=6.15{\rm eV}.
\ee
This relation remains valid with good accuracy even in presence of the large scale non-linear neutrino lumps that form and dissolve periodically after redshift $z\approx 2$ \cite{ABFPW}.

\subsection{Approximate analytic solution}

For a constant kinetial $k=1$ one finds for radiation or matter domination the standard scaling (tracker) solution for quintessence with an exponential potential \cite{CW3}. For slowly varying $k(\varphi)$ according to eq. \eqref{L2} we may therefore use the approximation of a solution in the vicinity of the scaling solution. The difference between the cosmon field according to a model with $\varphi$-dependent kinetial on one side, and the scaling solution on the other side, is denoted by $M\delta(y)$. We will derive next an approximate analytic solution for $\delta(y)$. The resulting time evolution of the early dark energy fraction is found as
\be\label{91A}
\Omega_h=\frac{nB(\chi)}{4},
\ee
instead of $\Omega_h =nB_\infty/4$ in case of a constant kinetial. Since $B(\chi)$ changes only very mildly for recent cosmology the results \cite{CWNEU} of investigations with constant $B_\infty$ continue to be a very good approximation.

For the evolution equations for $\varphi$ and $\ln\rho_h$ we make the ansatz
\be\label{F7}
\rho_h=f(\varphi)\rho_d~,~\varphi=M
\left(\frac{ny}{\alpha}+\delta(y)\right),
\ee
such that 
\ba\label{F8}
\partial_y\ln f&=&n-6+\frac{6}{f\bar\rho}\exp (-\alpha\delta),\\
\partial_y\delta&=&-\frac n\alpha+
\sqrt{\frac{6f}{k^2(1+f)}
\left (1-\frac{1}{f\bar\rho}\exp(-\alpha\delta)\right)}.\nn
\ea
For constant $k^2$ one recovers the scaling solution \cite{CW3,CW2,QU4} with a constant fraction of early dark energy $\Omega_e$, $\delta=0,~\partial_y f=0$,
\be\label{F9}
\bar \rho=\frac{6}{(6-n)f}~,~\Omega_e=\frac{f}{1+f}=
\frac{nk^2}{\alpha^2}.
\ee

For a smooth enough $\varphi$-dependence of $k^2$ we therefore expect a behavior close to this scaling solution. We employ 
\be\label{F10}
\frac{1}{f}=\left(1-\frac n6\right)\bar \rho\exp\big [-\alpha\zeta(y)\big]
\ee
and find 
\ba\label{F11}
&&\partial_y\zeta=\frac{6-n}{\alpha}
\Big(\exp \big[-\alpha(\delta+\zeta)\big]-1\Big),\\
&&\partial_y\delta=-\frac n\alpha\nn\\
&& +\sqrt{\frac{n\Omega_h}{k^2}}
\sqrt{1+\left(\frac 6n-1\right)\Big(1-\exp\big[-\alpha(\delta+\zeta)\big]\Big)},\nn
\ea
with 
\be\label{F12}
\Omega_h(y)=\frac{f(y)}{1+f(y)}.
\ee
For the variables
\ba\label{F13}
\Delta&=&\left(\frac 6n-1\right)\Big(1-\exp \big[-\alpha(\delta+\zeta)\big]\Big),\nn\\
u&=&1-\frac{\alpha^2\Omega_h}{nk^2}
\ea
one obtains
\ba\label{F14}
\partial_y\Delta&=&\big[6-n(1+\Delta)\big]
\big(\sqrt{(1+u)(1+\Delta)}-1-\Delta\big),\nn\\
\partial_y u&=&(1+u)
\left\{ \frac{M \partial\ln k^2}{\partial\varphi}
\left(\frac n\alpha +\partial_y\delta\right)-
\frac{\alpha\Omega_h}{1+f} \partial_y\zeta \right).\nn\\
\ea
At this point we assume that $(nM/\alpha)\partial\ln k^2/\partial\varphi$ is small. We can then expand in small $\Delta$ and $u$,
\ba\label{F15}
\partial_y\Delta&=&\frac{n-6}{2}(\Delta-u),\nn\\
\partial_y u&=&\frac{nM}{\alpha}
\frac{\partial\ln k^2}{\partial\varphi}-n\Omega_h(1-\Omega_h)\Delta.
\ea

With $\partial\ln k^2/\partial\varphi$ and $\Omega_h$ varying slowly the solution approaches the particular approximate solution
\be\label{F16}
\bar\Delta=\bar u=\frac{M}{\alpha \Omega_h(1-\Omega_h)}
\frac{\partial\ln k^2}{\partial\varphi}.
\ee
Indeed, if we neglect the $y$-dependence of $\bar\Delta$ and $\bar u$ one has for $u'=u-\bar u,\Delta'=\Delta-\bar\Delta$ the linear evolution
\ba\label{F17}
&&\partial_y{\Delta'\choose  u'}=A{\Delta'\choose u'},\nn\\
&&A=\frac{n-6}{2}
\left(\begin{array}{ccc}
1&,&-1\\
\frac{-2n\Omega(1-\Omega)}{n-6}&,&0
\end{array}\right).
\ea
The eigenvalues of the stability matrix $A$ are both negative, implying an exponential decrease of $\Delta'$ and $u'$ as $y$ increases. We conclude that cosmology approaches a solution with non-vanishing early dark-energy fraction decreasing with decreasing $k^2$, 
\be\label{F18}
\Omega_h=\frac{nk^2}{\alpha^2}(1-\bar u).
\ee
Solving eqs. \eqref{F18},\eqref{F16} for $\Omega_h$ we end with a general formula for slowly varying $k(\varphi)$,
\be\label{96A}
\Omega_h\approx\frac{nk^2}{\alpha^2}-\left(1-\frac{nk^2}{\alpha^2}\right)^{-1}
\frac{M}{\alpha}\frac{\partial\ln k^2}{\partial\varphi}.
\ee
Here we recall that for generic models of quintessence the formulation with exponential potential \eqref{I2} and possibly varying kinetial $k(\varphi)$ can be obtained by an appropriate rescaling of the scalar field.

Let us now turn to our model with $k^2=M\alpha/\big (2\kappa(\varphi-\bar\varphi)\big)$ and 
\be\label{F19}
\frac{\partial\ln k^2}{\partial\varphi}=-\frac{1}{\varphi-\bar\varphi}=-\frac{2\kappa k^2}{M\alpha}.
\ee
One has 
\be\label{F20}
\bar u=-\frac{2\kappa}{n(1-\Omega_h)},
\ee
and for $\Omega_h\ll 1$ a small parameter $\kappa\ll 1$ indeed implies $\bar u\ll 1$. In turn the early dark energy fraction
\be\label{F21}
\Omega_h=\frac{n}{2\kappa}
\frac{M}{\alpha(\varphi-\bar\varphi)}=
\frac{n}{2\kappa\ln \left(\frac{\chi^2}{m^2}\right)}=
\frac{nB(\chi)}{4}
\ee
decreases logarithmically for increasing $\chi$.

\subsection{Bounds on parameters}

Besides the determination of $\tilde\gamma m_\nu(t_0)$ by a measurement of the present dark energy fraction \eqref{106A} we can use bounds on early dark energy for an estimate of the parameter $\kappa$. 

For nucleosynthesis the dimensionless ratio $V/\chi^4=\mu^2/\chi^2=V'/M^4$ is of the order (MeV)$^4/(10^{18}$GeV)$^4\approx10^{-84}$. With $m^2/\mu^2\approx 10^7$ this implies $\ln(\chi/m)\approx 89$,
\be\label{F22}
\Omega^{(ns)}_h\approx\frac{1}{89\kappa}.
\ee
If we require $\kappa<1/2$ in order to maintain small $\bar u$ this yields a dark energy fraction larger than $2\%$ which could be detectable in the future \cite{CW2,BS,BHM}. 

For the present epoch one has 
\be\label{F23}
\frac{V'(\varphi_0)}{M^4}=\frac{\mu^2}{M^2}=
\left(\frac{2\cdot 10^{-3}{\rm eV}}{2.44\cdot 10^{18}{\rm GeV}}\right)^4
=(0.67\cdot 10^{-60})^2.
\ee
This sets the scale of our model
\be\label{F24}
\mu=1.64\cdot 10^{-33}{\rm eV},
\ee
similar to the present value of the Hubble parameter. With $\ln(M/\mu)\approx 138.55$, $\ln(m/\mu)\approx 16.1$ one has $\ln(M/m)=122.4$ and therefore 
\be\label{F25}
\alpha^2=490 \kappa.
\ee
Interestingly, for $\kappa<1/2$ one finds an upper bound on $\alpha,~\alpha < 15.6$. 

Over the restricted range since last scattering $k^2$ has changed only little
\be\label{F26}
k^{-2}(z)=1-\frac{6\kappa}{\alpha^2}\ln (1+z),
\ee
such that at last scattering the relation $\Omega_h\approx 3/\alpha^2\gtrsim 1/80$ is valid,
\be\label{104A}
\Omega^{ls}_h=\frac{3}{490\kappa}.
\ee
For $\kappa<0.5$ the model therefore predicts a lower bound on the fraction of dark energy at last scattering
\be\label{F27}
\Omega^{ls}_h\geq 0.012.
\ee

\smallskip\noindent
This is at the borderline of a possible detection with present observations \cite{A2a,A2b,Re,Sievers:2013wk,A2d,PEDE,PL}. One therefore infers the bound $\kappa\gtrsim 0.5$. We may take a value $\kappa=0.5$ which is compatible with observation and consistent with our approximate solution with small $\bar u$.

Formally, we may combine eqs. \eqref{F18}, \eqref{F20}, \eqref{F25} in order to obtain for $k^2\approx 1,~n\approx 3,~\Omega^{ls}_h\ll1$ the relation
\be\label{105A}
\Omega^{ls}_h=\frac{3}{490\kappa}+\frac{1}{245}.
\ee
This would imply a minimal value for $\Omega^{ls}_h$, close to the lower bound quoted in ref. \cite{PEDE}. (We recall, however, that the observational bound may change if the other ingredients of our model are included in the parameter estimation.) The minimum is reached, however, only for large $\kappa$ for which our approximation no longer holds. We do not expect a qualitative change for somewhat larger values of $\kappa$. While the analytical estimate becomes inaccurate for $\kappa\gtrsim 0.5$, a numerical solution can be extended easily to larger $\kappa$. It will be interesting to see if a saturation with a minimal value of $\Omega^{ls}_h$, as suggested by eq. \eqref{105A}, takes place for increasing $\kappa$. A lower bound on $\Omega^{ls}_h$ would make the present model distinguishable from $\Lambda CDM$ where $\Omega^{ls}_h$ is practically zero. 

In summary, the late cosmology of our model resembles closely growing neutrino quintessence with a variable cosmon-neutrino coupling $\beta$ \cite{CWNEU}. The interesting new features are an explanation of a large effective value for $\alpha$ in terms of the approach to the IR-fixed point, and the association of large positive $\tilde\gamma=-\beta/\alpha$ with a crossover affecting the neutrino masses in the recent and present cosmological epoch.

\section{Ultraviolet fixed point}
\label{Past fixed point}

The fixed point that is relevant for the infinite past $t\to-\infty$ (``past fixed point'') corresponds to $\chi\to 0$. It is characterized by an anomalous dimension $\sigma$ that appears in the scalar kinetic term. (In the language of universal critical exponents $\sigma$ corresponds to $-\eta$). The approach to the fixed point corresponds to $B^{-1}\to 0$, with limiting behavior of the flow equation \eqref{7A} given by 
\be\label{PF1}
\mu\partial_\mu B^{-1}=-\sigma B^{-1}.
\ee
At the fixed point scale symmetry is exact and not spontaneously broken. With all particle masses vanishing for $\chi\to 0$ the model contains only massless modes at the UV-fixed point. It is the existence of this fixed point that makes quantum gravity non-perturbatively renormalizable. We discuss here in more detail its possible properties. 

\subsection{Renormalized scalar field}

The solution of eq. \eqref{PF1},
\be\label{PF2}
B=\left(\frac m\chi\right)^\sigma,
\ee
contains an explicit mass scale $m$, in addition to the mass scale $\mu$. (The ratio $m/\mu$ can be considered as a dimensionless coupling that specifies $B$ besides the dimensionless parameter $\sigma$.) For $\sigma\neq 2$ one can absorb $m$ in the definition of a renormalized field 
\be\label{PF3}
\chi_R=\left(1-\frac\sigma 2\right)^{-1}
B^{\frac12}\chi=
\left(1-\frac{\sigma}{2}\right)^{-1}
m^{\frac \sigma 2}\chi^{1-\frac \sigma 2}.
\ee
For $\sigma<2$ the UV-limit $\chi\to 0$ corresponds to $\chi_R\to 0$, while for $\sigma<2$ this limit implies $\chi_R\to-\infty$. In terms of the renormalized field the effective action contains a scalar kinetic term with standard normalization
\ba\label{PF4}
\Gamma&=&\int_x\sqrt{g}
\left\{ \frac12\partial^\mu\chi_R\partial_\mu\chi_R+
e_\sigma|\chi_R|{^{\frac{4}{2-\sigma}}}
\left(\mu^2-\frac12 R\right)\right\},\nn\\
e_\sigma&=&\left|1-\frac\sigma 2\right|^{\frac{4}{2-\sigma}}m^{-\frac{2\sigma}{2-\sigma}}.
\ea

For the particular case $\sigma=1$ this yields $(\lambda=\mu^2/m^2)$
\ba\label{PF5}
\Gamma&=&\int_x\sqrt{g}
\left\{\frac12\partial^\mu\chi_R\partial_\mu\chi_R+\frac{\lambda}{16}\chi^4_R-
\frac{\chi^4_R}{32m^2}R\right\}.
\ea
The last term $\sim R$ vanishes for $\chi_R/m\to 0$, such that no mass scale remains in this limit. We may define a dimensionless coupling
\be\label{122A}
\tilde\lambda_R=\frac{V}{\chi^4_R}=\frac{\mu^2\chi^2}{\chi^4_R}=\frac{\mu^2}{m^2}
\left|\frac{\sigma}{2}-1\right|{^{-\frac{4}{\sigma-2}}}
\left(\frac{|\chi_R|}{m}\right)^{-\frac{4(\sigma-1)}{\sigma-2}}.
\ee
For $\sigma=1$ one has $\tilde\lambda_R=\lambda/16$. For $\sigma>1$ we distinguish two cases. For $\sigma<2$ the UV-fixed point is realized for $|\chi_R|\to 0$. In this limit $\tilde\lambda_R$ goes to zero. The potential term becomes subleading and can be neglected in the UV-limit. With both $V$ and $\chi^2 R$ neglected in the UV-limit the effective action contains indeed no mass scale. For $\sigma>2$ the UV-fixed point is approached for $|\chi_R|\to\infty$. Again $\tilde \lambda_R$ vanishes in this limit and $\Gamma$ does not involve a mass scale in the UV-limit. This behavior demonstrates scale invariance at the fixed point very explicitly. For the boundary case $\sigma=2$ one finds a logarithmic dependence of $\chi_R$ on $\chi$
\be\label{PF10}
\chi_R=m\ln\left(\frac\chi m\right)~,~\chi^2=m^2\exp \left(\frac{2\chi_R}{m}\right).
\ee
The fixed point is now realized for $\chi_R\to-\infty$ where both $V/\chi^4_R$ and $\chi^2 R/\chi^4_R$ vanish. We conclude that for the whole range $\sigma\geq 1$ the effective action describes an ultraviolet fixed point. 

The scale symmetry realized at the fixed point is of a non-standard type due to the non-vanishing anomalous dimension $\sigma$. While the renormalized scalar field $\chi_R$ scales proportional to mass, the original scalar field $\chi$ scales $\sim$ mass$^{2/(2-\sigma)}$. (For the example $\sigma=1$ one finds a scaling of $\chi\sim$ mass$^2$). Thus the effective action becomes invariant under the scaling
\be\label{PF6}
g_{\mu\nu}\to \alpha^2 g_{\mu\nu}~,~\chi\to \alpha^{-\frac{2}{2-\sigma}}\chi~,~\chi_R\to\frac 1\alpha \chi_R.
\ee
An interesting particular case is $\sigma=3$ where $\chi$ scales with the same factor as the metric. The term $\sim R$ involves a scale symmetry violation which vanishes in the limit $\chi/m\to 0$. It characterizes a relevant parameter for the deviation from the fixed point as $\chi$ increases. For $\sigma=1$ the term $\sim \mu^2\chi^2$ is invariant under the scaling \eqref{PF6}, as easily visible in eq. \eqref{PF5}. This situation changes for $\sigma>1$, where the potential becomes a relevant deviation from the fixed point as well. As an example we may consider $\sigma=3/2$ where $\chi$ scales $\sim$ mass$^4$ and 
\be\label{PF7} 
\Gamma=\int_x\sqrt{g}\left\{\frac12 \partial ^\mu\chi_R\partial_\mu\chi_R+2^{-16}
\mu^2\frac{\chi^8_R}{m^6}-2^{-17}
\frac{\chi^8_R}{m^6}R\right\}.
\ee
The two last terms vanish in the limit $\chi/m\to 0$ and correspond both to relevant parameters for deviations from the fixed point. A second example is $\sigma=3$,
\be\label{124A}
\Gamma=\int_x\sqrt{g}
\left\{\frac12\partial^\mu\chi_R\partial_\mu\chi_R+16\frac{m^6}{\chi^4_R}
\left(\mu^2-\frac12 R\right)\right\},
\ee
where $\chi/m\to 0$ is realized for $\chi_R\to-\infty$. At the fixed point we are left for both examples with a free massless scalar field. This simplicity makes the existence of such a fixed point rather plausible.

We observe that for $\sigma<1$ the coefficient $e_\sigma$ decreases less than $\sim m^{-2}$ for $m\to \infty$. As a consequence the dimensionless quantity $\tilde\lambda_R=V/\chi^4_R$ diverges for $\chi/m\to 0$. No fixed point is obtained in this case.

At the fixed point the term $\sim R$ vanishes. However, one may expect the presence of higher order invariants, as given by eq. \eqref{1A}. Such terms are scale invariant and therefore compatible with dilatation symmetry if the dimensionless quantities $C$ and $D$ are constant. (Slowly running $C$ and $D$ would be considered as marginal parameters for deviations from the fixed point.) In the limit $\chi\to 0$ these terms dominate the graviton propagator and the graviton-graviton scattering at nonzero momentum \cite{CWET}. 

It will be interesting to see by an actual calculation if a fixed point with the postulated properties exists. For the moment being our model only gives an illustration of the interesting cosmological consequences of such a fixed point. If an UV-fixed point is found to exist the important task will be the understanding of small deviations from the fixed point, as encoded in the behavior of $\beta$-functions close to their zeros. This will determine the coupling $\sim \chi^2 R$ and, for $\sigma>1$, the term $\sim \mu^2\chi^2$, as well as the flow of couplings of the standard model of particle physics. We emphasize that a fixed point with the simple effective action \eqref{2E} requires a substantial anomalous dimension $\sigma>1$. In turn, this induces an inflationary stage and its end.

\subsection{Gauge hierarchy}
We will next address possible interesting consequences of an UV-fixed point for particle physics, in particular the gauge hierarchy problem. This concerns the possibility that the effective coupling $\epsilon_H$ between the Higgs-doublet and the cosmon, which determines the Fermi scale, is driven to very small values by its flow in the vicinity of the UV-fixed point. 

The gauge hierarchy is related to the small value of the effective coupling $\epsilon_H$ which appears in the quantum effective potential for the Higgs doublet $\tilde h$ \cite{CI,VG}, 
\be\label{143A}
\tilde V_h=\frac12 \lambda_h(\chi/\mu) (\tilde h^\dagger\tilde h-\epsilon_h(\chi/\mu)\chi^2)^2~,~\epsilon_H=\lambda_h\epsilon_h.
\ee
For the present range of $\chi$ the function $\epsilon_h(\chi/\mu)$ must be (almost) independent of $\mu$ and have reached a very small value $\epsilon_h(\chi/\mu=M/\mu)=5\cdot 10^{-33}$. Besides their dependence on $\chi/\mu$ the functions $\lambda_h$ and $\epsilon_H$ also depend on $\hat h^\dagger\tilde h/\chi^2$. This latter dependence is described by the standard model $\beta$-functions. The running of $\epsilon_H$ with $\hat h^\dagger\tilde h/\chi^2$ is given by a perturbatively small anomalous dimension \cite{CWGH,CWFT}. We neglect this small effect and use $\epsilon_H\approx 10^{-32}$ independently of $\tilde h^\dagger\tilde h/\chi^2$. (A small value of $\epsilon_H$ for a Higgs field value of the order of the dynamical Planck mass $\chi$ remains small for a Higgs field value equal to the Fermi scale. This property reflects the (almost) second order character of the electroweak phase transition - the associated effective scale invariance of the non-gravitational physics protects a small value of the 
Higgs mass term \cite{CWFT,CWGH,CWMH,Bar,He,NiMe,NiMe2,FKV}.) We explore here if the small value of $\epsilon_H$ can be caused by the running of $\epsilon_H$ near the UV-fixed point, before it is stopped at the crossover for $\chi\approx m$. 

In order to understand this issue we first consider the effective renormalized quartic coupling for the cosmon for $\chi\ll m$,
\be\label{143B}
\lambda_R=\frac{1}{24}\frac{\partial^4V}{\partial\chi^4_R}~,~V=\mu^2e_\sigma(\mu)|\chi_R|{^{\frac{4}{2-\sigma}}},
\ee
which differs from $\tilde\lambda_R$ in eq. \eqref{122A} only by a multiplicative constant. 
From 
\be\label{143C}
\lambda_R\sim\left(\frac{|\chi_R|}{\mu}\right)^{\frac{4(\sigma-1)}{2-\sigma}}\sim\left(\frac{\chi^2}{\mu^2}\right)^{\sigma-1}
\ee
we extract the flow equation
\be\label{143D}
\mu\partial_\mu\lambda_{R|\chi}=-2(\sigma-1)\lambda_R=A_\lambda\lambda_R.
\ee
For $\sigma>1$ one finds $A_\lambda<0$ and the running coupling $\lambda_R$ is asymptotically free in the ultraviolet. We observe that the anomalous dimension $A_\lambda$ can be quite large.  A given trajectory (model) can be specified by the value of $\lambda_R$ at $\chi/\mu=1$. This is typically a rather small value, corresponding to the close vicinity to the fixed point. For larger values of $\chi/\mu$ the renormalized coupling $\lambda_R$ increases. 

We next turn to the cosmon-Higgs coupling that we define as
\be\label{144E}
\epsilon_H=-\frac{\partial^2\tilde V_h}{\partial(\chi^2)\partial(\tilde h^\dagger\tilde h)}.
\ee
The corresponding renormalized coupling 
\be\label{144F}
\epsilon_R=-\frac{\partial^2\tilde V_h}{\partial(\chi^2_R)\partial(h^\dagger_Rh_R)}
\ee
involves $\chi_R$ and the renormalized Higgs doublet $h_R$. Using $B=(\chi/m)^\sigma,\chi_R/\chi=\sqrt{B}/\left(1-\frac\sigma2\right)$, one has
\be\label{144G}
\frac{\partial(\chi^2_R)}{\partial(\chi^2)}=\frac{B}{1-\frac\sigma2}.
\ee
Similarly, the kinetic coefficient $B_h$ of the Higgs doublet may depend on $\mu$ in the vicinity of the UV-fixed point, resulting in 
\be\label{144H}
\frac{\partial(h^\dagger_Rh_R)}{\partial(\tilde h^\dagger\tilde h)}=
\frac{B_h}{1-\frac{\sigma_h}{2}}~,~
\mu\partial_\mu\ln B_h=\sigma_h.
\ee
This relates $\epsilon_R$ and $\epsilon_H$
\be\label{144I}
\epsilon_H=\frac{B B_h}{\left(1-\frac\sigma2\right)\left(1-\frac{\sigma_h}{2}\right)}\epsilon_R.
\ee

Let us now assume that $\epsilon_R$ is asymptotically free in the UV, similar to $\lambda_R$,
\be\label{144J}
\mu\partial_\mu\epsilon_R=A_\epsilon\epsilon_R~,~A_\epsilon<0.
\ee
This results in a flow of $\epsilon_H$ according to
\be\label{144K}
\mu\partial_\mu\epsilon_H=(A_\epsilon+\sigma+\sigma_h)\epsilon_H=\sigma_\epsilon\epsilon_H.
\ee
(We take $\sigma$ and $\sigma_h$ approximately constant here.) While $\epsilon_R$ decreases with increasing $\mu,\epsilon_H$ can increase if the sum $\sigma+\sigma_h$ overwhelms the negative contribution $A_\epsilon$ such that $\sigma_\epsilon>0$, 
\be\label{144L}
\epsilon_H\sim \left(\frac{\mu}{\chi}\right)^{\sigma_\epsilon}.
\ee
Turned around, $\epsilon_H$ will then decrease for increasing $\chi$ and fixed $\mu$. 

The behavior \eqref{144L} is valid only for the vicinity of the UV-fixed point for $\chi\lesssim m$. For the vicinity of the standard model-fixed point, $\chi\gg m$, we assume that $\epsilon_H$ reaches rapidly its constant fixed point value. (Formally $\sigma_\epsilon\approx0$ for $\chi\gg m$.) Specifying the trajectory at a given ratio $\chi\ri/\mu$, $\epsilon\ri=\epsilon_H(\chi\ri/\mu)$, the value of $\epsilon_H$ for $\chi^2\gg m^2$ is reduced by a factor
\be\label{144M}
\epsilon_H\approx \left(\frac{\chi\ri}{m}\right)^{\sigma_\epsilon}\epsilon\ri.
\ee
This  factor could explain the gauge hierarchy. For $\epsilon\ri$ of the order one one needs
\be\label{144N}
\frac{\chi\ri}{m}\approx 10^{-\frac{32}{\sigma_\epsilon}}.
\ee
For example, for $\chi\ri=\mu$ a value $\sigma_\epsilon\approx 5-6$ would be sufficient for a decrease of $\epsilon_H$ between $\chi=\mu$ and $\chi=m$ by around 30 orders of magnitude. This would relate the smallness of the ratio Fermi scale/Planck mass and the small amplitude of primordial density fluctuations, cf. eq. \eqref{25C},
\be\label{144O}
\frac{\kl h\kr}{M}\sim {\cal A}^{\sigma_\epsilon/4}.
\ee
(For $\chi\ri\ll\mu$ smaller values of $\sigma_\epsilon$ would be sufficient to achieve the suppression factor needed for the gauge hierarchy.) 

The possible emergence of a gauge hierarchy, expressed by the tiny coupling $\epsilon_H(\chi\gg m)\approx 10^{-32}$, can be viewed from different perspectives. While $\epsilon_H$ should be approximately constant for $\chi\gg m$, nothing prevents an increase of $\epsilon_H$ for $\chi\ll m$, such that values of the order one can be reached for small enough $\chi$. The increase of $\epsilon_H$ towards the UV-fixed point remains compatible with an asymptotically free renormalizable coupling $\epsilon_R$. For sufficiently small $\chi$ all asymptotically free renormalized couplings are very small. If the anomalous dimension $|A_\epsilon|$ for the coupling $\epsilon_R$ is smaller than the corresponding one for other couplings the coupling $\epsilon_R$ still remains small at the crossover scale where the flow effectively stops and $\epsilon_R$ roughly equals $\epsilon_H$. 

The coupling $\epsilon_H$ measures the distance from the electroweak phase transition which is of second order (up to small QCD-effects). This guarantees that its flow vanishes for $\epsilon_H=0$. Such a setting generalizes to a large class of models, including grand unified models. Then $\epsilon_H$ measures the distance from the hyperface in coupling constant space corresponding the phase transition. While the location of this hypersurface may be complicated in a given basis for the couplings (often associated with a ``fine tuning problem'') the general structure of the flow equation for $\epsilon_H$ remains the same \cite{CWWTL}.

The two steps in the flow of $\epsilon_H$, first a fast decrease for $\chi\ll m$ and then an almost constant behavior for $\chi\ll m$, would realize an old idea for a possible explanation of the gauge hierarchy \cite{CWGH}. The necessary large values of anomalous dimensions are often found in the gravitational contribution to the flow \cite{Rev,HPRW,Per,Ei}. In our scenario $\sigma$ has to be large in order to realize an UV-fixed point. Then also $\sigma_\epsilon$ will typically have a large value, unless some particular cancellation occurs in eq. \eqref{144K}. Without an explicit computation of the $\mu$-flow equation our discussion remains an educated guess. It clearly shows, however, that an ultraviolet fixed point with large anomalous dimensions could play an important role for the gauge hierarchy problem. This also applies for a possible understanding of the value of the Higgs boson mass. If the flow of the quartic Higgs coupling $\lambda_h(\chi/\mu)$ exhibits a large positive anomalous dimension the 
``asymptotic safety scenario for the Higgs boson mass'' is realized \cite{SW}, which has led to a predicted value $m_h\approx 126$GeV with a few GeV uncertainties, the best value today being around $129$GeV.

It is an attractive speculation that extensions of our model which comprise the standard model of particle physics exhibit an ultraviolet fixed point for which all renormalized dimensionless couplings are asymptotically free. The effective action at the fixed point comprises then only kinetic terms for the renormalized fields. Gravitational couplings as $C$ may be marginal. Graviton fluctuations could be responsible for large anomalous dimensions.

\section{Field relativity}
\label{Field relativity}

Once quantum fluctuations are included on the level of the quantum effective action the corresponding field equations can be solved with arbitrary field variables. Values and correlations of physical observables are independent of the choice of fields used to describe them \cite{CWVN}. This exact property may be called ``field relativity'' \cite{CWU}. Indeed, observables are expressed as functionals of fields. Again, they can be written in terms of arbitrary field variables. In general, the specific functional expression for a given physical observable will be changed under a change of field variables (see refs. \cite{CWVN,CWET} for the transformation of some quantities relevant for cosmology as temperature or proper time.) We stress that only dimensionless quantities can be physical observables \cite{CWVN}. Different choices of field variables are called different frames. A well known example for a frame transformation is the Weyl transformation from the Jordan to the Einstein frame \cite{We,Di} that we 
have employed in sect. \ref{Primordial cosmology and inflation}.

In the present section we employ frame transformations for several different purposes. We first show that a very large class of coupled scalar-gravity models can be brought to the form \eqref{1}, with $B(\chi/\mu)$ the only free function. Typically this holds if the field equations contain no more than two derivatives and the scalar potential is monotonic. A formal treatment of a large class of such models, including the ones of the Horndeski type \cite{Hor}, can be found in ref. \cite{CWMG}. Our discussion of a crossover between a past and future fixed point can therefore be carried over to a large class of models. 

We have described the crossover as a ``kinetial crossover'' where the relevant information is encoded in the scalar kinetic term, i.e. the function $B(\chi/\mu)$. Field transformations can be used to express the same physics as a ``potential crossover'' \cite{CWSF}, where the information is now contained in the shape of the scalar potential $V(\chi/\mu)$, while the kinetic term has a standard normalization. We also present a ``primordial flat frame'' for which the cosmological solution approaches flat space in the infinite past for models without higher order curvature invariants. Finally, we cast the effective action into the form of a free scalar field coupled to gravity. While the kinetic term is standard and the potential quadratic, the crossover information is now contained in a $\chi$-dependent function multiplying the curvature scalar. Having at hand the formulation of the ultraviolet and infrared fixed points in different frames may facilitate the search for such fixed points in a genuine quantum 
gravity calculation. 

We omit in this section higher order curvature invariants as in eq. \eqref{1A}. They would have to be transformed appropriately under field transformations. This section therefore deals with various expressions for the quantum effective action encoded in eq. \eqref{1}.

\subsection{Field transformations within Jordan frames} 

We will call ``Jordan frames'' the choice of fields for which the curvature scalar in the effective action \eqref{1} is multiplied by $\chi^2$. We allow for a general potential $V(\chi)$ instead of $\mu^2\chi^2$ in eq. \eqref{1}. In contrast, the ``Einstein frame'' or ``big bang frame'' \eqref{I2} has a constant coefficient $M^2$ in front of the curvature scalar. The Einstein frame is unique, up to a choice of the scalar field $\varphi$ which may be replaced by $\chi$ or a field $\sigma$ with standard normalization of the kinetic term. The Jordan frames, however, are not yet uniquely fixed, since there exist field transformations keeping the term $\sim \chi^2 R$ invariant, while changing $V(\chi)$ and $B(\chi)$. The particular choice of fields where $V(\chi\to\infty)=\mu^2\chi^2$ will be called ``freeze frame''. We may parametrize the Jordan frames by two dimensionless functions $B(\chi/\mu)$ and 
\be\label{J1}
v\left(\frac{\chi}{\mu}\right)
=\frac{V(\chi)}{\chi^4}.
\ee
Indeed, the most general quantum effective action with no more than two derivatives takes in the Jordan frame the form
\be\label{106B}
\Gamma=\int_x\sqrt{g}
\left\{-\frac12\chi^2 R+v(\chi)\chi^4 +\frac12\big(B(\chi)-6\big)\partial^\mu\chi\partial_\mu\chi\right\}.
\ee
The two functions $B$ and $v$ contain redundant information, since they can be changed by appropriate field transformations. 

Using appropriate field transformations we can bring a large class of effective actions with up to two derivatives into the generic form \eqref{106B}. For any positive and monotonically increasing function $F(\chi')$ multiplying the graviton kinetic term $-R$ we can choose a normalization of the scalar field $F= \chi^2$ in order to bring the system to the Jordan frame. We can then use the residual transformation within the Jordan frame in order to obtain $v=\mu^2/\chi^2$ such that $B(\chi)$ remains the only free function. Alternatively, we can obtain a constant scalar kinetic term at the prize of a more complicated function $v$. 

Consider the transformation
\be\label{J2}
\chi=h(\tilde \chi)~,~g_{\mu\nu}=\frac{\tilde\chi^2}{h^2(\tilde\chi)}\tilde g_{\mu\nu}.
\ee
This transforms the effective action \eqref{106B} to
\be\label{J3}
\Gamma=\int_x\sqrt{\tilde g}
\left\{-\frac12\tilde\chi^2\tilde R+\tilde v(\tilde\chi)\tilde\chi^4+\frac12
\big (\tilde B(\tilde \chi)-6\big)\partial^\mu\tilde\chi\partial_\mu\tilde\chi \right\},
\ee
leaving the coefficient of the curvature scalar form-invariant. In terms of the variables $\tilde\chi$ and $\tilde g_{\mu\nu}$ the new functions $\tilde B$ and $\tilde\chi$ read 
\ba\label{J4}
\tilde B&=&B\big(h(\tilde\chi)\big)
\left(\frac{\partial\ln h}{\partial \ln \tilde\chi}\right)^2,\nn\\
\tilde v&=&v\big(h(\tilde\chi)\big).
\ea
A pair of functions $(\tilde B,\tilde v)$ describes the same model as the pair $(B,v)$ if the two are related by eq. \eqref{J4} with a suitable choice of $h(\tilde \chi)$. The corresponding effective actions are related by a field transformation. A given model can be expressed by a whole family of Jordan frames, parametrized by $h(\tilde\chi)$. 

Of course, this equivalence also requires appropriate transformations in the particle physics sector. For example, preserving the canonical kinetic term for a fermion field $\tilde\psi$ requires
\be\label{141A}
\psi=\left(\frac{h}{\tilde \chi}\right)^{\frac32}\tilde \psi.
\ee
This implies that fermion masses $\sim\tilde\chi$ remain form-invariant under the rescaling, with the same dimensionless coupling $f$,
\be\label{141B}
f\sqrt{\tilde g}\tilde\chi\bar{\tilde \psi}\tilde\psi=f\sqrt{g}\chi\bar\psi\psi.
\ee
The rescaling leaves the dimensionless ratio between fermion mass and Planck mass, $m(\tilde\chi)/\tilde \chi=f$, invariant. 

\subsection{Kinetial crossover}

We can employ the field transformations \eqref{J2} in order to bring a large class of potentials $\tilde V(\tilde\chi)$ to the ``freeze form'' $V=\mu^2\chi^2$. Indeed, any function $\tilde v(\tilde\chi)$ which decreases monotonically with limits $\tilde v(\tilde \chi\to 0)\to\infty$, $\tilde v(\tilde\chi\to\infty)\to 0$ can be transformed to $v=\mu^2/\chi^2$ by choosing
\be\label{J15}
h(\tilde\chi)=
\frac{\mu}{\sqrt{\tilde v(\tilde\chi)}}.
\ee
The choice of the effective action \eqref{1} is therefore rather generic, since a large family of potentials can be brought to the particular form $V=\mu^2\chi^2$. 

A first example takes a constant potential
\be\label{142A}
\tilde V=\bar\lambda_c~,~\tilde v=\frac{\bar\lambda_c}{\tilde\chi^4}~,~h(\tilde\chi)=
\frac{\mu\tilde \chi^2}{\sqrt{\bar\lambda_c}}.
\ee
Eq. \eqref{J4} yields
\be\label{142B}
B=\frac{\tilde B}{4}.
\ee
This relates the large-$\chi$-behavior of the two models (A) and (B) in ref. \cite{VG}.

As a second example we may consider the models of ref. \cite{CWET} $(A,\alpha=$const.),
\be\label{121A}
\tilde v=\frac{\mu^2\tilde\chi^{-A}}{m^{2-A}+\tilde\chi^{2-A}}~,~\tilde B=\frac{4}{\alpha^2}.
\ee
With 
\be\label{121B}
h^2=(m^{2-A}+\tilde\chi^{2-A})\tilde\chi^A=\chi^2
\ee
one finds
\be\label{121C}
B=\tilde B
\left(1+\frac{(2-A)m^{2-A}}{Am^{2-A}+2\tilde\chi^{2-A}}\right)^2,
\ee
where $\tilde\chi$ is related to $\chi$ by eq. \eqref{121B}. For large $\chi/m$ one has $\chi=\tilde \chi,B=\tilde B$, while for $\chi\to 0$ the limiting behavior is
\be\label{121D}
\chi=m^{1-\frac A2}\tilde\chi^{\frac A2}~,~B=\frac{4\tilde B}{\tilde A^2}=\frac{4}{\tilde \alpha^2}.
\ee
For these models the flow equation for $B$ exhibits two fixed points with finite values of $B$, e.g. $4/\tilde\alpha^2$ for $\mu\to\infty$ and $4/\alpha^2$ for $\mu\to 0$. This is the type of models investigated in refs. \cite{VG,CWU}. For $\mu\to\infty$ one has $\sigma=-\partial \ln B/\partial \ln \chi\to 0$. The renormalized scalar field $\chi_R$ equals $\tilde \chi$ up to a constant factor, and $V/\chi^4_R$ diverges for $\chi_R\to 0$. This is not a fixed point in the sense of our previous discussion, but it could represent a possible fixed point in terms of different variables. 

As mentioned above, even much more general classes of models can be described by a kinetial crossover. In the appendix E we discuss models where the coefficient of the curvature scalar $\tilde \chi^2$ in eq. \eqref{J3} is generalized to $c_1\tilde\chi^2+c_2\mu^2$, and map such models to the kinetial crossover form \eqref{1}. This will shed light on the role of a possible term $\sim\mu^2 R$ for variable gravity models. 

\subsection{Potential crossover}

Alternatively, we may use the transformation \eqref{J4} in order to transform our models \eqref{1}, \eqref{7B} of a kinetial crossover to an equivalent model with a potential crossover. For this purpose we want to achieve a constant $\tilde B$, using for $h$ a solution of the differential equation
\be\label{J5}
\frac{\partial\ln h}{\partial\ln\tilde \chi}=
\sqrt{\frac{\tilde B}{B(h)}}.
\ee
Once $h(\tilde\chi)=\chi$ is computed in this way we can compute the associated scalar potential $V(\tilde\chi)=\tilde\chi^4 v\big(h^{-1}(\chi)\big)$.

As an example we consider the effective action \eqref{1} with $B(h)$ obeying eq. \eqref{5},
\be\label{J6}
\frac{1}{\kappa B(h)}-\ln B(h)=\ln
\left(\frac{h}{m}\right).
\ee
For small $h$ or large $B$ one has the limiting behavior
\be\label{J7}
B^{-1}=\frac hm
\ee
or 
\be\label{J8}
\frac{\partial \ln h}{\partial \ln \tilde \chi}=\sqrt{\frac{\tilde B h}{m}}.
\ee
The solution of eq. \eqref{J8} involves an integration constant $c_h$
\be\label{J9}
h=m\left(c_h-\frac12\sqrt{\tilde B}\ln \left(\frac{\tilde\chi}{m}\right)\right)^{-2}.
\ee
This yields the potential
\be\label{J10}
\tilde V=\tilde v\tilde\chi^4=
\frac{\mu^2\tilde\chi^4}{h^2}=\frac{\mu^2}{m^2}
\left[ c_h\tilde\chi-\frac12\sqrt{\tilde B}\tilde\chi\ln
\left(\frac{\tilde\chi}{m}\right)\right]^4,
\ee
which vanishes for $\tilde\chi\to 0$
\be\label{J11}
V(\tilde\chi\to 0)=\frac{\mu^2\tilde B^2}{16m^2}
\left[\tilde \chi\ln\left(\frac{m}{\tilde\chi}\right)\right]^4\to 0.
\ee

On the other hand, for large $h$ and small $B$ we use $B^{-1}=\kappa \ln \left(\frac hm\right)$ and therefore
\be\label{J12}
\frac{\partial\ln h}{\partial\ln \tilde \chi}=
\sqrt{\kappa\tilde B\ln \left(\frac hm\right)}.
\ee
The solution 
\be\label{J13}
h=m\exp 
\left\{\frac{\kappa\tilde B}{4}
\left[\ln \left(\frac{\tilde\chi}{m}\right)+\tilde c_h\right]^2\right\}
\ee
increases faster than $\tilde\chi$ for $\tilde\chi\to\infty$. The corresponding potential reads
\be\label{J14}
\tilde V=\frac{\mu^2}{m^2}\tilde\chi^4\exp 
\left\{-\frac{\kappa\tilde B}{2}
\left[\ln\frac{\tilde\chi}{m}+{\tilde c_h}\right]^2\right\}.
\ee
The full potential makes a crossover from eq. \eqref{J10} for $\tilde\chi\ll m$ to eq. \eqref{J14} for $\tilde \chi\gg m$. The two integration constants $c_h$ and $\tilde c_h$ are related in order to ensure a smooth matching, e.g. $c_h\approx \exp \{-\kappa\tilde B\tilde c^2_h/8\}$.

One may also choose a hybrid setting with a constant kinetic term $\tilde B$ for $\chi\to 0$, while for $\chi\to\infty$ one keeps the freeze frame $V=\mu^2\tilde\chi^2$. This is achieved by choosing $h(\tilde\chi\to\infty)=\tilde\chi$, while $h(\tilde \chi\to 0)$ is given by eq. \eqref{J9}.

\subsection{Primordial flat frame} 

Let us consider the frame where for $\chi\to 0$ the functions $\tilde v$ and $\tilde B$ are related by 
\be\label{P1}
\frac{\partial\ln \tilde v}{\partial \ln \tilde \chi}=-\tilde B(\tilde \chi)+
\frac{\partial \ln \tilde B}{\partial \ln \tilde \chi}.
\ee
This is the condition for finding for the infinite past flat space as a solution of the field equations derived from the action \eqref{J3} \cite{VG}. We can transform our crossover model \eqref{1}, \eqref{2B} to this ``primordial flat frame'' by a suitable choice of $h$ in eq. \eqref{J2}. The function $h(\tilde \chi)$ has to obey a differential equation which follows from 
\be\label{P2}
\tilde v=\frac{\mu^2}{h^2}~,~\tilde B=B(h)
\left(\frac{\partial \ln h}{\partial \ln \tilde\chi }\right)^2, 
\ee
namely
\ba\label{123A}
&&\left(2+\frac{\partial\ln B}{\partial\ln h}\right)
\frac{\partial\ln h}{\partial\ln \tilde\chi}-B
\left(\frac{\partial\ln h}{\partial\ln\tilde\chi}\right)^2\nn\\
&&\hspace{1.5cm} +2\frac{\partial}{\partial\ln\tilde\chi}\ln
\left(\frac{\partial\ln h}{\partial\ln \tilde\chi}\right)=0.
\ea
We are interested in $h\to 0$ where $B=(m/h)^\sigma$, $\partial\ln B/\partial\ln h=-\sigma$. In this approximation eq. \eqref{123A} is obeyed by
\be\label{123B}
\frac{\partial\ln h}{\partial\ln \tilde\chi}=(2+\sigma)
\left(\frac hm\right)^\sigma.
\ee
This yields the relation between $\chi$ and $\tilde\chi$,
\ba\label{123C}
\chi&=&h(\tilde\chi)=\hat m\left(\ln\frac{\bar m}{\tilde\chi}\right)^{-\frac1\sigma},\nn\\
\hat m&=&\big[\sigma(2+\sigma)\big]^{-\frac1\sigma}m,
\ea
with $\bar m$ an integration constant. One infers
\be\label{123D}
\tilde v=\frac{\mu^2}{\hat m^2}
\left(\ln\left(\frac{\bar m}{\tilde\chi}\right)\right)^{\frac2\sigma}
\ee
and
\be\label{123E}
\tilde B=\frac{2+\sigma}{\sigma\ln\left(\frac{\bar m}{\tilde\chi}\right)}.
\ee

In the primordial flat frame $\tilde B$ vanishes for $\tilde \chi\to 0$ (which corresponds to $\chi\to 0$), in contrast to the divergence of $B$ in the frame of eqs. \eqref{1}, \eqref{2B}. The dimensionless potential $\tilde v=\tilde V/\tilde\chi^4$ diverges with an inverse power of a logarithm instead of $v\sim\chi^{-2}$. Different frames can describe the same physical situation with rather different pictures. This extends to the form of the flow equations. For $\tilde \chi\to 0$ one has 
\ba\label{123F}
\frac{\partial\tilde B}{\partial\ln\tilde\chi}&=&\frac{\sigma}{2+\sigma}\tilde B^2,\nn\\
\frac{\partial\tilde v}{\partial\ln \tilde\chi}&=&-\frac{2}{2+\sigma}\tilde B\tilde v.
\ea
This transfers to the $\mu$-flow equation for fixed $\tilde\chi$ and $\tilde g_{\mu\nu}$ (instead of fixed $\chi$ and $g_{\mu\nu}$)
\be\label{123G}
\mu\partial_\mu\tilde B=-\frac{\sigma}{2+\sigma}\tilde B^2~,~\mu\partial_\mu\tilde v=\frac{2}{2+\sigma}
\tilde B\tilde v.
\ee

On the level of a given quantum effective action the flow equations in different frames can be obtained from each other by a simple transformation of field variables. The quantum computation of the flow equations needs more care. One may employ a field transformation which leaves the functional integral invariant. It will involve, however, a Jacobian from the functional measure. In practice, a computation is often done with the implicit assumption of a unit Jacobian. This singles out a particular frame. Two settings for which the ``classical action'' is related by a field transformation, while both use the same definition of the measure for the respective fields (e.g. unit Jacobian), result in different models that do not yield equivalent predictions for observations. In other words, it is sufficient for the UV-fixed point of our model that there exists a frame for which a quantum computation with unit Jacobian yields the flow equation \eqref{123G} or, equivalently, eq. \eqref{18A} supplemented with $\
partial v/\partial\ln \mu =2 v$.

\subsection{Asymptotic solution in the primordial flat frame}

We could have started our discussion of inflation with the quantum effective action \eqref{J3},
\ba\label{165A}
\Gamma&=&\int_x\sqrt{g}
\left\{-\frac12\chi^2 R+\bar\lambda\chi^4\ln\left(\frac{\bar m}{\chi}\right)\right.\nn\\
&&\left.+\left[\ln^{-1}\left(\frac{\bar m}{\chi}\right)-3\right]\partial^\mu\chi\partial_\mu\chi\right\},
\ea
where we take $\sigma=2$ for simplicity, $\bar\lambda=\mu^2/\hat m^2$, and we omit the tilde on the fields. In the absence of radiation and matter the field equations \eqref{C5a}, \eqref{C6a} read
\ba\label{165B}
&&\left(\frac{2}{\ln\left(\frac{\bar m}{\chi}\right)}-6\right)
\left(\frac{\ddot\chi}{\chi}+3H\frac{\dot\chi}{\chi}\right)
+\ln^{-2}\left(\frac{\bar m}{\chi}\right)\left(\frac{\dot\chi}{\chi}\right)^2\nn\\
&&+\bar\lambda\chi^2\left(4\ln\left(\frac{\bar m}{\chi}\right)-1\right)=12H^2+6\dot H,
\ea
and
\be\label{165C}
\left(H+\frac{\dot\chi}{\chi}\right)^2=
\frac{1}{3\ln\left(\frac{\bar m}{\chi}\right)}
\left(\frac{\dot\chi}{\chi}\right)^2+\frac{\bar\lambda}{3}\chi^2\ln
\left(\frac{\bar m}{\chi}\right).
\ee
In leading order the solution for $\chi\to 0,t\to -\infty$ corresponds to flat space with slowly increasing $\chi$ according to
\be\label{165D}
H=0~,~\dot\chi=\sqrt{\frac{\bar\lambda}{3}}\chi^2\ln^{\frac12}\left(\frac{\bar m}{\chi}\right),
\ee
such that the sum of scalar potential and (negative) kinetic term vanishes. The approximate time evolution of $\chi$ is given implicitly by
\be\label{165E}
\chi\ln^{\frac12}\left(\frac{\bar m}{\chi}\right)=\sqrt{\frac{3}{\bar\lambda}}(t_c-t)^{-1},
\ee
with $\chi\to 0$ for $t\to-\infty$. 

In the next to leading order the solution becomes
\be\label{165F}
H=\frac{c_H\dot\chi}{\ln\left(\frac{\bar m}{\chi}\right)\chi}~,~
\delta\chi=\frac{c_\chi\chi}{\ln\left(\frac{\bar m}{\chi}\right)},
\ee
with $\chi=\chi_0+\delta\chi$ and $\chi_0$ the leading order solution according to eq. \eqref{165D}. The field equations \eqref{165B}, \eqref{165C} are both obeyed for 
\be\label{165G}
c_H-c_\chi=\frac16,
\ee
such that we are left at this stage with two free integration constants $t_c$ and $c_H$. For $c_H\neq 0$ the qualitative evolution of the Hubble parameter reads
\be\label{165I}
H\approx \frac{c_H}{(t_c-t)\ln\big[\tilde m(t_c-t)\big]},
\ee
with $\tilde m$ varying only slowly with time. Geometry approaches flat space in the infinite past $(t\to-\infty)$, with a slowly vanishing or diverging Robertson-Walker scale factor depending on the sign of $c_H$,
\be\label{165J}
a=c_a\ln^{-c_H}\left(\frac{\bar m}{\chi}\right).
\ee
For the particular solution $c_H=0,c_\chi=-1/6$ one has the leading behavior
\be\label{165K}
H=\frac{\tilde c_H\dot\chi}{\ln^2\left(\frac {\bar m}{\chi}\right)\chi}.
\ee
In this case geometry approaches in the infinite past Minkowski space, with a constant scale factor $a_\infty$ according to 
\be\label{165L}
a=a_\infty \exp \left\{-\frac{\tilde c_H}{\ln \left(\frac{\bar m}{\chi}\right)}\right\}.
\ee

\subsection{Eternal universe}

The geometry given by the solution \eqref{165I} or \eqref{165K}, with $\chi$ according to eq. \eqref{165D}, is free of any singularity as long as $\chi$ remains finite. (The formal singularity of eq. \eqref{165E} for $t\to t_c$ is an artefact of the approximation, which is no longer valid for $t$ near $t_c$.) Space-time is geodesically complete. In this frame it is straightforward to see that a universe given by our solution is eternal. It has existed since the infinite past. As discussed in detail in ref. \cite{CWET}, physical time can be measured by the number of oscillations of wave functions. Physical time indeed goes to minus infinity in the limit $t\to-\infty$. Discrete oscillation numbers are the same in all frames, such that physical time is frame-independent. The eternity of the universe is therefore independent of the chosen frame. 

Unphysical singularities in the Einstein frame arise from a singularity of the field transformation. (See ref. \cite{CWET} for a more detailed discussion.) Indeed, a Weyl scaling $g'_{\mu\nu}=(\chi^2/M^2)g_{\mu\nu}$ brings the effective action \eqref{165A} to the form 
\ba\label{165M}
\Gamma&=&\int_x\sqrt{g'}
\Big\{-\frac{M^2}{2}R'+\bar\lambda M^4\ln \left(\frac{\bar m}{\chi}\right)\nn\\
&&+\frac{M^2}{\chi^2\ln \left(\frac{\bar m}{\chi}\right)}\partial^\mu\chi\partial_\mu\chi\Big\}.
\ea
This field transformation becomes singular for $\chi\to 0$, which corresponds to the infinite past in physical time or to the big bang singularity in the Einstein frame. With the identification 
\be\label{165N}
\bar\lambda\ln\left(\frac{\bar m}{\chi}\right)=\exp\left(-\frac{\alpha\varphi}{M}\right)
\ee
we recover the form of eq. \eqref{I2}, with kinetial 
\be\label{165O}
k^2=\frac{2\alpha^2}{\bar\lambda}\exp \left(-\frac{\alpha\varphi}{M}\right).
\ee
For $\sigma=2$ and $\bar\lambda=8\mu^2/m^2$ this agrees with eq. \eqref{33A}, providing for a direct link to the discussion of inflation in sect. \ref{Primordial cosmology and inflation}.

\subsection{Free scalar field coupled to gravity}

Another interesting frame change transforms the effective action \eqref{1} to a scalar field theory without self interactions,
\be\label{FF1}
\Gamma=\int_x\sqrt{\tilde g}
\left\{-\frac12 f(\tilde\chi)\tilde \chi^2\tilde R+\mu^2\tilde\chi^2+
\frac{\tilde K}{2}\partial^\mu\tilde\chi\partial_\mu\tilde\chi\right\},
\ee
with $\tilde\chi$-independent $\tilde K$. This  is achieved by transformations that leave $\sqrt{g}\chi^2$ invariant,
\be\label{FF2}
g_{\mu\nu}=\frac 1f\tilde g_{\mu\nu}~,~\chi= f\tilde \chi.
\ee
The transformed kinetic coefficient becomes
\be\label{FF3}
\tilde K=f\left\{B-6+(2B-6)
\frac{\partial \ln f}{\partial \ln \tilde \chi}+\left(B-\frac32\right)
\left(\frac{\partial \ln f}{\partial \ln \tilde \chi}\right)^2\right\}.
\ee
For example, one may obtain $\tilde K=0$ by solving for a given $B(f\tilde \chi)$ the differential equation for $f$
\be\label{FF4}
B(1+y)^2=6\left(1+\frac{y}{2}\right)^2~,~y=\frac{\partial \ln f}{\partial \ln \tilde \chi}.
\ee
In this frame the scalar field has only gravitational interactions. The cosmological field equation expresses $\tilde\chi$ as a function of $\tilde R$ by the implicit equation
\be\label{FF5}
f+\frac12\frac{\partial f}{\partial \ln \tilde \chi}=\frac{2\mu^2}{\tilde R}.
\ee

The transformation \eqref{FF2} can be used in both ways. An effective action with constant $\tilde K$ and non-trivial $f(\tilde \chi)$ can be mapped to the form \eqref{1}, with 
\be\label{FF6}
B=\left(1+\frac{\partial \ln f}{\partial \ln \tilde \chi}\right)^{-2}
\left[\frac{\tilde K}{f}+6\left(1+\frac12\frac{\partial \ln f}{\partial \ln \tilde \chi}\right)^2\right].
\ee
In particular, for $f=(\tilde \chi/\tilde m)^{\tilde\sigma}$ one finds
\ba\label{FF7}
B&=&(1+\tilde \sigma)^{-2}
\left[\tilde K\left(\frac{\tilde m}{\tilde \chi}\right)^{\tilde \sigma}+6
\left(1+\frac{\tilde \sigma}{2}\right)^2\right]\nn\\
&=&\left(\frac m\chi\right)^\sigma+\frac32(\sigma-2)^2,
\ea
with 
\be\label{FF8}
\sigma=\frac{\tilde\sigma}{1+\tilde \sigma}~,~
m=\big[(\sigma-1)^2\tilde K]^{\frac 1\sigma}\tilde m.
\ee
For $\sigma>1$ the asymptotic behavior $B=(m/\chi)^\sigma$ is therefore equivalent to a positive constant $\tilde K>0$ and $f$ diverging $\sim\chi^{\tilde \sigma}$, $\tilde \sigma=-\sigma/(\sigma-1)$. The limiting case of a constant coefficient of the curvature scalar, $\tilde \sigma=-2$, corresponds to $\sigma=2$. 

On the other hand, the behavior \eqref{L1} near the future fixed point for $\chi\to\infty$ can be cast into the form \eqref{FF1} for 
\be\label{FF9}
\tilde K=-6~,~f=1+\frac{1}{6\kappa \ln \frac{\tilde \chi}{m}}.
\ee
The fixed point corresponds to $f=1$, with flow equation
\be\label{FF10}
\partial_t(f-1)=6\kappa(f-1)^2.
\ee

There are two lessons to be learned from the discussion of this section. The first concerns the generality of our description of the crossover by a varying kinetic term. The second concerns the form of the $\mu$-flow equation underlying our approach. It depends on the choice of fields that are kept fixed as $\mu$ is varied, compare eqs. \eqref{123G} with eq. \eqref{18A} supplemented with $\mu\partial_\mu\ln v=2$. The form of the flow equation depends o the frame. It transforms according to a variable change in a differential equation.

\section{Conclusions}
\label{Conclusions}

We have investigated the cosmological consequences of a particle physics scenario for quantum gravity with an ultraviolet (UV) and infrared (IR) fixed point. The existence of an UV-fixed point renders quantum gravity non-perturbatively renormalizable (asymptotic safety). At this fixed point the exact scale symmetry is not spontaneously broken, such that all particles are massless. It seems possible that appropriate renormalized couplings obey asymptotic freedom. Their running is governed, however, by large (non-perturbative) anomalous dimensions. These large anomalous dimensions provide for a simple inflationary scenario. In the particle physics sector they could lead to a possible explanation of the gauge hierarchy for the electroweak symmetry breaking. 

For the IR-fixed point the exact scale symmetry is spontaneously broken, resulting in massive particles and a massless dilaton. The ratio between the effective scalar potential and the fourth power of the variable Planck mass vanishes at this fixed point. As the fixed point is approached this ratio decreases to tiny values. In the Einstein frame this leads to an asymptotically vanishing effective cosmological constant. Close to the fixed point the dilaton appears as a pseudo-Goldstone boson withe a very small mass - the cosmon. The potential energy of the cosmon field is responsible for dynamical energy. 

Dimensionless couplings can depend only on dimensionless ratios of quantities with dimension mass. In our case this is $\chi/\mu$, where $\chi$ is the value of a scalar singlet field (cosmon) and $\mu$ the intrinsic mass scale appearing in the flow equations for the running couplings. The UV-fixed point is reached for $\mu\to\infty$ or $\chi\to 0$, while the IR fixed point corresponds to $\mu\to 0,\chi\to \infty$. Cosmology describes a crossover from the UV-fixed point in the infinite past to the IR-fixed point in the infinite future. This is realized by a cosmological solution with $\chi(t\to -\infty)\to 0$, $\chi(t\to\infty)\to \infty$. 

The crossover between the two asymptotic fixed points is responsible for the different epochs in cosmology. We pursue models for which the crossover occurs in two distant steps, separated by a range of scales for which the flow of couplings is very slow. This range can be associated to the flow in the vicinity of an (approximate) ``standard model fixed point'' (SM), see Fig. \ref{IQOM1}. The range of $\chi$ and associated range in cosmological time where the SM-fixed point dominates describes the radiation and matter dominated epochs in cosmology. The UV-fixed point is responsible for the inflationary epoch, which ends at the first step of the crossover (UV$\to$SM). The IR-fixed point will correspond to an (unknown) future scaling solution. The second step of the crossover (SM$\to$IR) entails a transition period for the present cosmology for which dynamical dark energy (quintessence) dominates. 

The cosmology of our model involves four dimensionless parameters besides the masses and couplings of particles of the standard model and some dark matter candidate:

\begin{itemize}
 \item [$\sigma$:] anomalous dimension of the scalar at (or close to) the UV fixed point. It determines the spectral index $n$ and tensor ratio $r$ of the primordial fluctuations, cf. eq. \eqref{54C}.
\item[$\frac m\mu$:] scale of the first step of the crossover. It fixes the amplitude ${\cal A}$ of the primordial fluctuations, see. eq. \eqref{22AB}.
\item[$\kappa$:] coefficient of the approach to the IR-fixed point. It determines the fraction $\Omega^{ls}_h$ of early dark energy at last scattering, eq. \eqref{105A}.
\item[$\tilde\gamma$:] present growth rate of the ratio neutrino mass/electron mass. In the Einstein frame it leads to a sizeable neutrino-cosmon coupling $\beta=-\tilde\gamma\sqrt{4\kappa \ln (M/m)}$. The combination $\tilde\gamma m_{\nu 0}$ determines the present fraction of dark energy $\Omega_h$ (often called $\Omega_\Lambda$), eq. \eqref{106A}.
\end{itemize}

\noindent
In addition, the present average neutrino mass $m_{\nu 0}$ is not yet experimentally determined - only lower and upper bounds are established. We thus end with five unknown quantities that can be measured by cosmological tests. At present the model seems comparable with observations, with parameters $\sigma\approx 2.5,~\ln(m/\mu)\approx 12,~\kappa\approx 0.5,~\tilde \gamma m_{\nu 0}\approx 0.7.$

Our model has the same number of free parameters as the $\Lambda$CDM model (e.g. $n,r,{\cal A},\Omega_h,m_{\nu 0}$). A quantum gravity computation could aim for a determination of $\sigma$ and $\kappa$ which may not depend strongly on the particle physics content of the model. 

The overall description of cosmology by our model is simple. It describes all cosmological epochs by the dynamics of a single scalar field, the cosmon.  In the near future our model is subject to interesting tests: the details of inflation (relation between $n$ and $r$), early dark energy and possible consequences of large non-linear neutrino lumps. It is fascinating that a basic hypothesis about quantum gravity and the origin of mass, namely the existence of two fixed points and the necessary crossover between them, becomes testable by cosmology.

\section*{Appendix A: Flow equations for effective action}
\renewcommand{\theequation}{A.\arabic{equation}}
\setcounter{equation}{0}

A central ingredient for this paper are the dimensionless functions $B(\chi/\mu), ~C(\chi/\mu),~D(\chi/\mu),~E(\chi/\mu)$ and similar functions in the matter sector of the effective action. We will present here no computation of these quantities. The main line of this paper makes an ansatz and explores its cosmological consequences. Nevertheless, in view of a future quantum gravity computation of functions as $B(\chi/\mu)$, a few words are in order how their dependence on the ratio $\chi/\mu$ arises. This is the purpose of this appendix. 

The status of $B(\chi/\mu)$ is a renormalized function that appears in the quantum effective action for which all quantum fluctuations have been included. It contains the information on one-particle irreducible vertices that obtain by functional derivatives of the corresponding scalar kinetic term in the effective action \eqref{1}. These vertices are typically evaluated for external momenta given by the intrinsic mass scale $\mu$. (Sometimes momenta may be much smaller than $\mu$ and can effectively be set to zero.) One could also view $B$ as a $\mu$-dependent dimensionless coupling. However, the fact that $B$ can only depend on $\chi/\mu$ implies that any $\mu$-dependence translates to a field-dependent function $B(\chi/\mu)$. 

\subsection{Relative mass scales}

The key feature is the presence of two sets of scales in the quantum effective action, and therefore in the renormalized vertices. While $\mu$ denotes the set of all intrinsic mass scales, the field $\chi$ accounts for the mass scales associated to the spontaneous breaking of scale symmetry. In a certain sense $B(\chi/\mu)$ is an analogue to the universal Widom scaling function of magnetic systems, with $\mu$ standing for the explicit scale symmetry breaking away from the critical temperature related to $T-T_c$, while $\chi$ is the magnetization whose value can be dialed by a magnetic field. Another analogue are renormalized dimensionless couplings in the standard model of particle physics. They depend on external external momenta $\sim \mu$ and particle masses which are proportional to the value of the Higgs field $\tilde h\sim \chi$. (In our setting we may use $\tilde h=\sqrt{\epsilon_h}\chi$, with $\epsilon_h$ a constant for $\mu^2\ll \chi^2$, cf. sect. \ref{Past fixed point}.) The flow equation for 
$B(\chi/\mu)$ results from the relative shift of the system of intrinsic mass scales $\sim\mu$ as compared to the system of ``spontaneous'' mass scales $\sim \chi$. For any practical computation one has to specify these two systems. 

In our setting of variable gravity the spontaneous scale $\chi$ denotes the variable Planck mass. It acts as an effective cutoff for the contribution of graviton fluctuations to vertices with external momenta $Q^2\ll \chi^2$. While gravitons remain massless for arbitrary $\chi$, the contributions of graviton loops involve inverse powers of $\chi$ since they are proportional to the gravitational coupling $\sim\chi^{-2}$ (varying Newton's ``constant''). This leads to effective decoupling, with contributions to the flow suppressed by powers of $Q^2/\chi^2$. In a somewhat different context this has been found by the explicit computation in ref. \cite{HPRW}. An important exception for the effective decoupling of graviton fluctuations for $Q^2\ll\chi^2$ concerns situations where relevant graviton propagators in loops are close to poles. 

For loop momenta $q^2\gg \chi^2$ the higher order curvature terms $\sim C,D$ dominate the inverse propagators in the gravity sector, whose general form reads symbolically (omitting constants etc.) $Cq^4+\chi^2q^2$. Variation of $\chi$ can be seen as the variation of the transition scale from effective fourth-derivative gravity to effective second-derivative 
gravity. Furthermore, particle masses are proportional to $\chi$. For example, the electron mass is given by $m_e=h_e\tilde h=h_e\sqrt{\epsilon_h}\chi=f_e\chi$, with $h_e$ the Yukawa coupling of the electron to the Higgs doublet. Variation of the spontaneous scale $\chi$ therefore corresponds to a simultaneous change of the Planck mass and the particle masses. 

In the freeze frame the intrinsic scale $\mu$ enters directly the mass term for the cosmon. It induces an effective infrared cutoff for the cosmon fluctuations, as given by the squared renormalized mass $\mu^2_R\sim \mu^2/B$. (The cosmon mass is not proportional to $\chi$, in contrast to the other particle masses.) Furthermore, we evaluate all couplings at external momenta $Q^2\sim\mu^2$. This choice is motivated by our finding that the cosmological solutions of the field equations derived from the effective action \eqref{1}, \eqref{1A} are characterized by a Hubble parameter $H$ which is proportional to $\mu$. Fluctuations with wavelength larger than $H^{-1}\sim\mu^{-1}$ do not contribute to the effective action for the cosmologically relevant values and time derivatives of metric and scalar fields. This absence of long-wavelength fluctuations is mimicked by external momenta $Q^2\sim H^2\sim \mu^2$. A variation of the intrinsic scale $\mu$ therefore reflects a variation of external momenta combined with a 
variation of the scalar mass term. We should mention at this point that ``naive'' quantization depends on the frame. In a different frame the role of external momenta can look rather different, as we will discuss below. 

Having determined the two sets of intrinsic and spontaneous scales we can now establish the origin of the flow equations for functions as $B(\chi/\mu)$. We may either keep the particle masses and the Planck mass fixed (constant $\chi$) and vary external momenta and scalar mass term simultaneously (varying $\mu$). Equivalently, we may keep external momenta and scalar mass term fixed (constant $\mu)$, and vary simultaneously the effective cutoff for particle and graviton fluctuations (varying $\chi$).

In our approach we keep only two sets of scales, with all parameters of dimension mass either proportional to $\mu$ or proportional to $\chi$. This implies the simple relation between the $\mu$-flow at fixed $\chi$ and the $\chi$-flow at fixed $\mu$, 
\be\label{AAZ1}
\mu\frac{\partial B}{\partial \mu}_{|\chi}+\chi\frac{\partial B}{\partial \chi}_{|\mu}=0.
\ee
One could consider more generalized settings involving more than two sets of mass scales. For example, one could investigate models with an explicit ultraviolet cutoff $\Lambda$, such that $B$ depends on $\mu/\Lambda$ besides $\chi/\mu$. In our setting for renormalized $B$ the cutoff $\Lambda$ is considered as an intrinsic scale, with $\mu/\Lambda$ fixed. In the presence of an ultraviolet fixed point one can take the limit $\mu/\Lambda\to 0$. Furthermore, we have taken all relevant external momenta to be $\sim \mu$. Certain quantities may involve vastly different external momentum scales. 

\subsection{Functional flow equation for effective average}

\vspace{-0.12cm}
~{\bf action}

The functional flow equation for the effective average action \cite{CWFE} of dilaton quantum gravity \cite{HPRW} also involves two sets of mass scales: the infrared cutoff scale $k$ and the scale $\chi$ set by the value of the scalar field. Coupling functions as $B$ depend now also on the ratio $\chi/k$. Qualitatively we may associate $k$ with $\mu$, since both scales act effectively as an infrared cutoff - one explicitly and the other indirectly by virtue of non-zero external momenta. 

More precisely, the effective average action for our setting will involve the three sets of scales $\chi,\mu$ and $k$. Often only the highest effective infrared cutoff matters. Thus the flow of $B\left(\frac\chi\mu,\frac\chi k\right)$ is roughly independent of $\mu$ for $k\gg \mu$, and independent of $k$ for $k\ll\mu$. The quantum effective action corresponds to $k\to 0$. On the other hand, the presence of an UV-fixed point implies the existence of a scaling solution for $\Gamma_k$. The scaling form of the effective average action is independent of any intrinsic scale $\mu$. Setting $\mu=0$ the scaling function $B_*(\chi/k)$ only depends on the ratio $\chi/k$. By virtue of the above decoupling properties we may roughly identify
\be\label{AAZ2}
B\left(\frac\chi\mu;k=0\right)\approx B_*\left(\frac\chi k,k=\mu\right). 
\ee
This connects the coupling function $B(\chi/\mu)$ in the full quantum effective action $(k\to 0)$ to the scaling function of the effective average action $B_*(\chi/k)$.

The identification \eqref{AAZ2} is only approximate for several reasons. First, there are proportionality constants of order one, replacing $k=\mu$ by $k=c_i\mu$ on the r.h.s. of eq. \eqref{AAZ2}, with coefficients $c_i$ depending on particular loop contributions. They reflect the particular choice of cutoff as well as the ``final running'' in the region $k\lesssim\mu$ before the flow stops. Second, the $\mu$-flow equation needs to incorporate properly the simultaneous change of the scalar mass parameter which accompanies the change of scale of external momenta. Third, the form of the effective action \eqref{1} assumes a particular ``canonical'' choice of fields. The flow of the effective average action will typically not remain of the canonical freeze form \eqref{1}. One therefore needs to perform $k$-dependent field redefinitions \cite{GW,FW} in order to bring the effective average action to the canonical freeze form at every scale $k$.

Despite these shortcomings several important general features can be inferred from the association between dimensionless functions in the quantum effective action and scaling functions in the effective average action. An ultraviolet fixed point in the functional renormalization flow of the effective average action corresponds to $k$-independent renormalized dimensionless couplings in the limit $k\to \infty$. Such an UV-fixed point in the $k$-flow is typically reflected in an UV-fixed point for the $\mu$-flow for the quantum effective action. In the presence of an UV-fixed point the scaling function $B_*(\chi/k)$ is universal up to the dependence on ``renormalizable'' couplings. These free renormalizable couplings correspond to the relevant parameters for small deviations from the UV-fixed point. On the level of the $\mu$-flow equations for the quantum effective action these free parameters appear as integration constants for the solution of the flow equation.

\subsection{Scaling solutions and relevant parameters}

A simple example for the connection between relevant parameters and free integration constants in the scaling solution is the running of a non-abelian gauge coupling $g$ in the presence of spontaneous symmetry breaking which gives the gauge bosons a mass $\chi$. The qualitative form of the flow equation $(c>0)$,
\be\label{AAZ3}
k\partial_k g^2_{|\chi}=-cg^4\theta\left(1-\frac{\chi^2}{k^2}\right),
\ee
accounts for asymptotic freedom (UV-fixed point at $g=0$) and the stop of the flow once $k$ becomes smaller than the gauge boson masses. Here $\theta(x)$ is the heavy side function - smooth ``threshold functions'' approaching one for $\chi^2\ll k^2$ and zero for $\chi^2\gg k^2$ would lead to qualitatively similar results. The scaling solution of eq. \eqref{AAZ3},
\be\label{AAZ4}
g^{-2}=g^{-2}_*(y)=g^{-2}_0-\frac c2\ln y~\theta(1-y)~,~y=\frac{\chi^2}{k^2},
\ee
contains the free integration constant $g_0$. This corresponds to the value of the gauge coupling for $k\leq \chi$. One easily checks that the scaling function $g_*(y)$ is a fixed point of the flow equation for a fixed dimensionless ratio $y$
\be\label{AAZ5}
k\partial_kg^2_{|y}=k\partial_kg^2_{|\chi}+2y\partial_yg^2=0.
\ee
According to eq. \eqref{AAZ2} the scaling solution $g_*(y)$ can be taken over to the quantum effective action by the identification $y=\chi^2/\mu^2$. 

The IR-fixed point for the scaling solution \eqref{AAZ4} is rather trivial. For any finite positive $g^2_0$ the running of the coupling simply stops due to decoupling for $k^2<\chi^2$. The realization of an IR-fixed point by an effective stop of the flow due to excitations becoming heavy or interactions going to zero is a rather generic phenomenon. It is, however, not the only way how an IR-fixed point can be realized. 

The scaling solution \eqref{AAZ4} is not the most general solution of the flow equation \eqref{AAZ3}. For the general solution we can replace the constant $g^2_0$ by an arbitrary function of $\chi$, 
\be\label{AAZ6}
g^2_0(\chi^2)=g^2_0(yk^2).
\ee
This is no longer a scaling solution due to the explicit dependence on $k$ for fixed $y$. This dependence implies the presence of a further scale as the ultraviolet cutoff $\Lambda$, such that $g^2_0$ depends on the dimensionless ratios $y$ and $k/\Lambda$. In the limit $k\to 0$ at fixed $y$ the general solution approaches a scaling solution, with constant $g^2_0=g^2_0(0)$. This holds provided that the limit $g^2_0(\chi^2/\Lambda^2\to 0)$ is finite. 

We emphasize that a scaling solution is approached universally in the infrared limit $k\to 0$ despite the presence of a relevant parameter at the UV-fixed point. (We do not distinguish between marginal and relevant couplings here.) The relevant parameter rather manifests itself by the presence of a free parameter in the scaling solution. This feature can be understood by the observation that the limit $k\to 0$ at fixed $y$ and $\Lambda$ corresponds to the limit $\Lambda\to\infty$ at fixed $y$ and $k$. In the presence on an ultraviolet fixed point the limit $\Lambda\to\infty$ can be taken, removing any explicit dependence on $\Lambda$ for renormalized dimensionless quantities. As a consequence $g$ can only depend on $y$ in this limit and the scaling function has to be approached. There remains, however, some memory of the behavior close to the UV-fixed point. This corresponds to the relevant parameters or renormalizable couplings. The corresponding information has to appear in the form of free parameters 
for 
the scaling solution.

The characteristic features discussed here are not special for the case of asymptotic freedom. As an example of a dimensionless coupling $\lambda$ with UV- and IR-fixed point we consider the flow equation
\ba\label{AAZ7}
k\partial_k\lambda_{|\chi}&=&c(\lambda-\lambda_{IR})(\lambda-\lambda_{UV})\theta(\left(1-\frac{\chi^2}{k^2}\right),\nn\\
&&\lambda_{IR}>\lambda_{UV}~,~c>0.
\ea
The scaling solution reads for $y\leq 1$
\ba\label{AAZ8}
\lambda_*(y)&=&\frac{\lambda_{UV}\lambda_{IR}(1-z)+\lambda_0(\lambda_{IR}z-\lambda_{UV})}
{\lambda_{IR}-\lambda_{UV}z-\lambda_0(1-z)},\nn\\
z&=&y^{\frac c2(\lambda_{IR}-\lambda_{UV})}~,~\lambda_{UV}\leq \lambda_0\leq \lambda_{IR},
\ea
and $\lambda=\lambda_0$ for $y\geq 1$. In the UV-limit $y\to 0$ the scaling function $\lambda_*(y)$ approaches $\lambda_{UV}$, while the IR-limit is given by $\lambda_*(y\to\infty)=\lambda_0$. The free integration constant $\lambda_0$ denotes the value of the coupling when the crossover between $\lambda_{UV}$ and $\lambda_{IR}$ is stopped by the decoupling for $k^2<\chi^2$. For $\lambda_0$ very close to $\lambda_{IR}$ the value of $\lambda_*(y)$ remains close to this value for a large range of $y$. The discussion of the general solution is similar to our first example, with $\lambda_0$ depending on the combination $yk^2/\Lambda^2=\chi^2/\Lambda^2$.

\subsection{Stages of the flow in the freeze frame}

The analogy between the $\mu$-flow of the quantum effective action and the scaling solution of the effective average action helps to visualize which type of fluctuations contribute to quantities as $B(y)=B(\chi^2/\mu^2)$ for different ranges of $y$. For $y\lesssim 1$ the graviton fluctuations are effective. They are described by fourth order gravity for $y\to 0$. (The symbols $\lesssim,\gtrsim$ denote here order of magnitude estimates.) For $y\gtrsim 1$ gravity decouples unless the graviton fluctuations are dominated by a pole in the propagator. The flow will be dominated by particles with mass smaller than $\chi$. Those include the cosmon. More precisely, the propagators for the  scalar field $\chi$ and the scalar degrees of freedom in the metric mix and the cosmon is associated with a suitable eigenstate.

Beyond the cosmon many particles of the standard model have masses much smaller than $\chi$. In a grand unified theory there would be particles with mass around $(10^{-3}-10^{-2})\chi$ that decouple once $\mu$ gets smaller than this value. The particles of the standard model have masses substantially smaller than $\chi$ due to the electroweak gauge hierarchy and the small ratio between the QCD-scale and the Planck scale. They affect the running of the standard model couplings according to the well known perturbative $\beta$-functions. (The influence on the flow of $B$ is not known so far.) This standard model flow stops effectively once $\mu$ drops below the $\chi$-dependent electron mass $m_e(\chi)\approx 2\cdot 10^{-22}\chi$. Thus for $1\ll y\lesssim 2\cdot 10^{43}$ one expects a range of ``standard model flow''. In addition to the particles of the standard model also the cosmon and possibly the graviton and the scalar gravitational degree of freedom contribute to the flow. These contributions could 
actually dominate the flow of $B$. 

The standard model flow ends at a value of $y$ that corresponds to a cosmological epoch before the electroweak phase transition. (Recall that the value of $y$ relevant for present cosmology is around $10^{120}$.) In the Einstein frame the end of the standard model flow corresponds to a Hubble parameter of the order of the electron mass, cosmic time of the order $m^{-1}_e$, or temperature in the range $10^7$ GeV. For $y\gtrsim 10^{44}$ the running of the standard model couplings stops effectively and the flow is characterized by the standard model fixed point. 

In the vicinity of the standard model fixed point only neutrinos, photons, cosmon and possibly gravitational degrees of freedom as well as other light particles beyond the standard model (e.g. very light scalar fields for dark matter \cite{BW1,BW2}) contribute to the flow. Neutrinos will decouple once $\mu$ drops below the neutrino  mass. According to our assumption the flow in this range is unstable in the ``beyond standard model (BSM)-sector'', exhibiting a remaining (marginally) relevant coupling. The second step of the crossover occurs once this relevant deviation from the standard model fixed point becomes large. Neutrino masses are directly sensitive to the BSM-sector, such that the second stage of the crossover becomes first ``visible'' in these quantities. 

We end this short qualitative ``history'' of the flow with the remark that the effect on the kinetial $B$ may be indirect. For example, a contribution to the flow of the potential (e.g. a running $\chi$-independent term) translates to a flow of $B$ by field transformations, as discussed in sect. \ref{Field relativity} and appendix E.

\subsection{Flow equations in different frames}

Let us assume for this paragraph that the leading term in the potential for large $\chi$ is not a term quadratic in $\chi$ as in eq. \eqref{1}, but rather a constant $V=\tilde\mu^4$. A cosmological model of this type has been discussed in ref. \cite{VG}, model (B). Realistic cosmology requires $\tilde \mu\approx 2\cdot 10^{-3}$ eV. The characteristic size of the Hubble parameter is $H\approx \tilde\mu^2/\tilde\chi$, with $\tilde\chi$ the scalar field in this frame. We can take over the preceding discussion with the association $\mu=\tilde\mu^2/\tilde\chi$ or $\tilde y=\tilde\chi^2/\tilde\mu^2=\sqrt{y}$. 

Once the kinetial $\tilde B(\tilde y)$ is found in this setting, it can be transformed to the one in the freeze frame, $B(y)$, by the variable transformation discussed in sect. \ref{Field relativity}, e.g. eq. \eqref{142B}. This demonstrates a general feature. Typically, a quantum computation of the flow equation will be done in a particular frame, e.g. assuming implicitly a simple functional measure. The translation to the freeze frame can then be done on the level of the quantum effective action.

\section*{Appendix B: Fixed points and crossover for dimensionless couplings}
\renewcommand{\theequation}{B.\arabic{equation}}
\setcounter{equation}{0}

In this appendix we discuss the general structure of the crossover between two fixed points for a dimensionless coupling. For this purpose we use a simple example for the flow equations. We apply these findings to a discussion of the behavior of neutrino masses during the second stage of the crossover. For a dimensionless coupling $h$ depending on $\chi/\mu$ we formulate the flow equation directly in terms of the $\chi$-dependence for fixed $\mu$. 

Consider a dimensionless coupling $h$ whose dependence on $\chi$ obeys the flow equation
\be\label{C1}
\frac{\partial h^2}{\partial \ln \chi}=c(h^2-f_1)(h^2-f_2),
\ee
with $f_1$ and $f_2$ the values of two fixed points for $h^2$, with $0<f_1<f_2$. We take $c>0$ such that $f_2$ is approached for $\chi\to 0$ and $f_1$ for $\chi\to\infty$. The solution of eq. \eqref{C1} is given implicitly by
\be\label{C2}
\frac{h^2-f_1}{f_2-h^2}=\left(\frac{\chi}{\chi_0}\right)^{-c(f_2-f_1)}.
\ee
As $\chi$ increases from zero to infinity this describes for $h^2$ a crossover from the fixed point at $f_2$ to the one at $f_1$. The free integration constant $\chi_0$ is proportional to $\mu$. It determines at which value of $\chi/\mu$ the crossover occurs. This depends how ``close'' a given trajectory is to the UV-fixed point. The ratio $\chi_0/\mu$ can therefore be exponentially large or small.

For sufficiently large $\chi$ one has approximately
\be\label{C3}
h^2=f_1+(f_2-f_1)\left(\frac{\chi}{\chi_0}\right)^{-c(f_2-f_1)}.
\ee
Then the relative change of $h^2$,
\be\label{C4}
\frac{1}{h^2}\frac{\partial h^2}{\partial\ln \chi}=-\frac{c(f_2-f_1)^2}{f_1}
\left(\frac{\chi}{\chi_0}\right)^{-c(f_2-f_1)},
\ee
becomes tiny for
\be\label{C5}
\ln (\chi/\chi_0)\gg\big[c(f_2-f_1)\big]^{-1}.
\ee
Particle masses can be written as $m_A=h_A(\chi)\chi$, with an appropriate dimensionless coupling function $h_A(\chi)$. If $h_A$ obeys eqs. \eqref{C1}, \eqref{C5} the particle mass $m_A$ scales (almost) proportional to $\chi$. This is what we will assume for all particles of the standard model except for neutrinos. The scaling $m_A\sim\chi$ indicates dilatation or scale symmetry, as appropriate for a fixed point.

For these particles we associate $\chi_0$ with $m$. For nucleosynthesis and the subsequent epochs of cosmology the ratio $\chi/m$ is already huge. For sufficiently large $c(f_2-f_1)$ the $\chi$-dependence of couplings and mass ratios for standard model particles will be too small to be observable. For smaller $c(f_2-f_1)$, however, a tiny residual $\chi$-dependence of couplings could result in an observable time variation. The couplings during nucleosynthesis may then be slightly different from the present ones. Eq. \eqref{C4} relates the time variation during nucleosynthesis with the one in the present cosmological epoch.

Let us next discuss the crossover from $f_2$ to $f_1$, specializing to $f_1\ll f_2$. We may consider the crossover region $f_1\ll h^2\ll f_2$ where we approximate
\be\label{C6}
h^2=f_2\left(\frac{\chi}{\chi_0}\right)^{-cf_2}.
\ee
For the particular flow equation \eqref{C1} this region is characterized by a constant anomalous dimension 
\be\label{C7}
\frac{\partial\ln h^2}{\partial\ln\chi}=-cf_2.
\ee
We will assume that neutrino masses show this type of crossover behavior. 

Neutrino masses are characterized by the seesaw formula,
\be\label{C8}
m_\nu=\frac{h^2_\nu\varphi^2_H}{M_{B-L}},
\ee
with $\varphi_H$ the Fermi scale (expectation value of the Higgs doublet) and $M_{B-L}$ a characteristic high mass scale where $B-L$ symmetry is violated. Severe bounds on the time variation of electron over proton mass indicate that $\varphi_H/\chi$ must be close to a fixed point already for values of $\chi$ characteristic for nucleosynthesis. A first scenario may assume $h^2=M_{B-L}/\chi$, with $h^2$ obeying the flow eq. \eqref{C1}. With $\varphi_H\sim\chi$  eqs. \eqref{C7}, \eqref{C8} yield 
\be\label{C9}
\tilde \gamma=\frac12\frac{\partial\ln(m_\nu/\chi)}{\partial\ln\chi}=\frac{cf_2}{2}.
\ee
A non-trivial scaling $m_\nu\sim\chi^{1+2\tilde\gamma}$ may correspond to such a crossover situation. 

Other forms of a crossover, with $\tilde\gamma$ depending on $\chi$, are conceivable as well. With $\tilde m=m_\nu/\chi$, any positive continuous function $\tilde\gamma(\tilde m)$ with two different zeros at $\tilde m_1$ and $\tilde m_2$ describes a crossover between $\tilde m=\tilde m_1$ for $\chi\to 0$ and $\tilde m=\tilde m_2$ for $\chi\to\infty$. 

For our second scenario we take
\be\label{C10}
\tilde\gamma=\frac{\tilde m-\tilde m_1}{\tilde m_1}\frac{\tilde m_2-\tilde m}{\tilde m_2},
\ee
with constant fixed point values $\tilde m_1$ and $\tilde m_2$, $\tilde m_1\ll\tilde m_2$. Again, a non-zero value of $\tilde\gamma$ reflects the $\chi$-dependence of $M_{B-L}/\chi$. For the crossover region $\tilde m_1\ll \tilde m\ll \tilde m_2$ we can approximate
\be\label{C11}
\tilde\gamma=\frac{\tilde m}{\tilde m_1}.
\ee
The corresponding solution of eq. \eqref{C9}, $\tilde\gamma=\partial\ln\tilde m/(2\partial\ln\chi)$, namely
\be\label{C12}
\tilde m=\frac{m_\nu}{\chi}=\frac{\tilde m_1}{\ln\left(\frac{\bar\chi^2_\nu}{\chi^2}\right)},
\ee
diverges for $\chi$ approaching the constant $\bar\chi_\nu$. This is, however, outside the validity of the approximation. For $\tilde m$ approaching $\tilde m_2$ eq. \eqref{C10} implies that the increase with $\chi$ is stopped. In the crossover region, however, the fixed point at $\tilde m_2$ is not yet visible. We could also multiply the r.h.s. of eq. \eqref{C10} with a constant. Within the crossover region this constant can be absorbed into a redefinition of $\tilde m_1$. At this point the $\chi$-dependence of the average neutrino mass involves two parameters, $\tilde m_1$ and $\bar\chi_\nu$. We will see that it corresponds to the setting of ref. \cite{CWNEU}. 

The parameter $\tilde m_1$ is given by the ratio between the average neutrino mass and the Planck mass in earlier epochs of cosmology, before the crossover in the neutrino sector sets in. Taking in eq. \eqref{C8} an ``early value'' $M_{B-L}/\chi\approx 10^{-3}$, as appropriate for $B-L$ violation at some scale characteristic for grand unification, and $\varphi_H/\chi\approx 10^{-16}$, as given by the electroweak gauge hierarchy, we estimate 
\be\label{C13}
\tilde m_1=10^{-29}h^2_\nu,
\ee
with $h^2_\nu$ typically smaller than one. (For these estimates we employ the present value of $\chi$, namely $M=2.44 \cdot 10^{27}$eV.) For the present value of $\tilde m$ one has
\be\label{C14}
\tilde m(t_0)=\left(\frac{m_\nu(t_0)}{0.25{\rm eV}}\right)\cdot 10^{-28}.
\ee
\\
This is well compatible with our assumption that for the present cosmological epoch the neutrino masses are in the crossover region,
\ba\label{C15}
\tilde \gamma(t_0)&=&\frac{\tilde m(t_0)}{\tilde m_1}-1=
\frac{1}{\ln\left(\frac{\bar\chi^2_\nu}{M^2}\right)}\nn\\
&=&\frac{10}{h^2_\nu}\left(\frac{m_\nu(t_0)}{0.25{\rm eV}}\right)-1\gg 1.
\ea
For $\tilde\gamma(t_0)=9$ the ratio $M_{B-L}/\chi$ has decreased at present by a factor of ten as compared to its early value.

\section*{Appendix C: Field equations for variable gravity and asymptotic solutions}
\renewcommand{\theequation}{C.\arabic{equation}}
\setcounter{equation}{0}
\setcounter{subsection}{0}

In this appendix we discuss field equations and solutions of variable gravity. For the solutions we do not attempt to give a complete overview but rather concentrate on a few characteristic ones.

\subsection{Field equations}

In this appendix we investigate the field equations derived from the effective action
\be\label{C1a}
\Gamma=\int_x\sqrt{g}
\left\{-\frac12\chi^2 R-\frac C2 R^2+\frac{B(\chi)-6}{2}\partial^\mu\chi\partial_\mu\chi+V(\chi)\right\},
\ee
with constant $C$. We have omitted the term $\sim D$ in eq. \eqref{1A} since it does not contribute to the field equations for a spatially flat Robertson-Walker metric if $D$ is constant. The gravitational field equation is obtained by variation of the effective action \eqref{C1a} with respect to the metric,
\ba\label{C2a}
&&\chi^2(R_{\mu\nu}-\frac12 Rg\mn)+D^2\chi^2 g\mn-D_\mu D_\nu\chi^2\nn\\
&&+(B-6)\left(\frac12\partial^\rho \chi\partial_\rho\chi g\mn-\partial_\mu\chi\partial_\nu\chi\right)+V g\mn\\
&&+C(2RR\mn-\frac12 R^2 g\mn+2D^2 R g\mn-2 D_\mu D_\nu R)=T\mn.\nn
\ea
Here $D_\mu$ denotes the covariant derivative, $D^2=D^\mu D_\mu$, and $T\mn$ is the energy momentum tensor. The cosmon field equation is given by
\be\label{C3a}
(B-6)D^2\chi +\frac12\frac{\partial B}{\partial \chi}\partial^\mu\chi\partial_\mu\chi=\frac{\partial V}{\partial\chi}-\chi R- q_\chi,
\ee
with $q_\chi$ the contribution from particles with $\chi$-dependent mass. Contracting eq. \eqref{C2a} yields a differential equation for the curvature scalar if $C\neq 0$,
\be\label{C4a}
6CD^2R-\chi^2R+6\chi D^2\chi+B\partial^\mu\chi \partial_\mu\chi+4V=T^\mu_\mu.
\ee

For a Robertson-Walker metric with vanishing spatial curvature a time dependent homogeneous scalar field obeys
\be\label{C5a}
(B-6)(\ddot{\chi}+3H\dot{\chi})+\frac12 \frac{\partial B}{\partial\chi}\dot{\chi}^2+\frac{\partial V}{\partial \chi}-\chi 
(12H^2+6\dot{H})=q_\chi,
\ee
while the $(0,0)$-component of eq. \eqref{C2a} yields $(T_{00}=\rho)$
\be\label{C6a}
3(\chi H+\dot{\chi})^2=\frac{B}{2}\dot{\chi}^2+V+18 C
(\dot{H}^2-2H\ddot{H}-6H^2\dot{H})+\rho.
\ee
Taking a time-derivative of eq. \eqref{C6a}, adding eq. \eqref{C5a} multiplied by $\dot{\chi}$ and using eq. \eqref{C4a} yields the generalized conservation equation $(T^\mu_\mu=-\rho+3p)$
\be\label{C7a}
\dot{\rho}+3H(\rho+p)+q_\chi\dot{\chi}=0.
\ee
We will use eqs. \eqref{C5a}, \eqref{C6a} and \eqref{C7a} as the three independent equations which determine the time evolution. We observe that the contribution from the term $\sim CR^2$ vanishes for constant $H$.

\subsection{Asymptotic solutions}

For primordial cosmology we neglect $T\mn$ and $q_\chi$. In the absence of matter and radiation eqs. \eqref{C5a} and \eqref{C6a} constitute two non-linear second order differential equations for $\chi$ and $H$. For $V=\mu^2\chi^2$ and $B=(m/\chi)^\sigma$ they admit the simple family of solutions
\be\label{B.7A}
\chi=0~,~H=H_0,
\ee
with $H_0$ an arbitrary constant. For positive $H\dot\chi$ these asymptotic solutions can be approached in the infinite past $t\to -\infty$ provided
\be\label{B.7B}
H_0>\frac{\mu}{\sqrt{6}}.
\ee
Small deviations from the asymptotic solution \eqref{B.7A} grow as time increases.

For the case $H_0=\mu/\sqrt{3}$ we recover the scaling solution \eqref{12}, \eqref{25Bb}, which reads for arbitrary $\sigma$
\be\label{B.7C}
H=\frac{\mu}{\sqrt{3}}~,~\chi=m
\left(\frac{\sqrt{3}}{2\sigma\mu(t_c-t)}\right)^{\frac1\sigma}.
\ee
For $C\neq 0$ and arbitrary $H_0>\mu/\sqrt{6}$ one can find more general solutions 
\ba\label{B.7D}
H&=&H_0+\delta H,\nn\\
\chi&=&m
\left(\frac{3H_0}{2\sigma(6H^2_0-\mu^2)(t_c-t)}\right)^{\frac1\sigma},
\ea
provided that $\delta H$ vanishes for $t\to -\infty$ according to the solution of the equation
\be\label{C.11A}
\dot H=\frac{\chi^2}{36C}
\left(\frac{\mu^2}{3H^2_0}-1\right).
\ee
Eq. \eqref{C.11A} is necessary for the approximate solution of eq. \eqref{C6a}. It is compatible with $|\delta H(t\to-\infty)|\to 0$ only for $\sigma<2$. Then typical solutions that can be extended towards the infinite past switch from an asymptotic solution \eqref{B.7D} for $t\to-\infty$ towards the scaling solution \eqref{B.7C} as time increases. We recall that such solutions with $H^2_0\neq \mu^2/3$ exist only for $C\neq 0,\sigma<2$. 

In order to estimate the influence of the higher curvature term $\sim CR^2$ we evaluate its contribution to equation \eqref{C6a}. For $t\to-\infty$ we investigate solutions of the type
\be\label{C8a}
H=b\mu+f(t_c-t)^{-\eta}~,~\chi=\bar\chi (t_c-t)^{-\zeta}.
\ee
The term $\sim C$ in eq. \eqref{C6a} contributes in leading order $-108CH^2\dot{H}=108 C\eta f b^2\mu^2(t_c-t)^{-(1+\eta)}$, to be compared with $V=\mu^2\bar\chi^2(t_c-t)^{-2\zeta}$ and $(B/2)\dot{\chi}^2=(m^\sigma/2)\zeta^2\bar\chi^{2-\sigma}(t_c-t)^{-2+\zeta(\sigma-2)}$. For $2\zeta<1+\eta$ the potential $V$ dominates the r.h.s of eq. \eqref{C6a} for $t\to-\infty$, implying $b=1/\sqrt{3}$. For the vicinity of the scaling solution \eqref{B.7C} this turns out to be indeed the leading behavior for $H$ for $t\to-\infty$. On the other hand, for the vicinity of the asymptotic solutions \eqref{B.7D} for $H_0\neq \mu/\sqrt{3}$ we will find $2\zeta=1+\eta$ such that the term $\sim C$ is of equal importance as $V$ on the r.h.s. of eq. \eqref{C6a}. We will discuss the two cases separately. 

For $H=\mu/\sqrt{3}$ the cosmon field equation reads in leading order
\be\label{C9a}
\dot{\chi}=\frac{2}{\sqrt{3}}\mu m^{-\sigma}\chi^{1+\sigma},
\ee
such that eq. \eqref{C8a} is obeyed with 
\be\label{C10a}
\zeta=\frac 1\sigma~,~\bar\chi=m
\left(\frac{\sqrt{3}}{2\sigma\mu}\right)^{\frac1\sigma}.
\ee
This is the scaling solution \eqref{B.7C}. In the infinite past $t\to-\infty$ one has $\chi=0$ and realizes eq. \eqref{2I}.

For the next to leading contribution to $H$ one finds
\be\label{C11a}
f(t_c-t)^{-\eta}=-\frac{5}{6\sigma}(t_c-t)^{-1}-f F_c(t_c-t)^{-1-\eta+\frac 2\sigma},
\ee
with 
\be\label{C12a}
F_c=\frac{6\sqrt{3}C\eta\mu}{m^2}
\left(\frac{2\sigma\mu}{\sqrt{3}}\right)^{\frac2\sigma}.
\ee
For $\sigma<2$ the l.h.s. of eq. \eqref{C11a} can be neglected for $t\to -\infty$, such that the term $\sim C$ dominates the next to leading correction to $H$. A solution for $t\to-\infty$ is therefore given by eq. \eqref{C8a} with 
\be\label{C13a}
\eta=\frac 2\sigma ~,~f=-\frac{5}{6\sigma F_c}.
\ee

As $t$ increases towards $t_c$ the l.h.s of eq. \eqref{C11a} increases faster than the r.h.s. if $\sigma < 2$. It equals the term $\sim C$ at a transition time $ t_{tr}$ given by 

\be\label{C14a}
t_c-t_{tr} = |F_c|{^{-{\frac{\sigma}{2-\sigma}}}} = 
\frac{\sqrt{3}}{2\sigma\mu}
\left(\frac{24|C|\mu^2}{m^2}\right)^{-\frac{\sigma}{2-\sigma}}.
\ee 
After the transition, for $t_c-t \ll t_c-t_{tr}$, the next to leading order contribution to $H$ switches to 
\be\label{B14A}
\eta=1~,~f=-5/(6\sigma),
\ee
and the contribution of the higher curvature term $\sim C$ becomes negligible. We observe that for $C$ of the order one the dimensionless ratio $\mu(t_c-t_{tr})$ is large due to the small ratio $\mu^2/m^2$. If we associate $t_c$ roughly with the end of inflation this implies that the higher order curvature term $\sim C$ becomes negligible long before the observable fluctuations cross the horizon. 

We conclude that for $\sigma < 2$ the higher curvature term $\sim C R^2$ does not influence the leading behaviour of the  scaling solution for $H$ and $\chi$. 
Only the next to leading terms in the solution for $t\rightarrow -\infty$ are influenced by $C\neq 0$. This influence ends effectively at $t_{tr}$, long before
the observable primordial density fluctuations leave the horizon. For the observable properties of inflation the role of the term $\sim C R^2$ is negligible.

For $\sigma>2$ the switch between the two types of solutions for $\delta H$ occurs in the inverse order. For $t\to -\infty$ one finds the solution \eqref{B14A}, while for $t\to t_c$ the term involving $C$ becomes dominant in eq. \eqref{C11a}. For $t>t_{\rm tr}$ the solution formally switches to eq. \eqref{C13a} and $C$ seems to matter. For $\sigma>2$ the formal transition time $t_{\rm tr}$ is very close to the end of inflation at $t_c$, however. At this time the approximation \eqref{C8a} is no longer valid. For $\sigma>2$ the higher curvature invariant $\sim C$ is negligible for all $t$, with 
\be\label{C21-A}
\delta H=-\frac{5}{6\sigma}(t_c-t)^{-1}.
\ee

Finally, we may also consider the boundary case $ \sigma = 2$ for which all three terms in eq. \eqref{C11a} have the same time dependence,
\be\label{C15a}
\eta =1~,~f =-\frac{5}{12(F_c+1)}~,~F_c=\frac{24C\mu^2}{m^2}.
\ee  
The quantitative influence of the term $\sim C$ is negligible for all $t$ due to the tiny ratio $\mu^2/m^2$. 

We next turn to the asymptotic solutions \eqref{B.7D} with $H_0\neq \mu/\sqrt{3}$ or $b\neq 1/\sqrt{3}$. For $\sigma<2$ they realize $\zeta=1/\sigma,\eta=2/\sigma-1$ according to eq. \eqref{C.11A}, such that the next to leading behavior for $H$ reads 
\be\label{B.16A}
H=H_0+f(t_c-t)^{-\frac{2-\sigma}{\sigma}}.
\ee
The constant $f$ is given by
\be\label{B.16B}
f=\frac{m^2(\mu^2-3H^2_0)\sigma}{108CH^2_0(2-\sigma)}
\left(\frac{3H_0}{2\sigma(6H^2_0-\mu^2)}\right)^{\frac2\sigma}.
\ee
This type of solution is only valid for $\sigma<2$. 

The family of solutions \eqref{B.7D} is not the only possible behaviour in the infinite past for $ t \rightarrow - \infty$. For example, a solution with $\chi = 0, \dot{\chi} = 0$ can be found for 
\be\label{C16a}
2H\ddot{H} = \dot{H}^2-6 H^2\dot{H}.
\ee
This differential equation admits solutions of the type \eqref{C8a} that approaches flat space in the infinite past $(b=0)$. With $\eta=1, f= -\frac{1}{2}$ the evolution of the scale factor describes a shrinking universe
\be\label{C17a}
H=-\frac{1}{2}(t_c-t)^{-1}~,~a= a_0\sqrt{t_c-t}
\ee
In this case the past infinite universe is flat with infinite scale factor, similar to the future in the radiation dominated Friedman universe,  $a=a_0\sqrt{t-t_c}$. (The time reflected solution $a=a_0\sqrt{t-t_c}$ also solves eq. \eqref{C16a}.) This type of solution with negative $H$ seems not to be connected with solutions that lead to realistic cosmologies for later times. 

\subsection{Renormalized scalar field}

We are interested in the general behavior of solutions close to the UV-fixed point, e.g. for small $\chi$. For $\chi$ close to zero it is advantageous to write the field equations in terms of the renormalized scalar field
\be\label{AF1}
\chi_R=\frac{2}{2-\sigma}\sqrt{B}\chi~,~B=\left(\frac m\chi\right)^\sigma.
\ee
With 
\ba\label{AF2}
&&\dot\chi=B^{-1/2}\dot\chi_R~,~\ddot{\chi}=B^{-\frac12}
\left(\ddot{\chi}_R+\frac{\sigma}{2-\sigma}\frac{\dot{\chi}^2_R}{\chi_R}\right),\nn\\
&&B\ddot{\chi}+\frac12\frac{\partial B}{\partial \chi}\dot\chi^2=B^{1/2}\ddot{\chi}_R,
\ea
and approximating $B-6\approx B$, the field equations \eqref{C5a}, \eqref{C6a} read in the absence of matter and radiation
\be\label{AF3}
\ddot{\chi}_R+3H\dot{\chi}_R=\frac{2-\sigma}{2B}
(12H^2+6\dot H-2\mu^2)\chi_R
\ee
and 
\ba\label{AF4}
36CH\ddot H&=&18C(\dot H^2-6H^2\dot H)\nn\\
&&+\left(\frac{2-\sigma}{2}\right)^2B^{-1}(\mu^2-3H^2)\chi^2_R\nn\\
&&+\frac12\dot\chi^2_R-3(2-\sigma)B^{-1}H\dot\chi_R\chi_R.
\ea
Once $B^{-1}$ is expressed in terms of $\chi_R$,
\be\label{AF5}
B^{-1}=\left(\frac{(2-\sigma)\chi_R}{2m}\right)^{\frac{2\sigma}{2-\sigma}},
\ee
this form is suitable for numerical solutions. (At a point where $H$ vanishes (or for $C=0$) the r.h.s. of eq. \eqref{AF4} has to be equal to zero.) 

We concentrate on $\sigma<2$ where $\chi\to 0$ corresponds to $\chi_R\to 0$. One recovers the simple solution $\chi_R=0$ or $\chi=0$ with an arbitrary constant value of the Hubble parameter $H=H_0$. The stability of this solution depends on the value of $H_0$. For $H_0<\mu/\sqrt{6}$ a small initial value of $\chi_R$ decreases towards zero. On the other hand, for $H_0>\mu/\sqrt{6}$ a perturbation in $\chi$ grows and the solution with $\chi=0$ is unstable. We are mainly interested in the second type of solution where $\chi$ moves away from the UV-fixed point at $\chi=0$ as time increases.

It is useful to express the field equations in terms of the potential for the renormalized field, 
\be\label{AF6}
V=\mu^2 m^2\left(\frac{(2-\sigma)\chi_R}{2m}\right)^{\frac{4}{2-\sigma}}.
\ee
The scalar field equation \eqref{AF3} reads
\be\label{AF7}
\ddot{\chi}_R+3H\dot\chi_R=-\frac{\partial V}{\partial\chi_R}
\left(1-\frac{6H^2+3\dot H}{\mu^2}\right).
\ee
We observe that for $H^2\approx \mu^2/3$ the prefactor of the potential derivative has a sign opposite to the case of a scalar field in a flat background, such that $\chi_R$ increases with increasing time. 

Eq. \eqref{AF4} can be written in the form 
\ba\label{AF8}
36 CH\ddot{H}&=&18C(\dot{H}^2-6H^2\dot{H})+V
\left(1-\frac{3H^2}{\mu^2}\right)\nn\\
&&+\frac12\dot{\chi}^2_R-\frac{3H\dot\chi_R}{\mu^2}\frac{\partial V}{\partial \chi_R}.
\ea
We observe that eqs. \eqref{AF7} and \eqref{AF8} can be directly derived from the effective action \eqref{2E}, \eqref{2F} for the renormalized scalar field,
\be\label{AF9}
\Gamma=\int_x\sqrt{g}
\left\{\frac12\partial^\mu\chi_R\partial_\mu\chi_R+V(\chi_R)-\frac{V(\chi_R)}{2\mu^2}R-\frac C2 R^2\right\}.
\ee
For a general effective action \eqref{1}, \eqref{1A} the relation between $\chi_R$ and $\chi$ can be extended to arbitrary $B$ in the form 
\be\label{AF10}
\frac{\partial \chi_R}{\partial\chi}=\sqrt{B-6}.
\ee
This coincides with eq. \eqref{AF1} if we replace $B$ by $B-6$.

\subsection{Numerical solutions}

We have investigated numerically the coupled system of differential equations \eqref{AF7}, \eqref{AF8}, \eqref{AF6} for constant $\sigma<2$, starting at some time $t_0$ with ``initial conditions'' for $\chi_R,\dot\chi_R, H$ and $\dot H$. For positive $H(t_0)$ we find that for a large range of initial values for $\dot\chi_R$ and $\dot H$ both $\chi_R$ and $H$ approach very rapidly almost constant values. After this first stage one observes a second stage of slow evolution according to the approximate equations
\ba\label{AF11}
\dot\chi_R&=&-\frac{1}{3H}\frac{\partial V}{\partial\chi_R}
\left(1-\frac{6H^2}{\mu^2}\right),\nn\\
\dot H&=&\frac{V}{108CH^2}\left(1-\frac{3H^2}{\mu^2}\right).
\ea
Part of the memory of the initial conditions has been lost at this stage. Finally, the evolution speeds up if $V$ and $\partial V/\partial \chi_R$ become large enough, typically for $\chi_R$ of the order $2m/(2-\sigma)$. At this point $B$ is of the order one, roughly corresponding to the end of an inflationary epoch. 

This behavior is illustrated in Figs. 4 and 5 which show the time evolution of $\chi_R$ and $H$ for two different initial values. Parameters are $\sigma=1.5,~m/\mu=3,~C=0.05$. All quantities are expressed in units of appropriate powers of $\mu$. We observe an extremely slow evolution, recalling that the time unit $\mu^{-1}$ amounts to $10^{10}$ yr. For realistic larger ratios of $m/\mu$ this would be even more extreme. We have compared our numerical results with solutions of the approximate equations \eqref{AF11}. The difference is not visible in these plots. The green curve corresponds to the scaling solution \eqref{B.7C} with $H_0=\mu/\sqrt{3}$, visible by the constant value of $H$. The other two solutions move towards this solution.

\begin{figure}[h!tb]
\centering
\includegraphics[scale=0.5]{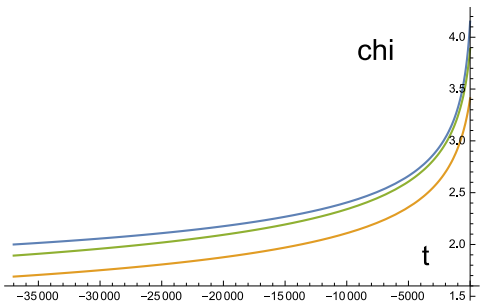}
\caption{Slow time evolution of the renormalized scalar field $\chi_R$. For a large range of initial parameters the family of such asymptotic solutions is approached very rapidly.}
\label{chi1}
\end{figure}

\begin{figure}[h!tb]
\centering
\includegraphics[scale=0.5]{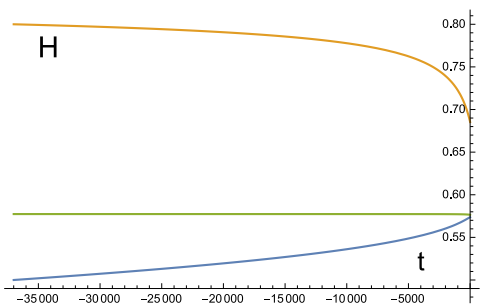}
\caption{Slow time evolution of the Hubble parameter.}
\label{H1}
\end{figure}

For an analytic discussion of eq. \eqref{AF11} we may introduce a new evolution variable $z$ by 
\be\label{AF12}
\frac{dz}{dt}=V,
\ee
such that the equations \eqref{AF11} read
\ba\label{AF13}
\frac{\partial\chi^2_R}{\partial z}&=&-\frac{8}{3(2-\sigma)}
\left(\frac{1}{H}-\frac{6H}{\mu^2}\right),\nn\\
\frac{\partial H}{\partial z}&=&\frac{1}{108C}
\left(\frac{1}{H^2}-\frac{3}{\mu^2}\right).
\ea
The evolution of $H$ has a fixed point at $H=\mu/\sqrt{3}$ that is approached for increasing $z$. Close to the fixed point $\chi_R$ increases monotonically with $z$,
\be\label{AF14}
\chi_R=\left(\frac{8z}{\sqrt{3}(2-\sigma)\mu}\right)^{\frac12}.
\ee
On the other hand, for $H<\mu/\sqrt{6}$ one observes a decrease of $\chi_R$ with increasing $z$.

In particular, for $C>0$ and $H\ll \mu$ one finds approximately
\be\label{AF15}
H(z)=\left(\frac{z}{36C}\right)^{\frac13},
\ee
and
\be\label{AF16}
\chi^2_R=\chi^2_{R0}-\frac{4}{2-\sigma}(36C)^{\frac13}z^{\frac23}.
\ee
Since $\chi^2_R$ must be positive eq. \eqref{AF16} holds only for $z<z_{\rm max}$, 
\be\label{AF17}
\z =\left(\frac{2-\sigma}{4}\right)^{3/2}(36C)^{-\frac12}\chi^3_{R0}.
\ee
For $z$ approaching $\z$ the time diverges according to 
\ba\label{AF18}
\frac{dz}{dt}=V&=&A\left(\z^{\frac23}-z^{\frac23}\right)^{\frac{2}{2-\sigma}}\nn\\
&=&A\left(\frac23\z^{-\frac13}\right)^{\frac{2}{2-\sigma}}
(\z-z)^{\frac{2}{2-\sigma}},
\ea
which implies
\be\label{AF19}
t=t_0+\tilde A(\z-z)^{-\frac{\sigma}{2-\sigma}}.
\ee
Accordingly, the Hubble parameter reaches a maximal value
\be\label{AF20}
H_{\rm max}=\left(\frac{\z}{36C}\right)^{1/3}=\left(\frac{2-\sigma}{144 C}\right)^{\frac12}\chi_{R0}.
\ee
Of course, this type of solution is a reasonable approximation only for $H_{\rm max}\ll\mu$. This type of solution corresponds to the case of a stable solution $\chi=0$, $H=H_0$, which is approached for $t\to\infty$. Alternatively, $H$ may exceed the value $H_{\rm cr}=\mu/\sqrt{6}$ for $z<\z$. From there on $\chi_R$ increases and both $\chi_R$ and $H$ move towards the scaling solution \eqref{B.7C}. 

At this point we can classify the behavior of the numerical solutions of the field eqs. \eqref{AF7}, \eqref{AF8} in terms of the asymptotic solutions \eqref{B.7D}, \eqref{B.16A}. These solutions form a family of attractor solutions provided $H_0>\mu/\sqrt{6}$. For a very large class of initial conditions the general solution approaches very rapidly this family of attractor solutions. The attractor solutions can be  extended to the infinite past, while this does not hold for neighboring solutions. The attractor solutions obey the approximate field equations \eqref{AF11}. The slow evolution according to these equations entails an approach of all attractor solutions towards the particular scaling solution \eqref{B.7C}. This slow evolution ends once $t$ is sufficiently close to $t_c$ such that $B$ reaches a value around one for $\chi$ around $m$. 

We show the fast approach to the family of attractor solutions in figs. 6 and 7 which compare the time evolution of $\chi_R$ and $H$ for initial conditions given at some arbitrary time $t_0$ which the scaling solution \eqref{B.7C}. (Parameters are again $\sigma=1.5,~ m/\mu=3,~C=0.05$.) As compared to the slow subsequent evolution, the approach to the family of scaling solutions is very rapid. 

\subsection{Crossing smoothly the big bang}

For fig. 6 we have chosen initial conditions such that $\chi_R(t)$ switches from negative to positive values. From eq. \eqref{AF1} we infer that also $\chi$ crosses zero at this moment $(\sigma<2)$. A vanishing value of $\chi$ corresponds to the big bang singularity in the Einstein frame. We have therefore established an explicit solution of field equations which cross from a pre-big-bang regime to an after-big-bang regime. In the freeze frame this crossing is completely regular. No singularity appears in the field equations for a vanishing field value of $\chi$. The big bang singularity in the Einstein frame is a pure artefact of the choice of fields which becomes singular for $\chi=0$. 

In the freeze frame it takes only a finite interval in physical time (as measured by the number of oscillations of wave functions) in order to cross from negative to positive values of $\chi$. Since physical time does not depend on the frame, we conclude that the big bang singularity in the Einstein frame is reached for a finite physical time for this type of solution. This contrasts to the asymptotic solutions where the big bang is reached only at an infinite interval of physical time in the past.

\begin{figure}[h!tb]
\centering
\includegraphics[scale=0.5]{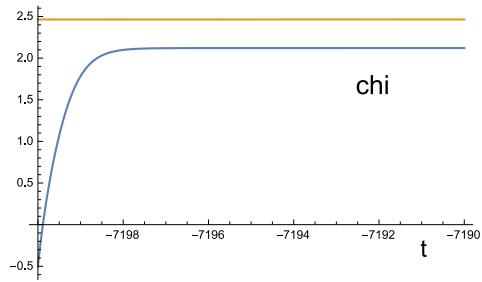}
\caption{Fast approach of the renormalized scalar field $\chi_R$ to the family of asymptotic solutions. In the Einstein frame the big bang singularity is crossed when $\chi_R$ goes through zero.}
\label{chi2in}
\end{figure}

\begin{figure}[h!tb]
\centering
\includegraphics[scale=0.5]{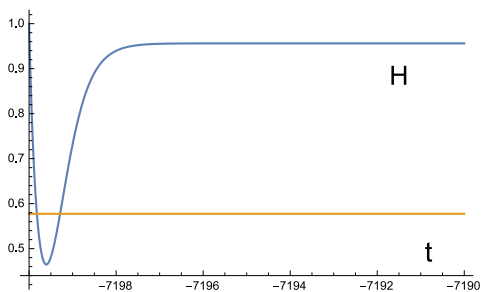}
\caption{Fast approach of the Hubble parameter to the almost constant value of the asymptotic solution.}
\label{H2in}
\end{figure}

\section*{Appendix D: Two scalar field description}
\renewcommand{\theequation}{D.\arabic{equation}}
\setcounter{equation}{0}

In this appendix we explain several features of the solutions discussed in appendix C in terms if an equivalent formulation with two scalar fields. We also discuss the possibility of a periodic crossing of the ``big bang singularity'' for the initial cosmology of our model. 

\setcounter{subsection}{0}
\subsection{Formulation with two scalar fields}

The higher curvature invariant $\sim C$ in eq. \eqref{C1a} entails an additional physical scalar degree of freedom if $C$ is a positive constant. (For the model \eqref{1}, \eqref{1A} we specialize to constant $C>0$, while we omit a constant $D$ since it does not contribute to the field equations.) Some aspects of the behavior of primordial cosmology become more apparent if we transform the effective action \eqref{C1a} to an equivalent model with two scalar fields. 

For this purpose we add to $\Gamma$ a term 
\be\label{A1}
\Delta\Gamma=\frac{C}{2}\int_x\sqrt{g}(\phi-R)^2,
\ee
such that
\ba\label{A2}
\Gamma'&=&\Gamma+\Delta\Gamma=\int_x\sqrt{g}
\left\{\frac{K}{2}\partial^{\mu}\chi\partial_{\mu}\chi+\mu^2\chi^2\right.\nn\\
&&\left.-\frac{1}{2}(\chi^2+2C\phi)
R+\frac{C}{2}\phi^2\right\}, 
\ea
where
\be\label{A2A}
K=B-6.
\ee
Inserting the solution of the field equation for $\phi$, namely $\phi=R$, into $\Gamma'$ we recover $\Gamma$. The field equations derived from $\Gamma$ and $\Gamma'$ are therefore equivalent. The term $\sim CR^2$ is effectively replaced by the interactions of the new scalar field $\phi$. The field equations are now of second order.

Employing a transformation of the metric,
\be\label{A3}
g_{\mu\nu} = \frac{M^2}{\chi^2+2C\phi}g'_{\mu\nu}=\bar{w}^2g'_{\mu\nu},
\ee
one has
\be\label{A4}
R=\bar{w}^{-2}\left\{R'-6D^2\ln\bar{w}-6\partial^{\mu}\ln\bar{w}~\partial_{\mu}\ln\bar{w}\right\},
\ee
where $R'$ and the covariant derivatives on the r.h.s. are formed with the metric $g'_{\mu\nu}$. In the new frame the effective action reads
\be\label{A5}
\Gamma'=\int_x\sqrt{g'}\left\{-\frac{M^2}{2}R'+V'+{\cal L}'_{kin}\right\},
\ee
with
\be\label{C6A}
V'=\frac{M^4}{(\chi^2+2C\phi)^2}
\left(\mu^2\chi^2+\frac{C}{2}\phi^2\right),
\ee
and
\ba\label{A6}
{\cal L}'_{kin}&=&\frac{M^2}{2(\chi^2+2C\phi)^2} \left\{ \big((K+6)\chi^2+2C\phi\big)\partial^{\mu}\chi\partial_{\mu}\chi \right. \nn\\
&&+\left.12C\chi\partial^{\mu}\chi\partial_{\mu}\phi+6C^2\partial^{\mu}\phi\partial_{\mu}\phi \right\}.
\ea
It describes standard Einstein gravity (with $M$ the reduced Planck mass) coupled to two scalar fields $\chi$ and $\phi$. 

It is instructive to study the potential $V'(\chi,\phi)$ as a function of $\phi$ for fixed $\chi$. For $\phi\rightarrow\infty$ it approaches the constant
\be\label{A7}
V'_{\infty} = \lim\limits_{\phi\rightarrow\infty} V'(\chi,\phi)=\frac{M^4}{8C},
\ee
while for $\phi\rightarrow 0$ one has
\be\label{A8}
\lim\limits_{\phi\rightarrow0}V'(\chi,\phi)=\frac{M^4\mu^2}{\chi^2}.
\ee
With
\be\label{A9}
\frac{\partial V'}{\partial \phi}= \frac{CM^4\chi^2(\phi-4\mu^2)}{(\chi^2+2C\phi)^3}
\ee
we observe a partial minimum for 
\be\label{A10}
\phi_{min}=4\mu^2.
\ee
We will see that this partial minimum corresponds to the scaling solution \eqref{B.7C}.

For $\phi=4\mu^2$ one has a valley in the potential $V'(\chi,\phi)$. Along this valley the potential
\be\label{A11}
V'(\chi,\phi_{min})=\frac{\mu^2M^4}{\chi^2+8C\mu^2}
\ee
vanishes for $\chi\rightarrow\infty$. As time increases $\phi$ typically settles at $\phi_{min}$, such that the kinetic term becomes 
\be\label{A12}
{\cal L}'_{kin}[\chi,\phi_{\rm min}]
=\frac{M^2((K+6)\chi^2+8C\mu^2)}{2(\chi^2+8C\mu^2)^2}\partial^{\mu}\chi\partial_{\mu}\chi.
\ee
As a result of the decaying potential \eqref{A11} the behaviour of the cosmological solution is then characterized by an increase of $\chi$ to values $\chi^2\gg\mu^2$. Asymptotically,
for $t\rightarrow\infty$, $\chi$ increases to infinity and the effective cosmological constant $V'$ vanishes.

For small $|C\phi/\chi^2|\ll1,$ and $C\mu^2/\chi^2\ll1$, we can expand 
\be\label{A12A}
V'= \frac{M^4\mu^2}{\chi^2}+\frac{CM^4}{2\chi^4}(\phi^2-8\mu^2\phi),
\ee
as well as
\ba\label{A12B}
{\cal L}'_{kin}&=&\frac{(K+6)M^2}{2\chi^2}\partial^{\mu}\chi\partial_{\mu}\chi\\
&&+\frac{3C^2M^2}{\chi^4}\partial^{\mu}\phi\partial_{\mu}\phi
+\frac{6CM^2}{\chi^3}\partial^{\mu}\chi\partial_{\mu}\phi.\nn
\ea
In the limit where derivatives of $\chi$ can be neglected $\phi$ describes a stable scalar field. Depending on the relative size of the kinetic term it will perform damped oscillations around its minimum at $\phi=4\mu^2$ and finally settle there, or it will reach this minimum in an overdamped approach. Realizing that $\phi$ equals the curvature scalar in the freeze frame, $R\approx12H^2$, we recover indeed the leading behaviour of the scaling solution in the freeze frame \eqref{B.7C}, $H\approx \sqrt{R/12}=\sqrt{\phi_{min}/12} = \mu/\sqrt{3}$. The time evolution of $\chi$ induces additional terms in the field equation for $\phi$ which do not, however, affect the stability of the model. For $\chi\gg\sqrt{C}\mu$ we can to a good approximation neglect the influence of the term $\sim CR^2$ and set $C=0$.

We next turn to solutions in primordial cosmology that feature small values of $\chi$. Expanding $V'$ in powers of $\chi^2/2C\phi$
\be\label{A13}
V'=\frac{M^4}{8C}+\frac{M^4}{8C^2\phi^2}(2\mu^2-\phi)\chi^2+\cdots,
\ee
we find a positive quadratic term $\sim\chi^2$ for $\phi<2\mu^2$, while it turns negative for $\phi>2\mu^2$. The field equations admit the solution
\be\label{C.12}
\chi=0~,~\phi=\phi_0=\text{ const.}~,~H^2=\frac{M^2}{24C}.
\ee
Recalling the relation between $\phi$ and the curvature scalar in the freeze frame this corresponds to the family of asymptotic solutions \eqref{B.7A}. For a stable kinetic term this solution is stable for $\phi_0<2\mu^2$ and unstable for $\phi_0>2\mu^2$. In accordance with the findings of appendix C (cf. eq. \eqref{B.7B}), the scalar field $\chi$ increases (for increasing time) for $\phi_0=12H^2_0>2\mu^2$, while it goes to zero for $\phi_0=12H^2_0<2\mu^2$. We observe the difference of the Hubble parameter in the freeze frame (denoted here by $H_0$) from the one in the Einstein frame with two scalars, as in eq. \eqref{C.12}. All values of $H_0$ in the freeze frame correspond to the same $H$ in eq. \eqref{C.12}, while the corresponding value of $\phi_0$ reflects $H_0$. 

The detailed discussion of the dynamics for small $\chi$ requires an understanding of the kinetic term. Its qualitative behavior depends on $\sigma$. For $\sigma<2$ we can neglect in eq. \eqref{A6} the term $\sim (K+6)\chi^2$ as compared to $2C\phi$. In the region of small enough $\chi$ the kinetic term becomes then block diagonal
\be\label{A14}
{\cal L}'_{kin}=\frac{3M^2}{4\phi^2}\partial^{\mu}\phi\partial_{\mu}\phi + \frac{M^2}{4C\phi}\partial^{\mu}\chi\partial_{\mu}\chi.
\ee
In this form we can find a field basis with canonical kinetic terms rather easily. 

For $\phi>0$ we define
\ba\label{A15}
&&\varphi=\sqrt{\frac{3}{2}}M\ln(\phi/\mu^2), \nn\\
&&\tilde{\chi}=\frac{M}{\sqrt{2C\phi}}\chi,
\ea
such that the kinetic term \eqref{A14} takes a canonical form,
\be\label{A16}
{\cal L}'_{kin}=\frac{1}{2}\partial^{\mu}\varphi\partial_{\mu}\varphi +\frac{1}{2}\partial^{\mu}\tilde{\chi}\partial_{\mu}\tilde{\chi},
\ee
with mixed derivatives $\sim\tilde{\chi}\partial^{\mu}\varphi\partial_{\mu}\tilde{\chi}$ again subleading. In terms of these fields the potential reads 
\be\label{A17}
V'=\frac{M^4}{8C}+\frac{M^2}{4C}\left[2\exp\Big(-\sqrt{\frac{2}{3}}\frac{\varphi}{M}\Big)-1\right]\tilde{\chi}^2.
\ee
For $\varphi/M<\sqrt{3/2}\ln 2$ the quadratic terms $\sim\tilde\chi^2$ is positive, while it turns negative for larger values of $\varphi$. The cosmologies discussed in the main text, where the solution $H=H_0,\chi=0$ is unstable such that any perturbation in $\chi$ increases, correspond to a negative quadratic term.

For $\sigma>2$ the term $\sim (K+6)\chi^2$ in eq. \eqref{A6} diverges for $\chi\to 0$. The kinetic term takes now for $\chi\to 0$ the approximate form
\be\label{D22A}
{\cal L}'_{\rm kin}=\frac{3M^2}{4\phi^2}\partial^\mu\phi\partial_\mu\phi+
\frac{M^2 m^\sigma}{8C^2\phi^2}
\chi^{2-\sigma}\partial^\mu\chi\partial_\mu\chi.
\ee
The field basis with canonical kinetic terms retains eq. \eqref{A15} for $\varphi$, while the other degree of freedom is given by 
\be\label{D22-A}
\hat \chi=\frac{Mm^{\frac\sigma2}}{|C\phi|(4-\sigma)}\chi^{\frac{4-\sigma}{2}}.
\ee
For $\sigma>4$ small values of $\chi$ correspond to large negative values of $\hat \chi$, similar to $\chi_R$ in eq. \eqref{PF3}, while for $2<\sigma<4$ one has $\hat\chi\to 0$ for $\chi\to 0$. The potential \eqref{A13} reads
\ba\label{D22B}
&&V'=\frac{M^4}{8C}+\frac{M^4}{8C^2}\big|C(4-\sigma)\big|^{\frac{4}{4-\sigma}}
\left(\frac{\mu}{m}\right)^{\frac{2\sigma}{4-\sigma}}\\
&&\exp 
\left(\frac{\sigma}{4-\sigma}\sqrt{\frac23}\frac\varphi M\right)
\left[2\exp \left(-\sqrt{\frac23}\frac\varphi M\right)-1\right] 
\left|\frac{\hat\chi}{M}\right|^{\frac{4}{4-\sigma}}.\nn
\ea
The stability in the $\hat\chi$-direction depends on $\varphi$ in a way similar to the case $\sigma<2$, but the dependence on $\hat \chi$ is no longer harmonic. For the example $\sigma=3$ the $\hat\chi$-dependent part is quartic, $\Delta V'\sim \hat \chi^4$. For $\sigma>4$ one encounters negative powers of $\hat \chi$. 

\subsection{Periodic crossing of ``big bang singularity''?}

In the Einstein frame an oscillation of $\chi$ corresponds to a periodic crossing of the big bang singularity. One may wonder if such a behavior is possible within our model. For this purpose it is instructive to discuss solutions where $\varphi$ is small enough such that $\chi=0$ becomes attractive as time increases. We focus on $\sigma<2$ with a harmonic potential \eqref{A17} for $\tilde \chi$. We  are interested to find out under which conditions there are oscillations around $\tilde \chi=0$. For this purpose we investigate cosmological solutions that are specified by fixing $\tilde{\chi},\dot{\tilde{\chi}}, \varphi$ and $\dot{\varphi}$ at some time $t_0$. We choose solutions for which $\tilde{\chi}^2(t_0)/M^2\ll1$, such that the approximation \eqref{A13} is valid, for example $\tilde{\chi}(t_0)=0,\dot{\tilde\chi}(t_0)=c_0M^2$. We also consider $0<\phi(t_0)<\mu^2$, corresponding to negative $\varphi(t_0)$, and take $\dot{\varphi}(t_0)=0$. For not too large $c_0$ the potential $V'$ is essentially constant, 
corresponding to an exponential expansion of the scale factor with 
\be\label{A18}
H=\pm\frac{M}{\sqrt{24C}}.
\ee
The scalar field $\tilde\chi$ could oscillate around zero, according to the effective field equation
\ba\label{A19}
&&\ddot{\tilde \chi }+ 3H\dot{\tilde\chi}+\tilde m^2(\varphi)\tilde \chi=0,\nn\\
&&\tilde m^2(\varphi)=\frac{M^2}{C}
\left[\exp \Big(-\sqrt{\frac23}\frac{\varphi}{M}\Big)-\frac12\right],
\ea
provided that $\tilde m$ changes only slowly on the time scale of the inverse frequency. One obtains
\ba\label{A20}
&&\tilde\chi=\frac{c_0M^2}{\omega_0}\exp 
\left(-\frac{3H(t-t_0)}{2}\right)\sin \big(\omega(\varphi)(t-t_0)\big),\nn\\
&&\omega^2(\varphi)=\frac{M^2}{C}
\left[\exp \Big(-\sqrt{\frac23}\frac{\varphi}{M}\Big)-\frac{9}{32}\right].
\ea
with $\omega_0=\omega(\varphi_0)=\omega\big(\varphi(t_0)\big)$. For negative $\varphi$ these oscillations are fast compared to the Hubble parameter
\be\label{A21}
\frac{\omega^2}{H^2}=24\exp 
\Big(-\sqrt{\frac23}\frac{\varphi}{M}\Big)-\frac{27}{4}.
\ee
They are damped for positive $H$ and increase for negative $H$. 

During the oscillations the average value of $\tilde\chi^2$ is given by 
\be\label{A22}
\langle\tilde\chi^2\rangle =\frac{c^2_0M^4}{2\omega^2_0}\exp \big(-3H(t-t_0)\big).
\ee
Inserting this into the potential \eqref{A17} yields an effective exponential potential for $\varphi$ which can be approximated for sufficiently negative $\varphi$ by 
\be\label{A23}
V'_{\rm eff}(\varphi)=\frac{M^4}{8C}+\frac{c^2_0M^4}{4}\exp 
\left[-\sqrt{\frac23}\frac{\varphi-\varphi_0}{M}-3H(t-t_0)\right].
\ee
The field equation for $\varphi$,
\be\label{C.29}
\ddot{\varphi}+3H\dot{\varphi}=\frac{c^2_0M^3}{\sqrt{24}}\exp 
\left(-\sqrt{\frac23}\frac{\varphi-\varphi_0}{M}-3H(t-t_0)\right),
\ee
can be solved approximately by neglecting the term $\sim \ddot{\varphi}$. This results in a relative change of $\tilde m^2(\varphi)$,
\ba\label{C.30}
&&\partial_t\ln\tilde m^2(\varphi)=-\frac{\sqrt{6C}c^2(t)M}{9}\exp 
\left(-\sqrt{\frac23}\frac{\varphi-\varphi_0}{M}\right)-3H,\nn\\
&&c^2(t)=c^2_0\exp\big(-3H(t-t_0)\big).
\ea
For consistency of our computation of oscillations the relative change per oscillation period, $\omega^{-1}\partial_t\ln \tilde m^2$, has to be small compared to one. Near $t_0$ it is indeed small provided that $c^2_0$ is small enough. We may take $\varphi\approx \varphi_0$ and use $\exp(\varphi_0/\sqrt{6}M)\ll \sqrt{3}$, as required by $\omega\gg H$. For a given value of $\varphi_0$ one has an upper bound on $c^2_0$ such that oscillations take place. This bound gets higher if $\varphi_0$ moves to smaller values.

Our discussion shows that a periodic crossing of the big bang singularity may indeed occur in our model. It is not clear under what circumstances an initial oscillatory behavior can turn into a realistic cosmology at later times. For positive $H$ the amplitude of the oscillation of $\tilde\chi$ decreases with increasing time. If this behavior continues scale symmetry is asymptotically realized for large $t$ and all particles are massless. The oscillatory behavior with decreasing amplitude could be stopped, however, if $\varphi$ increases towards positive values such that $V(\varphi,\tilde \chi$) becomes unstable in the $\tilde \chi$-direction. 

\section*{Appendix E: Variable gravity with additional constant coefficient of curvature term}
\renewcommand{\theequation}{E.\arabic{equation}}
\setcounter{equation}{0}
In this appendix we investigate a general class of variable gravity models with effective action 
\ba\label{B1}
\Gamma&=&\int_x\sqrt{\tilde g}
\Big\{-\frac12\xi
(\tilde\chi^2+\tilde m^2)\tilde R\nn\\
&&+\frac{\tilde K(\tilde\chi)}{2}\partial^\mu\tilde \chi\partial_\mu\tilde\chi+\tilde V(\tilde\chi)\Big\},
\ea
with constant coefficients $\xi$ and $\tilde m^2$. This addresses the question to what extent the omission of a curvature term with constant coefficient in the action \eqref{1} is a crucial ingredient for our scenario or rather a matter of convenience. For easy comparison we will map the action \eqref{B1} to the kinetial crossover form \eqref{1} by suitable field transformations. In particular, we will find that the simple case of a constant potential $\tilde V(\tilde\chi)=\tilde V_0$ or, more generally, $\tilde V(\tilde\chi\to 0)\to\tilde V_0$, can realize a crossover scenario very similar to the one discussed in the main text. 

An important feature of the action \eqref{B1} is most easily understood in the Einstein frame, where the potential becomes
\be\label{B2}
V_E(\tilde\chi)=\frac{M^4\tilde V(\tilde \chi)}{\xi^2(\tilde \chi^2+\tilde m^2)^2}.
\ee
Depending on the form of $\tilde V(\tilde \chi)$ the potential $V_E(\tilde \chi)$ may not be a monotonic function. For the example $\tilde V=\mu^2\tilde\chi^2$ one finds a maximum of $V_E$ for $\tilde \chi=\tilde m$. Such models  can still provide for a satisfactory description of inflation if the initial value of $\tilde \chi$ is close to the maximum of $V_E$. Nevertheless, such a setting destroys somewhat the simplicity and beauty of a crossover from a fixed point at $\tilde \chi=0$ to another one for $\tilde \chi\to\infty$. 

For a large class of potentials $\tilde V$ no maximum of $V_E$ occurs and $V_E(\tilde \chi)$ is a monotonic function. For example, this happens for a constant $\tilde V(\tilde \chi)=\tilde V_0$ or for $\tilde V=\tilde V_0+\mu^2\chi^2$ if $\tilde V_0$ is large enough, cf. ref. \cite{CI}. We will see that for this class of models the kinetial crossover is very similar to the one discussed in the main text. 

Using the transformation
\ba\label{B3}
\tilde g\mn&=&\frac{\chi^2}{\xi(\tilde \chi^2+\tilde m^2)}
g\mn=
\frac{\chi\mu}{\sqrt{\tilde V}}g\mn,\nn\\
\tilde \chi^2&=&\frac{\sqrt{\tilde V}\chi}{\xi\mu}-\tilde m^2,
\ea
one finds indeed the effective action \eqref{1}, where
\be\label{B4}
B=\frac32+
\left(\frac{\tilde K\tilde X}{\xi}+\frac38\tilde A^2\right)\tilde D^2+\frac32\tilde A\tilde D,
\ee
with 
\ba\label{B5}
\tilde A&=&\frac{\partial \ln \tilde V}{\partial \ln \tilde \chi}~,~\tilde X=\frac{\tilde \chi^2}{\tilde\chi^2+\tilde m^2}=
1-\frac{\xi\mu \tilde m^2}{\sqrt{\tilde V}\chi},\nn\\
\tilde D&=&\frac{\partial \ln \tilde \chi}{\partial \ln\chi}=\frac{2}{4\tilde X-\tilde A}.
\ea
To be specific, we take
\be\label{B6}
\tilde V=\tilde \mu^{4-\tilde A}\tilde \chi^{\tilde A}
\ee
with constant $\tilde A$. 

For $0<\tilde A< 4$ we observe that $\tilde D$ and therefore $B$ becomes singular for $\tilde X_c=\tilde A/4$. This singularity corresponds to the maximum of the potential in the Einstein frame $V_E\sim \tilde V/(\tilde \chi^2+\tilde m^2)^2$. One may employ the freeze frame for solutions where $\tilde X\neq \tilde X_c$ and note the natural occurrence of large values for $B$ for $\tilde X$ near $\tilde X_c$. Inflationary models of this type have been discussed in ref. \cite{CWET}.

We observe that $B$ vanishes for $\tilde \chi\to 0,~\tilde X\to 0$, according to 
\ba\label{B7}
B&=&\frac32\left(1+\frac{\tilde A\tilde D}{2}\right)^2+\frac{\tilde K\tilde X\tilde D^2}{\xi}\nn\\
&=&\left (24\tilde X^2+\frac{4\tilde K\tilde X}{\xi}\right)
(4\tilde X-\tilde A)^{-2}.
\ea
The limit $\tilde \chi\to 0$ corresponds to $\chi\to\infty$, 
\be\label{B8}
\chi =\xi\mu^{\frac{\tilde A}{2}-1}\tilde \chi^{-\frac{\tilde A}{2}}
(\tilde \chi^2+\tilde m^2).
\ee
Using
\ba\label{B9}
\frac{\partial \ln\tilde X}{\partial \ln \chi}&=&\frac{4(1-\tilde X)}{4\tilde X-\tilde A},\\
\frac{\partial \ln B}{\partial \ln \tilde X}&=&1+\frac{6\tilde X}{6\tilde X+\tilde K/\xi}-
\frac{8\tilde X}{4\tilde X-\tilde A},\label{B10}
\ea
we may compute $\sigma=-\frac{\partial \ln B}{\partial \ln \chi}$. For $\tilde\chi\to 0$ one finds for $\tilde K(\tilde \chi\to 0)>0$ a rather large anomalous dimension $\sigma=4/\tilde A$, which increases to $\sigma=8/\tilde A$ if $\tilde K(\tilde \chi\to 0)=0$. Solutions where $\tilde \chi$ approaches zero for increasing time do not provide for an acceptable description of dark energy. 

On the other hand, for $\tilde A<4$ also the limit $\tilde \chi\to \infty$ corresponds to $\chi\to\infty$. In this limit one has $\tilde X\to 1,~\tilde D\to\frac{2}{4-\tilde A}$ and therefore
\be\label{B11}
B=\frac{4(6+\tilde K/\xi)}{(4-\tilde A)^2}.
\ee
Small values of $B$, as required for realistic dark energy, can be achieved if $\tilde K(\tilde \chi\to\infty)$ reaches $-6\xi$ or a value slightly larger. Thus realistic cosmologies correspond to solutions where $\tilde \chi$ increases from a value close to $\tilde \chi_c=m\sqrt{\tilde A/(4-\tilde A)}$ to infinity. 

For a constant $\tilde V$, i.e. $\tilde A=0$, one finds
\be\label{B12}
B=\frac32+\frac{\tilde K}{4\xi\tilde X}.
\ee
The freeze frame is valid for the range $\chi>\chi_c,\chi_c=\xi\mu \tilde m^2\sqrt{\tilde V}$, which corresponds to $\tilde\chi^2>0$. Again, one finds large $B$ in the vicinity of the divergence for $\tilde \chi^2\to 0$, and corresponding models of inflation \cite{VG}. The leading behavior of $B$ for $\tilde\chi\to 0$ is given by
\ba\label{D.12A}
B&=&\frac{\tilde K}{4\xi}\frac{\tilde m^2+\tilde \chi^2}{\tilde\chi^2}=\frac{\tilde K}{4\xi}\frac{\chi}{\chi-\chi_c},\nn\\
\chi_c&=&\frac{\xi\mu\tilde m^2}{\tilde V}.
\ea
The corresponding value for $\sigma$ diverges for $\chi\to\chi_c$,
\be\label{D.12B}
\sigma=-\frac{\partial\ln B}{\partial\ln \chi}=\frac{\chi_c}{\chi-\chi_c}-\frac{\partial\ln \tilde K}{\partial\ln \chi}=
\frac{\chi_c-\frac{\chi}{2}\frac{\partial\ln\tilde K}{\partial\ln\tilde\chi}}{\chi-\chi_c}.
\ee
For $\chi>\chi_c$ it drops rather rapidly, however. Estimating for constant $\tilde K$ the value $\sigma(N)$ at horizon crossing according to eq. \eqref{23C} yields
\be\label{D.12C}
\sigma(N)=\left(\frac{8\zeta N}{\tilde K}-1\right)^{-1}.
\ee
For suitable values of $\tilde K/\xi$ one may obtain values of $\sigma(N)$ comparable to the ones discussed in the main text even for constant $\tilde K$. On the other hand, for $\partial\ln \tilde K/\partial\ln\tilde \chi=-\tilde\sigma$ and horizon crossing in the region $\chi(N)\gg\chi_c$, one finds
\be\label{D.12D}
\sigma=\frac{\tilde\sigma}{2}.
\ee
This value constitutes a lower bound for $\sigma$. 

The properties of the UV-fixed point for $\tilde m^2>0$ and constant $\tilde V$ differ from the ones discussed in sect. \ref{Fixed points and crossover}. For $\xi\tilde m^2=c_1\mu^2$ and $\tilde V=c_2\mu^4$ the dimensionless ratios $c_1$ and $c_2$ take finite constant values for $\mu/\chi\to \infty$. This type of UV-fixed point resembles the one found in the flow of the effective average action \cite{Rev,Per} if one identifies $\mu$ with the renormalization scale $k$. 

For $\tilde A<0$ the potential $\tilde V$ diverges for $\tilde \chi\to 0$. The limit $\tilde \chi\to 0$ corresponds to $\chi\to 0$, while for $\tilde \chi\to\infty$ one has $\chi\to \infty$. This setting is very close to the one discussed in the main text. A divergence of $B$ for $\chi\to 0$ requires an increase of $\tilde K$ for $\tilde \chi\to 0$ stronger than $\tilde \chi^{-2}$. Large values of $B$ arise rather naturally for small $\chi$ if $|\tilde A|$ is small. On the other hand, small values of $B$ for $\chi\to\infty$ occur if $\tilde K(\tilde \chi\to\infty)/\xi$ is close to $-6$. 

We conclude that for the models of eq. \eqref{B1} a realistic description of dark energy in the limit of large $\tilde \chi$ requires a kinetial close to the stability bound at $\tilde K/\xi=-6$. In the limit $\tilde \chi\to\infty$ one simply may neglect $\hat m^2$ and absorb $\xi$ by a suitable rescaling of $\tilde \chi$. For $\tilde \chi\to 0$ stability requires $\tilde K(\tilde \chi\to 0)\geq 0$ if $\hat m^2>0$. A variation of the kinetial $\tilde K(\tilde \chi)$ is therefore always required for realistic models. The large values of $B(\tilde \chi\to 0)$ (or $B(\tilde \chi\to\tilde\chi_{{\rm max}})$ in case of a maximum of $V_E$) needed for inflation can partially be induced as an effect of the term $\sim\tilde m^2 R$. The crossover scenario discussed in the main text is realized if $\tilde V(\tilde\chi\to 0)$ diverges, $\tilde A<0$. Also the behavior $\tilde V(\tilde\chi\to 0)\to$ const. $(\tilde A=0)$ can describe a crossover from an UV-fixed point at 
$\tilde \chi=0$ to an IR-fixed point for $\tilde \chi\to\infty$. In the freeze frame the UV-fixed point occurs at $\chi=\chi_c\neq 0$ in this case.

\vspace{2.0cm}
\noindent

\bibliography{inflation_quintessence_origin_of_mass}

\end{document}